\documentclass[amsmath,amssymb,aps,prd,nofootinbib]{revtex4-2}
\usepackage{amsmath}
\usepackage{color}
\usepackage{graphicx}
\usepackage{epsfig}
\usepackage{subfig}
\usepackage{bm}
\usepackage{hyperref}
\usepackage{natbib}
\usepackage{pgfplots,mathtools}
\usepackage{hyperref}
\usepackage{amsmath}
\usepackage{braket}
\usepackage{slashed}
\usepackage[compat=1.0.0]{tikz-feynman}
\usepackage{physics}
\usepackage{xcolor}

\usepackage{xfrac}
\usepackage{algorithm}
\newcommand{\be}{\begin{equation}}
\newcommand{\ee}{\end{equation}}
\newcommand{\bea}{\begin{eqnarray}}
\newcommand{\eea}{\end{eqnarray}}
\newcommand{\ba}[1]{\begin{array}{#1}}
	\newcommand{\ea}{\end{array}}
\newcommand{\nn}{\nonumber}

\newcommand{\del}{\partial}
\newcommand{\al}{\alpha}
\newcommand{\la}{\lambda}
\newcommand{\na}{\nabla}
\newcommand{\D}{\Delta}
\newcommand{\Ep}{\mathcal{E}}
\newcommand{\orcid}[1]{\href{https://orcid.org/#1}{\includegraphics[width=8pt]
		{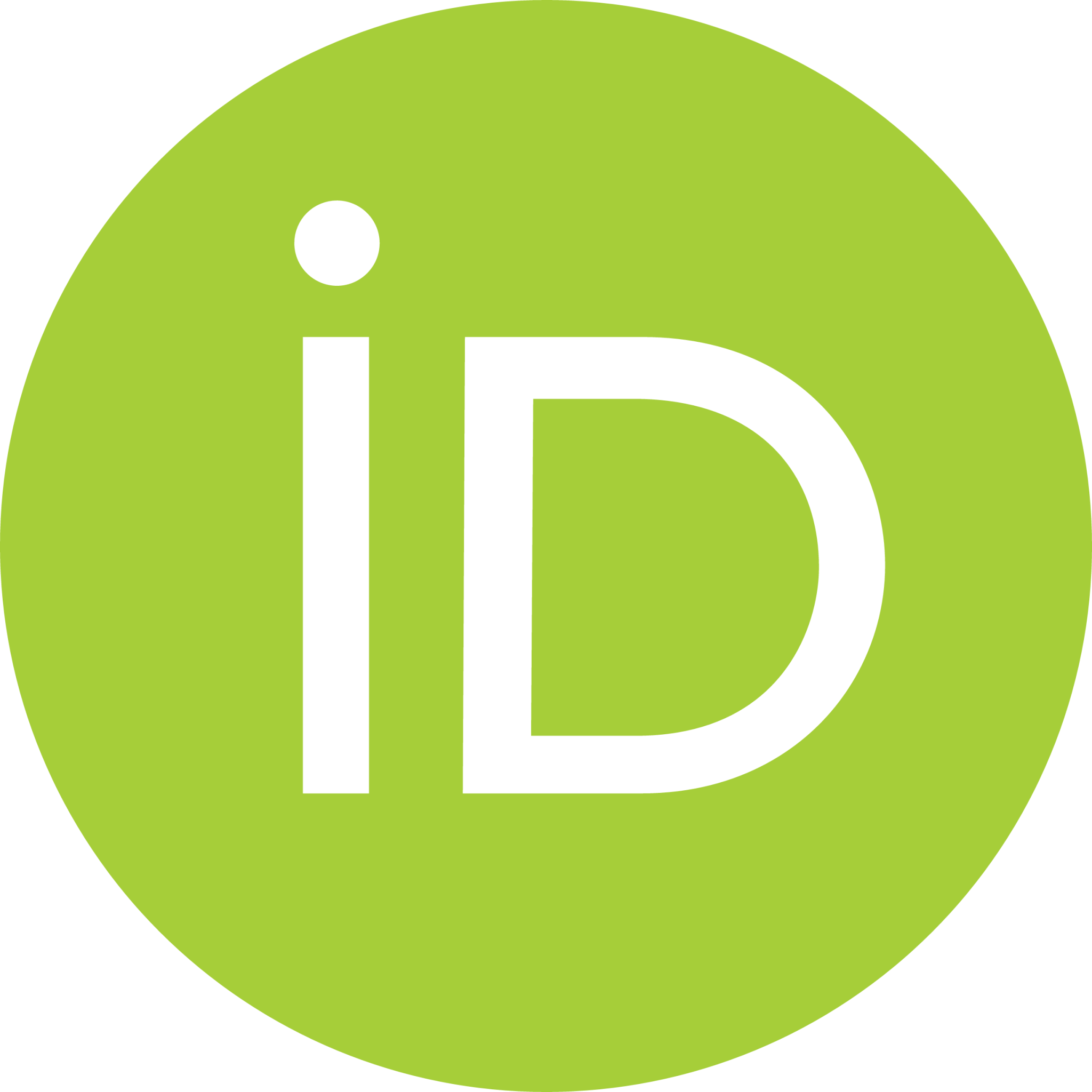}}}
\begin{document}
\title{On the Wiedemann–Franz law violation in Graphene and quark-gluon plasma systems}
\author{Ashutosh Dwibedi\orcid{0009-0004-1568-2806}$^1$, Subhalaxmi Nayak\orcid{0009-0001-0145-6785
	}$^1$, Sathe Subodh Kiran\orcid{0009-0009-4711-8755}$^2$, Sabyasachi Ghosh\orcid{0000-0003-1212-824X}$^1$, Sesha Vempati\orcid{0000-0002-0536-7827}$^1$}
\affiliation{$^1$Department of Physics, Indian Institute of Technology Bhilai, Kutelabhata, Durg, 491002, Chhattisgarh, India}
\affiliation{$^2$Department of Physics, Sardar Vallabhbhai National Institute of Technology,
	Ichchhanath, Surat, 395007, Gujarat, India}

\begin{abstract}
 A comparative study of the thermodynamic and transport properties of the ultra-relativistic quark-gluon plasma produced in heavy ion collisions with the "quasi-relativistic" massless electron-hole plasma in graphene sample has been performed. We observe that the enthalpy per net charge carriers emerges as a useful physical quantity determining the transport variables in hydrodynamic domain. Lorenz ratio is defined as thermal to electrical conductivity ratio, normalized by temperature and Lorenz number $L_{0}=\frac{\pi^{2}}{3}\left(\frac{k_{B}}{e}\right)^{2}$. The validity of the Wiedemann-Franz law can be checked by evaluating the Lorenz ratio, which is expected to be unity. We investigate the validity of the Wiedemann–Franz law by examining whether the Lorenz ratio equals unity or deviates from it. Our findings indicate that, within the fluid-based framework, the Lorenz ratio consistently leads to a violation of the Wiedemann–Franz law. This is attributed to the proportional relation between Lorenz ratio and enthalpy per net charge carriers in the fluid. Based on the experimental observation, graphene and quark-gluon plasma, both systems at a low net carrier density, violate the Wiedemann–Franz law due to their fluidic nature. However, graphene at a relatively high net carrier density obeys the Wiedemann–Franz law, followed by metals with high Fermi energy or electron density. It indicates a fluid to the non-fluid transition of the graphene system from low to high carrier density. In this regard, the fluid or non-fluid aspect of quark-gluon plasma at high density is yet to be explored by future facilities like Compressed Baryonic Matter and Nuclotron-based Ion Collider fAcility experiments.
\end{abstract}
                        
\maketitle

\section{Introduction}
	Understanding the charge carrier/quasi-particle transport under the influence of external perturbation such as electric field and/or thermal gradient is fundamentally vital in the context of device physics. The corresponding transport coefficients are electrical ($\sigma$) and thermal ($\kappa$) conductivities. Although $\sigma$ and $\kappa$ appear to be intrinsic properties, they crucially depend on various other parameters such as carrier and defect density, energy dispersion relation, Coulombic, and other scatterers, mean free path with reference to the size of the sample, temperature ~ (\textit{T}) \textit{etc.}~\cite{novoselov2004,berger2004,novoselov2005two,zhang2005experimental,ssarma2007,chen2007,dassarmaJa2008,ChenJ2008,sungjae2008,martin2008observation,GeimPonomarenko2009,Monteverde2010,Ando1998,Ando2006,DonaldN2006,Donald2007,ADAM20081022,NPeres2006,Ziegler2006,Ostrovsky2006,Schliemann2007,adam2007self,SarmaDas2007,PeresLo2007,Novikov2007,katsnelson2008electron,Kumazaki2006,Peres2006,Efetov2010,Tayari2013,EHwang2008,MunozP2012,Park2014,Sohier2014,Kim2016}. Notably, the $\sigma$ and $\kappa$ can be connected \textit{via} the Wiedemann–Franz (WF) law, which yields a constant for conventional metals and known as Lorentz number (${L_0}$)~\cite{win2024wied,ashcroft2011solid,ziman2001electrons}. Violations to the WF law are observed in graphene~\cite{AnLucas2016,Zarenia_2019,ma14112704,tu2023wiedemann,PhysRevB.106.205126,PhysRevResearch.4.043107} and quark-gluon plasma (QGP) systems~\cite{JAISWAL2015548,Sahoo:2019xjq,Rath:2019nne,singh2023effect,pradhan2023conductivity}, where the transport behavior of the carrier is not similar to that of conventional metals. Although these two systems are at the extreme ends of the chemical potential (${\mu}$) spectrum, the constituents share an exotic common property \textit{viz.} 'hydrodynamic transport'~\cite{Bandurin2016} that causes the violation of WF law. Furthermore, various physical characteristics such as masslessness, linear dispersion of energy-momentum relation ($E_{\pm}=\pm \hbar|\vec{k}| v_{F}\equiv \pm |\vec{p}| v_{F}$ and $E= |\vec{p}| c$ for graphene and QGP respectively) of charge carriers are notably similar. For graphene, the effective mass of the charge carriers vanish, and they, in many aspects, behave like ultra-relativistic Dirac particles ~\cite{neto2009electronic,RevModPhys.83.407}. This is similar to that of the constituents of QGP at the core of neutron star or in heavy ion collision experiments. On the other hand, in graphene, the ${\mu}$ can be tuned \textit{via} doping~\cite{novoselov2004} while in QGP the baryonic-${\mu}$ can be altered by changing the energy of the collision-beam~\cite{Jaiswal:2016hex}. In the context of hydrodynamics behavior, $\mu-T$ plane can be broadly classified
	into Dirac fluid (DF) or Fermi liquid (FL) regions~\cite{Landau1,Landau2,Landau3}. For the domain $\frac{\mu}{k_B T}\gg 1$, the electrons in graphene obey the well-known Fermi liquid theory~\cite{Landau1,Landau2,Landau3}, whereas, in QGP, the same may be expected at the core of neutron stars. Here ${k_B}$ is Boltzmann constant and $T$ is temperature. For $\frac{\mu}{k_B T}\ll 1$, both QGP and lightly doped graphene are identified to follow fluidic behavior. In the case of QGP, ultra-relativistic hydrodynamics is highly successful in explaining the particle spectra and anisotropic flow coefficients extracted from the experimental data, see Ref.~\cite{Jaiswal:2016hex} and the references therein. In the context of graphene various studies~\cite{levitov2016electron,krishna2017superballistic,BandurinF18,Patrick2019,sulpizio2019visualizing,ABerdyugin2019,Ella2019,ku2020imaging,crossno2016observation,block2021observation} have explored the experimental aspects of electron hydrodynamics. From theoretical viewpoints the electron hydrodynamics has also been explored focusing on various aspects like shear, hall viscosities and corresponding \textit{KSS} bound~\cite{MuSchmalian2009,NarozhnyBN2015,Hartnoll_2007}. More recently, there are also experimental evidences~\cite{crossno2016observation,majumdar2025universalityquantumcriticalflow} on the determination of $L\equiv \kappa/(\sigma T)$  depicting a strong deviation of Lorentz ratio (LR) $L/L_{0}$ from unity in the DF regime of graphene. This leads to intensive theoretical investigation seeking the explanation of enhanced LR near the Dirac point of graphene. Lucas \textit{et al.}~\cite{AnLucas2016} have used the relativistic fluid dynamics formalism in the DF regime to explain the experimentally observed enhancement in LR, whereas Zarenia \textit{et al.}~\cite{Zarenia_2019}, developed a ”disorder enabled hydrodynamics” to explain the same. Tu \textit{et al.}~\cite{tu2023wiedemann}, on the other hand, explained the enhancement of LR with the inclusion of bipolar diffusion effect and a band gap at the Dirac point. 
	
	In this paper, by considering the similarity between QGP and graphene, we have developed a microscopic theory for the latter that takes care of hydrodynamic transport in the domain of low charge impurity densities/charge puddles. We calculated the $\kappa$, $\sigma$ and then LR for graphene (QGP) systems by explicitly considering the contribution to $L$ from electrons and holes (quarks and anti-quarks). LR for graphene with respect to net carrier density is compared with that of experimental data~\cite{crossno2016observation} where a good agreement in the DF domain is obtained. 
	
The article is arranged as follows. In Sec.~(\ref{GT}) and Sec.~(\ref{QGPT}), we derive respectively the LR for graphene and QGP starting from the Boltzmann transport equation (BTE) in relaxation time approximation (RTA). Next, we compare the results of graphene and QGP with reference to net number density, energy density, pressure, enthalpy, electrical conductivity, thermal conductivity, and LR in Sec.~(\ref{RD}). In Sec.~(\ref{sum}), we conclude by summarizing our investigations.

\section{Formalism}
\subsection{Thermoelectric transports in Graphene}\label{GT}
The carrier transport in the materials can be described with the help of the BTE, which determines the fate of charge carriers in different energy bands~\cite{ashcroft2011solid,ziman2001electrons}. In graphene, the low-energy electron excitations can be modeled by a two-band electronic system comprised of the valence band and conduction band. The BTE for the two-band electronic system in the presence of electric field $\vec{\tilde{E}}$ can be written as~\cite{Narozhny:2022ncn,Narozhny2019uib}:
\bea
\frac{\del f_{\la}}{\del t} + \Vec{v}_{\la}\cdot \frac{\del f_{\la}}{\del \Vec{r}}-e\vec{\tilde{E}}\cdot \frac{\del f_{\la}}{\del \Vec{p}_{\la}}= C_{\la}[f_{\la}]~,\label{f1}
\eea
where the band index $\la=+$ for conduction band and $-$ for valence band. The microscopic variables quasi-momentum (or crystal momentum), energy, and the group velocity of the electrons are respectively defined as $\Vec{p}_{\la}\equiv \hbar \Vec{k}_{\la}$, $E_{\la}\equiv E_{\la}(\Vec{k}_{\la})$, and $\Vec{v}_{\la}\equiv \frac{1}{\hbar} \frac{\del E_{\la}}{\del \Vec{k}_{\la}}$ with $\Vec{k}_{\la}$ being the wave vector in the reciprocal space. The collision kernel $C_{\la}[f_{\la}]=  \big(\frac{\del f_{\la}}{\del t}\big)_{coll}$ gives rise to changes in the distribution function due to random incessant collisions. The $C_{\la}[f_{\la}]$ contains all the information about the momentum conserving and non conserving interactions of the electrons with other electrons, phonons, and lattice defects~\cite{Lucasfong2018}. We will consider a temperature window where the momentum non conserving scatterings are negligible compared to momentum conserving electron-electron scatterings. In this temperature regime, the hydrodynamic electron flow is expected, contrary to the diffusive flow of electrons observed in metals in normal conditions~\cite{Lucasfong2018}. Therefore, in the temperature window under consideration, the collision kernel $C_{\la}[f_{\la}]$ ensures energy, momentum, and charge conservation. 
Using the dispersion relation $E_{\la}=\la v_{F}\hbar |\Vec{k}_{\la}|$ the group velocity of the electrons in the vicinity of Dirac point is obtained as $\Vec{v}_{\la}= \frac{1}{\hbar} \frac{\del E_{\la}}{\del \Vec{k}_{\la}}=\la v_{F}\frac{\Vec{k}_{\la}}{|\Vec{k}_{\la}|}$. Rewriting Eq.~(\ref{f1}) with the substitution of the group velocity $\Vec{v}_{\la}=\la v_{F}\frac{\Vec{k}_{\la}}{|\Vec{k}_{\la}|}$ we have,
\bea
&&\frac{E_{\la}}{v_{F}^{2}}\frac{\del f_{\la}}{\del t}+ \hbar \Vec{k}_{\la}\cdot \frac{\del f_{\la}}{\del \Vec{r}}-e \vec{\tilde{E}} \cdot \bigg(\hbar \Vec{k}_{\la} \frac{\del f_{\la}}{\del E_{\la}}+\frac{E_{\la}}{v_{F}^{2}} \frac{\del f_{\la}}{\del \Vec{p}_{\la}}\bigg)=\frac{E_{\la}}{v_{F}^{2}}C[f_{\la}]~.\label{f2}
\eea
Eq.~(\ref{f2}) describes the dynamics of electrons in both the conduction band and valence band in graphene near the Dirac cone. Since it is customary to describe the valence band carriers by holes, we will write down the BTE for valence band carriers with the following change of variables $f_{h}\equiv 1-f_{-}$, $E_{h}\equiv -E_{-}$ and $\Vec{k}_{h}\equiv -\Vec{k}_{-}$ as, 
\bea
&& \frac{E_{h}}{v_{F}^{2}}\frac{\del f_{h}}{\del t}+ \hbar\Vec{k}_{h}\cdot \frac{\del {f_{h}}}{\del \Vec{r}} + e \vec{\tilde{E}}\cdot \bigg(\hbar\Vec{k}_{h}\frac{\del f_{h}}{\del E_{h}} + \frac{E_{h}}{v_{F}^{2}} \frac{\del f_{h}}{\del \Vec{p}_{h}}\bigg)=-\frac{E_{h}}{v_{F}^{2}}C_{h}[1-f_{h}]~,\label{f3}
\eea
where we defined $C_{-}[{1-f_{h}}]\equiv C_{h}[1-f_{h}]$.
Similarly, the BTE for the conduction band electrons in graphene can be rewritten by calling $f_{+}\equiv f_{e}$, $E_{+}=E_{e}$, $\Vec{k}_{+}=\Vec{k}_{e}$ and $C_{+}[f_{+}]\equiv C_{e}[f_{e}]$ as:
\bea
&& \frac{E_{e}}{v_{F}^{2}}\frac{\del f_{e}}{\del t}+ \hbar\Vec{k}_{e}\cdot \frac{\del {f_{e}}}{\del \Vec{r}} + e \vec{\tilde{E}}\cdot \bigg(\hbar\Vec{k}_{e}\frac{\del f_{e}}{\del E_{e}} + \frac{E_{e}}{v_{F}^{2}} \frac{\del f_{e}}{\del \Vec{p}_{e}}\bigg)=\frac{E_{e}}{v_{F}^{2}}C_{e}[f_{e}]~.\label{f4}
\eea
Eq.~(\ref{f4}) and ~(\ref{f3}) form the basis of the dynamics of the electrons and holes in the graphene, respectively. The relations: $\Vec{p}_{e,h}=\hbar\Vec{k}_{e,h}$, $E_{e,h}=(\hbar k_{e,h}) v_{F}$, and $\Vec{v}_{e,h}=v_{F}\frac{\Vec{k}_{e,h}}{k_{e,h}}$ is similar to a massless relativistic (ultra-relativistic) particle  where the limiting speed is $v_{F}$. Notably, there exists a large set of literature~\cite{Hartnoll_2007,MMuller2008,Fritz2008,Markus2008,Foster2009,mendoza2013hydrodynamic,AnLucas2016,ALucas2016,Lucasfong2018} in the field of graphene where the relativistic behavior of electrons has been explored by considering $v_{F}$ as the limiting speed. Recently, in Ref.~\cite{Aung:2023vrr}, the authors have calculated the ratio of shear viscosity to entropy density of the electron-fluid in graphene by drawing an analogy with the relativistic hydrodynamics used in the context of QGP. The similarities and differences of the hydrodynamics practiced in the literature QGP and graphene were described in detail there~\cite{Aung:2023vrr}. Here we will follow the same method to write the Eq.~(\ref{f4}) and ~(\ref{f3}) in a covariant form where the speed of light $c$ is replaced with the limiting speed $v_{F}$. By defining the four position vector $x^{\mu}$, four-momentum of electrons $p_{e}^{\mu}$ and four-momentum of holes $p_{h}^{\mu}$ (with $\mu=0 \text{ to } 2$) as $x^{\mu}\equiv (v_{F}t,x^{i})$, $p^{\mu}_{e}\equiv(\frac{E_{e}}{v_{F}}, p_{e}^{i}=\hbar k_{e}^{i})$ and $p^{\mu}_{h}\equiv(\frac{E_{h}}{v_{F}}, p_{h}^{i}=\hbar k_{h}^{i})$, Eq.~(\ref{f4}) and ~(\ref{f3}) can be rewritten as:
\bea
&& p^{\mu}_{e} \del_{\mu}f_{e} -e F^{\mu\nu}p_{e\nu}\frac{\del f_{e}}{\del p_{e}^{\mu}}=\Tilde{C}_{e}[f_{e}]~,\label{f5}\\
&& p^{\mu}_{h} \del_{\mu}f_{h} + e F^{\mu\nu}p_{h\nu}\frac{\del f_{h}}{\del p_{h}^{\mu}}=\Tilde{C}_{h}[f_{h}]~,\label{f6}
\eea
where we adopted the relativistic notation $\del_{\mu}\equiv \frac{\del}{\del {x^{\mu}}}$ and redefined the collision terms as  $\Tilde{C}_{e}[f_{e}]\equiv \frac{E_{e}}{v_{F}^{2}} C_{e}[f_{e}]$ and $\Tilde{C}_{h}[f_{h}]\equiv -\frac{E_{h}}{v_{F}^{2}} C_{h}[1-f_{h}]$. The $F^{\mu\nu}=\frac{1}{v_{F}^{2}}(\tilde{E}^{\mu}u^{\nu}-\tilde{E}^{\nu}u^{\mu})$ is the Faraday tensor with the fluid four-velocity $u^{\mu}=\gamma_{u}(v_{F},\Vec{u})$, where $\gamma_{u}=1/\sqrt{1-u^{2}/v_{F}^{2}}$. $\tilde{E}^{\mu}$ is the comoving electric field perpendicular to $u^{\mu}$, \textit{i.e.}, $\tilde{E}^{\mu}u_{\mu}=0$. One can observe the similarity of the electron-hole BTE of graphene given in Eq.~(\ref{f5}) and~(\ref{f6}) with the BTE of electron-positron plasma~\cite{Janicke_1977,Palmroth_2018,Elskens_2020} or quark-antiquark plasma~\cite{Kushwah:2024zgd,Panda:2020zhr,Panda:2021pvq}. Electron-positron plasma can be realized in astrophysics and quark-antiquark plasma is created in Large Hadron Collider (LHC) or Relativistic Heavy Ion Collider (RHIC). This similarity will be used to define particle flow, stress-energy tensor, and the corresponding thermodynamic variables in a covariant manner. The net particle four flow $N^{\mu}$ for the electron hydrodynamics in graphene is defined as the electron flow minus the hole flow: 
\bea
N^{\mu}\equiv N^{\mu}_{e}-N^{\mu}_{h}= 4 v_{F}^{2}\bigg[\int \frac{d^{2}\Vec{p_e}}{h^{2}E_{e}}p^{\mu}_{e}f_{e}-\int \frac{d^{2}\Vec{p}_{h}}{h^{2}E_{h}}p^{\mu}_{h}f_{h} \bigg]~.\label{f7}
\eea
where $4$ is the degeneracy factor, which is the product of spin and valley degeneracies, \textit{i.e.}, total degeneracy $=$ spin degeneracy ($2$) $\times$ valley degeneracy ($2$)$=4$. Similarly, the total stress-energy tensor for the electron hydrodynamics in graphene is defined as:
\bea
T^{\mu\nu}\equiv T^{\mu\nu}_{e}+ T^{\mu\nu}_{h}=4 v_{F}^{2}\bigg[\int \frac{d^{2}\Vec{p_e}}{h^{2}E_{e}}p^{\mu}_{e}p^{\nu}_{e}f_{e}+\int \frac{d^{2}\Vec{p}_{h}}{h^{2}E_{h}}p^{\mu}_{h}p^{\nu}_{h}f_{h} \bigg]~.\label{f8}
\eea
In the out-of-equilibrium situation, the stress-energy tensor and the net particle flow can be split into two parts: $T^{\mu\nu}=T^{\mu\nu}_{0}+ T^{\mu\nu}_{D}$ and $N^{\mu}= N^{\mu}_{0}+ N^{\mu}_{D}$. The $T^{\mu\nu}_{0}$ and $N^{\mu}_{0}$ are the ideal parts that give information about the thermodynamics of the system. $T^{\mu\nu}_{D}$ and $N^{\mu}_{D}$ are the dissipative parts containing information about the various dissipative flows like shear flow, energy diffusion, particle diffusion, \textit{etc.} The ideal parts $N^{\mu}_{0}$ and $T^{\mu\nu}_{0}$ can be expressed as integrals over the local equilibrium distribution function $f^{0}_{e}=1/(e^{(u_{\mu}p^{\mu}_{e}-\mu_{e})/k_{B}T}+1)$ and $f^{0}_{h}=1/(e^{(u_{\mu}p^{\mu}_{h}-\mu_{h})/k_{B}T}+1)$ with $ \mu_{e}= \mu ~\text{and}~ \mu_{h}= -\mu $ : 
\bea
&& N^{\mu}_{0}\equiv 4 v_{F}^{2}\bigg[\int \frac{d^{2}\Vec{p_e}}{h^{2}E_{e}}p^{\mu}_{e}f^{0}_{e}-\int \frac{d^{2}\Vec{p}_{h}}{h^{2}E_{h}}p^{\mu}_{h}f^{0}_{h} \bigg]=n~u^{\mu}~,\label{f9}\\
&& T^{\mu\nu}_{0}\equiv 4 v_{F}^{2}\bigg[\int \frac{d^{2}\Vec{p_e}}{h^{2}E_{e}}p^{\mu}_{e}p^{\nu}_{e}f^{0}_{e}+\int \frac{d^{2}\Vec{p}_{h}}{h^{2}E_{h}}p^{\mu}_{h}p^{\nu}_{h}f^{0}_{h} \bigg]= v_{F}^{-2}~\Ep u^{\mu} u^{\nu}-P~\D^{\mu\nu}~,\label{f10}
\eea
where $\D^{\mu\nu}=\eta^{\mu\nu}-v_{F}^{-2}u^{\mu}u^{\nu}$ and $n$, $\Ep$ and $P$ are respectively the net number density, energy density, and pressure of the electron-hole fluid. It is easy to see from Eq.~(\ref{f9}) and~(\ref{f10}) that the thermodynamic variables $n$, $\Ep$ and $P$ can be expressed as:
\bea
&& n= v_{F}^{-2}u_{\mu}N^{\mu}_{0}=n_{e}-n_{h}=4\bigg[\int \frac{d^{2}\Vec{p}_{e}}{h^{2}} f^{0}_{e}-\int \frac{d^{2}\Vec{p}_{h}}{h^{2}} f^{0}_{h}\bigg]~,\label{f11}\\
&& \Ep=v_{F}^{-2}u_{\mu}u_{\nu}T^{\mu\nu}_{0}=\Ep_{e}+\Ep_{h}= 4\bigg[\int \frac{d^{2}\Vec{p}_{e}}{h^{2}} E_{e} f^{0}_{e}+\int \frac{d^{2}\Vec{p}_{h}}{h^{2}} E_{h}f^{0}_{h}\bigg]~,\label{f12}\\
&& P=-\frac{1}{2}\D_{\mu\nu}T_{0}^{\mu\nu}=P_{e}+ P_{h}=4v_{F}^{2}\bigg[\int \frac{d^{2}\Vec{p}_{e}}{h^{2}} \frac{p_{e}^{2}}{2E_{e}}  f^{0}_{e}+\int \frac{d^{2}\Vec{p}_{h}}{h^{2}} \frac{p_{h}^{2}}{2E_{h}}f^{0}_{h}\bigg]~.\label{f13}
\eea
The above phase space integrals for the thermodynamic variables have been evaluated in Appendix~(\ref{ape2}) in terms of Fermi integral functions. Now, the dissipative part of the particle flow $N^{\mu}_{D}$ and stress-energy tensor $T^{\mu\nu}_{D}$ can be expressed as integral over $\delta f_{e}\equiv f_{e}-f^{0}_{e}$ and $\delta f_{h}\equiv f_{h}-f^{0}_{h}$ as,
\bea
&& N^{\mu}_{D}\equiv 4 v_{F}^{2}\bigg[\int \frac{d^{2}\Vec{p_e}}{h^{2}E_{e}}p^{\mu}_{e}\delta f_{e}-\int \frac{d^{2}\Vec{p}_{h}}{h^{2}E_{h}}p^{\mu}_{h}\delta f_{h} \bigg]~,\label{f14}\\
&& T^{\mu\nu}_{D}\equiv 4 v_{F}^{2}\bigg[\int \frac{d^{2}\Vec{p_e}}{h^{2}E_{e}}p^{\mu}_{e}p^{\nu}_{e}\delta f_{e}+\int \frac{d^{2}\Vec{p}_{h}}{h^{2}E_{h}}p^{\mu}_{h}p^{\nu}_{h}\delta f_{h} \bigg]~.\label{f15}
\eea
The thermoelectric coefficients connected with the electron hydrodynamics in graphene can be obtained from $N^{\mu}_{D}$. For the evaluation of $N^{\mu}_{D}$ from Eq.~(\ref{f14}) one needs to determine the unknowns $\delta f_{e}$ and  $\delta f_{h}$ with the help of Eq.~(\ref{f5}) and~(\ref{f6}). For the determination of $\delta f_{e}$ and  $\delta f_{h}$ we will rewrite Eq.~(\ref{f5}) and~(\ref{f6}) in the RTA of Anderson-Witting type~\cite{ANDERSON1974466}, \textit{i.e.}, $C_{e}[f_{e}]=-\frac{u_{\mu}p^{\mu}_{e}}{v_{F}^{2}}\frac{\delta f_{e}}{\tau_{c}}$ and $C_{h}[f_{h}]=-\frac{u_{\mu}p^{\mu}_{h}}{v_{F}^{2}}\frac{\delta f_{h}}{\tau_{c}}$, where we assumed same collision time $\tau_{c}$ for electrons and holes. At this juncture, it is necessary to discuss our choice of hydrodynamic frame and the energy dependence of the relaxation time. In the theory of relativistic dissipative hydrodynamics, the choice of hydrodynamic frame is of primary importance in which one defines the fluid velocity and thermodynamic variables through the use of matching conditions~\cite{Rocha:2023ilf,Rocha:2021lze,Dwibedi:2024mff}. In Anderson-Witting type RTA models one usually works from the Landau-Lifshitz hydrodynamic frame\footnote{In Landau-Lifshitz frame the dissipative part of the energy flow $W^{\la}$ vanishes~\cite{DeGroot:1980dk,Cercignani2002}, \textit{i.e.}, $W^{\la}\equiv\Delta_{\mu}^{\la}T^{\mu\nu}u_{\nu}=0$.} with energy independent relaxation time. Nevertheless, the actual relaxation time for a system is energy-dependent, and its energy dependence can be calculated for given interactions using quantum mechanical or quantum field theoretical methods. Since the present article focuses on the effect of the fluid dynamic nature of the electron flow on the thermoelectric coefficients, our assumption of a constant relaxation time is justifiable. The Eq.~(\ref{f5}) and~(\ref{f6}) in the RTA with the substitution of $f_{e,h}=f^{0}_{e,h}+\delta f_{e,h}$ yield,
\bea
&& p^{\mu}_{e} \del_{\mu}f^{0}_{e} -\frac{e}{v_{F}^{2}}(\tilde{E}^{\mu}u^{\nu}-\tilde{E}^{\nu}u^{\mu})p_{e\nu}\frac{\del f^{0}_{e}}{\del p_{e}^{\mu}}=- \frac{u_{\mu}p^{\mu}_{e}}{v_{F}^{2}}\frac{\delta f_{e}}{\tau_{c}}~,\label{f16}\\
&& p^{\mu}_{h} \del_{\mu}f^{0}_{h} +  \frac{e}{v_{F}^{2}}(\tilde{E}^{\mu}u^{\nu}-\tilde{E}^{\nu}u^{\mu})p_{h\nu}\frac{\del f^{0}_{h}}{\del p_{h}^{\mu}}=-\frac{u_{\mu}p^{\mu}_{h}}{v_{F}^{2}}\frac{\delta f_{h}}{\tau_{c}}~,\label{f17}
\eea
where we neglected the space-time gradients of $\delta f_{e,h}$ from the LHS of the equations~\cite{Jaiswal:2013npa}. Evaluating the space-time gradients of $f^{0}_{e,h}$ by using $f^{0}_{e}=1/(e^{(u_{\mu}p^{\mu}_{e}-\mu_{e})/{k_{B}T}}+1)$ and $f^{0}_{h}=1/(e^{(u_{\mu}p^{\mu}_{h}-\mu_{h})/{k_{B}T}}+1)$ in Eq.~(\ref{f16}) and ~(\ref{f17}) with $\mu\equiv\mu_{e}=-\mu_{h}$ we have,
\bea
\delta f_{e}&=&\frac{\tau_{c}v_{F}^{2}}{u_{\mu}p^{\mu}_{e}}\Bigg[v_{F}^{-2}(u_{\al}p^{\al}_{e})\left((u_{\beta}p^{\beta}_{e})D\frac{1}{k_{B}T}+\frac{p_{e}^{\al}Du_{\al}}{k_{B}T}-D \frac{\mu}{k_{B}T}\right) +\frac{p^{\al}_{e}p^{\beta}_{e}}{k_{B}T}\na_{\al}u_{\beta} +(u_{\al}p^{\al}_{e}) p^{\beta}_{e}\na_{\beta}\frac{1}{k_{B}T}\nn\\
&& -p_{e}^{\al}\na_{\al}\frac{\mu}{k_{B}T} + \frac{e p_{e}^{\al}}{k_{B}T}\tilde{E}_{\al}\Bigg]f^{0}_{e}(1-f^{0}_{e})~,\label{f18}\\
\delta f_{h}&=&\frac{\tau_{c}v_{F}^{2}}{u_{\mu}p^{\mu}_{h}}\Bigg[v_{F}^{-2}(u_{\al}p^{\al}_{h})\left((u_{\beta}p^{\beta}_{h})D\frac{1}{k_{B}T}+\frac{p_{h}^{\al}Du_{\al}}{k_{B}T}+D \frac{\mu}{k_{B}T}\right) +\frac{p^{\al}_{h}p^{\beta}_{h}}{k_{B}T}\na_{\al}u_{\beta} +(u_{\al}p^{\al}_{h}) p^{\beta}_{h}\na_{\beta}\frac{1}{k_{B}T}\nn\\
&& +p_{h}^{\al}\na_{\al}\frac{\mu}{k_{B}T} - \frac{e p_{h}^{\al}}{k_{B}T}\tilde{E}_{\al}\Bigg]f^{0}_{h}(1-f^{0}_{h})~,\label{f19}
\eea
where the spatial and temporal derivative operators $\na^{\mu}$ and $D$ are defined as $\na^{\mu}\equiv \D^{\mu\nu}\del_{\nu}\xrightarrow[]{LRF}(0,-\Vec{\na})$ and $D\equiv u^{\mu}\del_{\mu}\xrightarrow[]{LRF}\frac{\del}{\del t}$. Eq.~(\ref{f18}) along with Eq.~(\ref{f19}) can be readily used for the calculation of the thermoelectric coefficients of graphene. We consider a full dynamical scenario in which the fluid velocity profile possesses a space-time gradient. Since the dissipative fluxes like shear flow, thermal flow, \textit{etc.}, are proportional to spatial gradients, the temporal derivatives of $1/T$, $\mu/T$, and $u^{\mu}$ occurring in Eq.~(\ref{f18}) and~(\ref{f19}) should be eliminated with the help of the conservation equation of ideal electron hydrodynamics in graphene,
\bea
&&Dn=-n\na_{\mu}u^{\mu}~,\label{f20}\\
&&Du^{\mu}=\frac{v_{F}^{2}}{\Ep+P}\bigg[\na^{\mu}P+\rho \tilde{E}^{\mu}\bigg]~,\label{f21}\\
&&D\Ep=-(\Ep+P)\na_{\mu}u^{\mu}~,\label{f22}
\eea
where $\rho\equiv -en=-e(n_{e}-n_{h})$ is the charge carrier density. The Eq.~(\ref{f20}) ensures charge conservation, and Eq.~(\ref{f21}) and ~(\ref{f22}) provide the energy-momentum conservation for the electron hydrodynamics. After all the temporal derivatives have been eliminated from Eq.~(\ref{f18}) and~(\ref{f19}) the $\delta f_{e,h}$ can be easily expressed as a linear combination of three independent spatial gradient terms: $\na_{\mu}u^{\mu}$, $\frac{1}{2}(\na^{\mu}u^{\nu}+\na^{\nu}u^{\mu})-\frac{1}{2}(\na_{\al}u^{\al})\D^{\mu\nu}$, and $-\na^{\al}\frac{\mu}{k_{B}T}+\frac{e}{k_{B}T}\tilde{E}^{\al}$. The first  ($\na_{\mu}u^{\mu}$) and second term ($\frac{1}{2}(\na^{\mu}u^{\nu}+\na^{\nu}u^{\mu})-\frac{1}{2}(\na_{\al}u^{\al})\D^{\mu\nu}$) give rise to bulk and shear stresses in the fluid (see Appendix~(\ref{ape1})). They are of primary importance for the calculation of bulk and shear viscosity. Since the present article is structured for the calculation of thermoelectric coefficients, we will neglect viscous stresses to write,
\bea
&&\delta f_{e}=-\frac{\tau_{c} v_{F}^{2}}{u_{\mu}p^
	{\mu}_{e}}p^{\al}_{e}\bigg[\frac{n}{\Ep+P}u_{\beta}p^{\beta}_{e}-1\bigg]\left(-\na_{\al}\frac{\mu}{k_{B}T}+\frac{e\tilde{E}_{\al}}{k_{B}T} \right)f^{0}_{e}(1-f^{0}_{e})~,\label{f23}\\
&&\delta f_{h}=-\frac{\tau_{c} v_{F}^{2}}{u_{\mu}p^
	{\mu}_{h}}p^{\al}_{h}\bigg[\frac{n}{\Ep+P}u_{\beta}p^{\beta}_{h}+1\bigg]\left(-\na_{\al}\frac{\mu}{k_{B}T}+\frac{e\tilde{E}_{\al}}{k_{B}T} \right)f^{0}_{h}(1-f^{0}_{h})~,\label{f24}
\eea
The expression for the current density (dissipative part of charge flow) $J^{\mu}$ can be written as,
\bea
J^{\mu}\equiv-e \D^{\mu}_{\nu}N^{\nu}_{D}&=& -4e ~v_{F}^{2} \D^{\mu}_{\nu}\bigg[\int \frac{d^{2}\Vec{p}}{h^{2}E_{e}}p^{\nu}_{e}\delta f_{e}-\int \frac{d^{2}\Vec{p}_{h}}{h^{2}E_{h}}p^{\nu}_{h}\delta f_{h} \bigg]~,\nn
\eea
substituting $u^{\mu}=(v_{F},\Vec{0})$ in the above expression we have,
\bea
&&J^{i}=-4e~ v_{F}^{4}\tau_{c}\bigg[\int \frac{d^{2}\Vec{p}_{e}}{h^{2}} \frac{p_{e}^{i}p_{e}^{j}}{E_{e}^{2}}\left( \frac{nE_{e}}{\Ep+P}-1\right)f^{0}_{e}(1-f^{0}_{e})-\int\frac{d^{2}\Vec{p}_{h}}{h^{2}} \frac{p_{h}^{i}p_{h}^{j}}{E_{h}^{2}}\left( \frac{nE_{h}}{\Ep+P}+1\right)f^{0}_{h}(1-f^{0}_{h})\bigg] X^{j}~,\label{f25}
\eea
where we defined $X^{i}\equiv\del_{i}\frac{\mu}{k_{B}T}+\frac{eE^{i}}{k_{B}T}$. We can simplify Eq.~(\ref{f25}) (see Appendix~(\ref{ape2})) to obtain,
\bea
J^{i}&=&\frac{4e\pi\tau_{c}}{h^{2}}~\frac{nk_{B}T}{\Ep+P}\bigg[ 2(k_{B}T)^{2}(f_{2}(A^{-1})-f_{2}(A))+\frac{\Ep+P}{n}k_{B}T (f_{1}(A^{-1})+f_{1}(A))\bigg]X^{i}\nn\\        
\text{or, }J^{\al}&=&\frac{4e\pi\tau_{c}}{h^{2}}~\frac{nk_{B}T}{\Ep+P}\bigg[ 2(k_{B}T)^{2}(f_{2}(A^{-1})-f_{2}(A))+\frac{\Ep+P}{n}k_{B}T (f_{1}(A^{-1})+f_{1}(A))\bigg] \left(-\na^{\al}\frac{\mu}{k_{B}T}+ \frac{e\tilde{E}^{\al}}{k_{B}T}\right)~.\label{f26}
\eea
The heat flow $q^{\mu}$ for relativistic fluid is defined as the difference between dissipative part of energy flow $W^{\mu}\equiv \Delta_{\al}^{\mu}T^{\al\beta}u_{\beta}$ and enthalpy flow $h^{\mu}\equiv \mathfrak{h} \Delta^{\mu}_{\al}N^{\al}_{D}$~\cite{DeGroot:1980dk,Cercignani2002}, \textit{i.e.}, $q^{\mu}=W^{\mu}-h^{\mu}=\Delta_{\al}^{\mu}(T^{\al\beta}u_{\beta}-\mathfrak{h}N^{\al}_{D})$, where $\mathfrak{h}\equiv \frac{\Ep+P}{n}$ is the enthalpy per particle. In Anderson-Witting type of RTA model, the dissipative part of energy flow vanishes~\cite{DeGroot:1980dk,Cercignani2002} and the expression of heat flow becomes: $q^{\mu}=-h^{\mu}=-\frac{\Ep+P}{n}\Delta_{\al}^{\mu}N^{\al}_{D}=-(\Ep+P)\frac{J^{\mu}}{\rho}$. The preceding definition of heat flow with Eq.~(\ref{f26}) give the following expression for $q^{\al}$:
\bea
&&q^{\al}=\frac{4\pi\tau_{c}k_{B}T}{h^{2}}\bigg[ 2(k_{B}T)^{2}(f_{2}(A^{-1})-f_{2}(A))+\frac{\Ep+P}{n}k_{B}T (f_{1}(A^{-1})+f_{1}(A))\bigg] \left(-\na^{\al}\frac{\mu}{k_{B}T}+ \frac{e\tilde{E}^{\al}}{k_{B}T}\right)~.\label{f27}
\eea
We can rewrite the  expression of current density and heat flow with the help of Gibbs-Duhem relation $n ~d\big(\frac{\mu}{k_{B}T}\big)=\frac{1}{k_{B}T}~dP+(\Ep+P)~d\big(\frac{1}{k_{B}T}\big)$ as:
\bea
J^{\mu}&=&\frac{4e\pi\tau_{c}}{h^{2}}(k_{B}T)^{2}~\bigg[ 2(f_{2}(A^{-1})-f_{2}(A))+\frac{\Ep+P}{nk_{B}T} (f_{1}(A^{-1})+f_{1}(A))\bigg]\left[-\frac{\rho}{\Ep+P}\tilde{\Ep}^{\mu}+\frac{1}{T}\na^{\mu}T\right]~,\label{fd1}\\
q^{\mu}&=&\frac{4\pi\tau_{c}}{h^{2}}(k_{B}T)^{2}\frac{\Ep+P}{n}~\bigg[ 2(f_{2}(A^{-1})-f_{2}(A))+\frac{\Ep+P}{nk_{B}T} (f_{1}(A^{-1})+f_{1}(A))\bigg]\left[-\frac{\rho}{\Ep+P}\tilde{\Ep}^{\mu}+\frac{1}{T}\na^{\mu}T\right]~,\label{fd2}
\eea
where we used the definition $\tilde{\Ep}^{\mu}\equiv \tilde{E}^{\mu}+\frac{1}{\rho}\na^{\mu}P$.
The electrical and thermal conductivity of the electron hydrodynamics can be identified by comparing the microscopically derived expressions of vectorial dissipative flows (charge and heat) in Eq.~(\ref{fd1}) and~(\ref{fd2}) with the macroscopic expressions:
\bea
J^{\mu}&=&a_{11} \tilde{\Ep}^{\mu}+a_{12}\na^{\mu} T\nn\\
\text{and, }q^{\mu}&=&a_{21} \tilde{\Ep}^{\mu}+a_{22}\na^{\mu}T~.\nn
\eea
In this paper, we will focus only on the diagonal components of the matrix $"\bf a"$ since the electrical conductivity $\sigma$ and thermal conductivity $\kappa$ are identified with the diagonal elements, \textit{i.e.}, $\sigma=a_{11}$ and $\kappa=a_{22}$. The off-diagonal element $a_{12}$ gives rise to the electric current due to the spatial variation in the temperature $T$. Similarly, the element $a_{21}$ gives rise to the heat current due to the presence of an electric field $\tilde{E}^{\mu}$. The electrical and thermal conductivity of graphene is given by,
\bea
&&\sigma= 4\pi\tau_{c}e^{2}\left(\frac{k_{B}T}{h}\right)^{2}~\frac{n}{\Ep+P}\bigg[ 2(f_{2}(A^{-1})-f_{2}(A))+\frac{\Ep+P}{nk_{B}T} (f_{1}(A^{-1})+f_{1}(A))\bigg]~,\label{f28}\\
&&\kappa=4\pi \tau_{c} k_{B}\left(\frac{k_{B}T}{h}\right)^{2}\frac{\Ep+P}{nk_{B}T}\bigg[ 2(f_{2}(A^{-1})-f_{2}(A))+\frac{\Ep+P}{nk_{B}T} (f_{1}(A^{-1})+f_{1}(A))\bigg]~,\label{f29}
\eea
The $L$ is given by,
\bea
&&L=\frac{\kappa}{\sigma T}=\left(\frac{\Ep+P}{nk_{B}T}\right)^{2}\frac{k_{B}^{2}}{e^{2}}=\left(\frac{\mathfrak{h}}{k_{B}T}\right)^{2}\frac{k_{B}^{2}}{e^{2}}~.\label{f30}
\eea
Now, in the next section we address the transport coefficients of the QGP system. 
\subsection{Thermoelectric transport in QGP}\label{QGPT}
The inherent properties of the quantum chromodynamics (QCD) make two distinct phases of the quark-matter possible: confined quark-matter as we observe in hadrons and deconfined quark matter which is supposed to be present in the early stages of universe and the core of neutron stars. This is due to the infrared slavery and asymptotic freedom of the QCD interactions, respectively~\cite{10.1143/PTP.44.291,PhysRevD.9.2291,PhysRevLett.34.1353}. The deconfined stage of the quark matter known as QGP can be created in the heavy ion collision experiments in RHIC and LHC. In the early stage of heavy ion collision experiments, the QGP formed can be effectively modeled by relativistic fluid dynamics with the transport coefficients obtained from BTE~\cite{Jaiswal:2016hex}. For simplicity we will take a single flavored quark-anti quark $(q, \bar{q})$ system and write down the BTE as~\cite{Panda:2020zhr,Panda:2021pvq},
\bea
&& p^{\mu} \del_{\mu}f_{q,\bar{q}} -Q_{q,\bar{q}} F^{\mu\nu}p_{\nu}\frac{\del f_{q,\bar{q}}}{\del p^{\mu}}=-(u_{\alpha}p^{\alpha})\frac{f_{q,\bar{q}}-f^{0}_{q,\bar{q}}}{\tau}~,\label{q1}
\eea
where for notational convenience we ignored subscripts on momentum variables. $F^{\mu\nu}$ and $\tau$ are the electromagnetic Faraday tensor and average collision time, respectively. Here, we will identify the quark-antiquark system with up-quark and anti-up-quark, \textit{i.e.},  $(q, \bar{q})$=$(u, \bar{u})$. $f^{0}_{q}=1/(e^{(u_{\mu}p^{\mu}_{q}-\mu_{q})/k_{B}T}+1)$ and $f^{0}_{\bar{q}}=1/(e^{(u_{\mu}p^{\mu}_{\bar{q}}+\mu_{q})/k_{B}T}+1)$ are the local equilibrium distribution functions for the quarks and anti-quarks, where $u^{\mu}$ is the fluid four-velocity and $\mu_{q}$ is the chemical potential of quark. The charge $Q_{q}=Q_{u}=\frac{2e}{3}$ and $Q_{\bar{q}}=Q_{\bar{u}}=-\frac{2e}{3}$. The derivation of thermoelectrical transport coefficients of this $(u, \bar{u})$ plasma runs similar to Sec.~(\ref{GT}). The expressions of thermodynamic variables of this system are also similar to Sec.~(\ref{GT}). The main differences in the expressions of QGP and graphene arise because of the following two reasons. $(i)$ The universal speed $c$ in QGP and Fermi velocity $v_{F}$ in graphene and $(ii)$ the dimensionality of QGP (3D system) and graphene (2D system). Keeping these two differences in mind, one can easily write down the expressions for QGP as follows:
\bea
&& \tilde{n}=2\bigg[\int \frac{d^{3}\Vec{p}}{h^{3}} f^{0}_{q}-\int \frac{d^{3}\Vec{p}}{h^{3}} f^{0}_{\bar{q}}\bigg]=16\pi \left(\frac{k_{B}T}{hc}\right)^{3} \left(f_{3}(A)-f_{3}(A^{-1})\right), (A\equiv e^{\mu_{q}/k_{B}T})~,\label{q2}\\
&& \tilde{\Ep}=3 P=2\bigg[\int \frac{d^{3}\Vec{p}}{h^{3}} Ef^{0}_{q}+\int \frac{d^{3}\Vec{p}}{h^{3}} E f^{0}_{\bar{q}}\bigg]=48 \pi \frac{(k_{B}T)^{4}}{(hc)^{3}} \left(f_{4}(A) + f_{4}(A^{-1}) \right)~,\label{q3}\\
&& \tilde{\sigma} = \frac{16 \pi \tau Q_{u}^{2}}{3c} \left(\frac{k_{B}T}{h}\right)^{3} \frac{\tilde{n}}{\tilde{\Ep}+\tilde{P}}\left[3(f_{3}(A^{-1})-f_{3}(A))+\frac{\tilde{\Ep}+\tilde{P}}{\tilde{n}k_{B}T}(f_{2}(A)+f_{2}(A^{-1}))\right]~,\label{q4}\\
&& \tilde{\kappa} = \frac{16 \pi \tau k_{B}}{3c}\left(\frac{k_{B}T}{h}\right)^{3} \frac{\tilde{\Ep}+\tilde{P}}{\tilde{n}k_{B}T} \left[3(f_{3}(A^{-1})-f_{3}(A))+\frac{\tilde{\Ep}+\tilde{P}}{\tilde{n}k_{B}T}(f_{2}(A)+f_{2}(A^{-1}))\right]~,\label{q5}\\
&& \tilde{L}=\frac{\tilde{\kappa}}{\tilde{\sigma} T}= \left(\frac{\tilde{\Ep}+\tilde{P}}{\tilde{n}k_{B}T}\right)^{2}\frac{k_{B}^{2}}{Q_{u}^{2}}=\left(\frac{\tilde{\mathfrak{h}}}{k_{B}T}\right)^{2}\frac{k_{B}^{2}}{Q_{u}^{2}}~,\label{q6}
\eea
where for the ease of the presentation, we used the same notations as that of Sec.~(\ref{GT}) but with a tilde to distinguish QGP variables from that of graphene.
    
\section{Results and discussion} \label{RD}
\begin{figure*}  
	\centering 
	\includegraphics[scale=0.3]{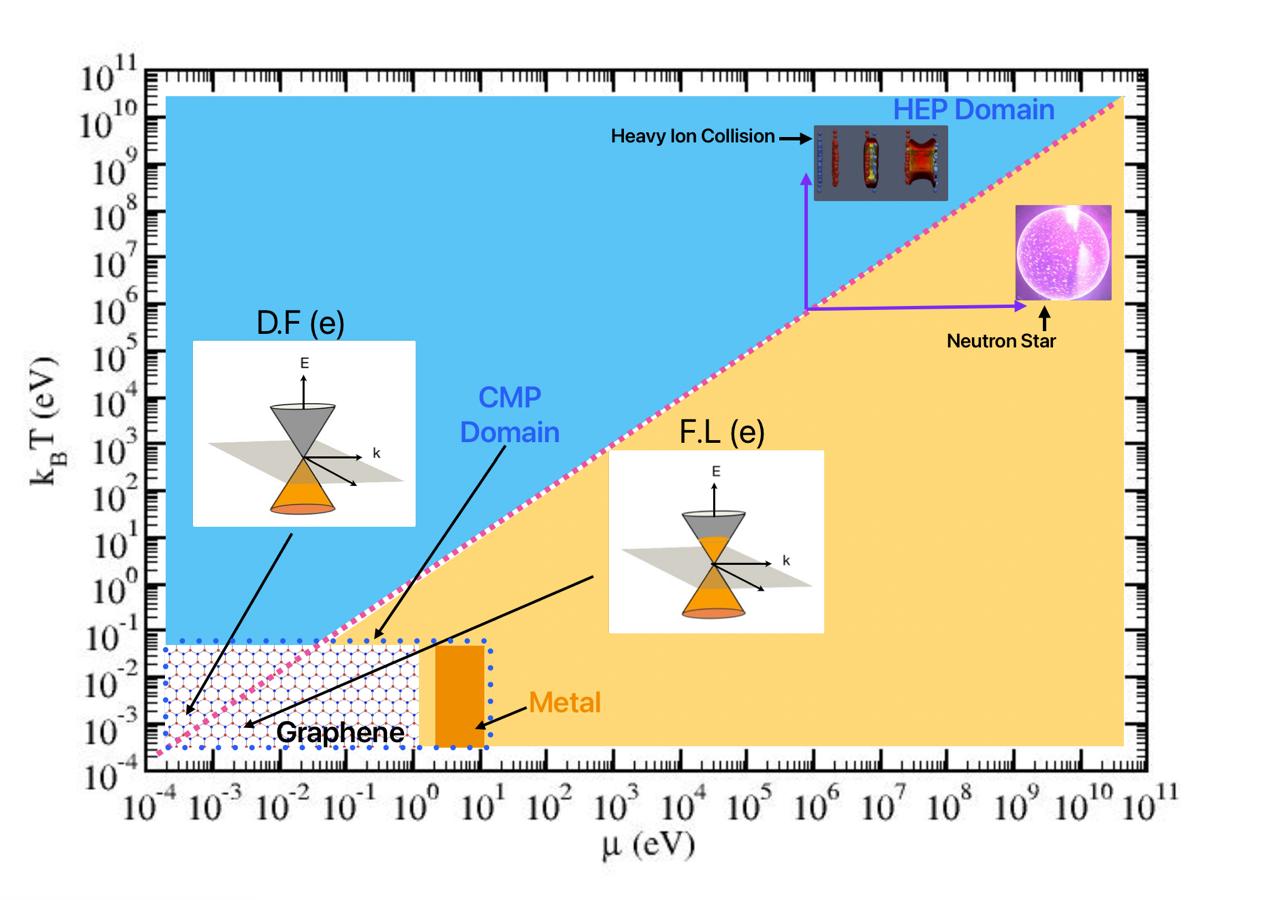}
	\caption{Representation of different physical systems starting from CMP system, graphene, and metals to HEP system, QGP and Neutron star in $\mu-T$ plane} 	\label{fig:phase}
\end{figure*}

Let us first qualitatively discuss the condensed matter physics (CMP) and high energy physics (HEP) systems in $\mu-T$ plane to get acquainted with the order of magnitude of the $\mu$ and $k_{B}T$ values in different physical systems like graphene, metals, QGP and neutron stars. For this purpose, we have first shown the domain of CMP and HEP in the $\mu-T$ plane in Fig.~(\ref{fig:phase}). We have separated the whole plane in two parts by the line $\mu/k_{B}T=1$, where the regions, $\mu/k_{B}T<1$ and $\mu/k_{B}T>1$ may be identified with the DF and FL region, respectively. From the figure, it is apparent that the chemical potential $\mu$ for CMP lies within $0-10$ eV. For a typical graphene sample, it varies on the meV scale, whereas for metals, it can be from $2-10$ eV. Similarly, the temperature scale for the CMP systems is of the few eVs, whereas in HEP, it is of the order of a few MeVs. It can be seen from Fig.~(\ref{fig:phase}) that electrons in conventional metal show the FL behavior, whereas those in the monolayer of graphene can have both DF and FL behavior. Similarly, for HEP-heavy ion collision systems, we may assume the existence of  DF and FL regimes depending on the energy of the collisions. In very high-energy heavy ion collisions, the energy of the colliding nuclei is of the order of TeV. The resulting QGP acts as a DF, comprising ultra-relativistic quarks. On the other hand, the heavy ion collision energies of the CBM and NICA (in GeV order) try to reproduce the Neutron star environment. At such relatively lower beam energies, the resulting quark matter corresponds to FL.

\begin{figure*}  
	\centering 
	\includegraphics[scale=0.5]{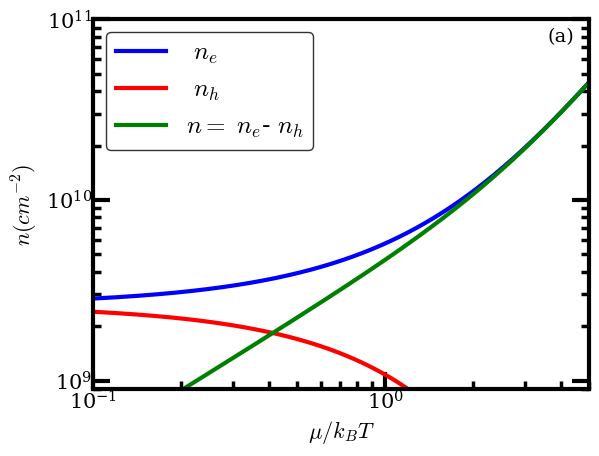}
	~ ~\includegraphics[scale=0.5]{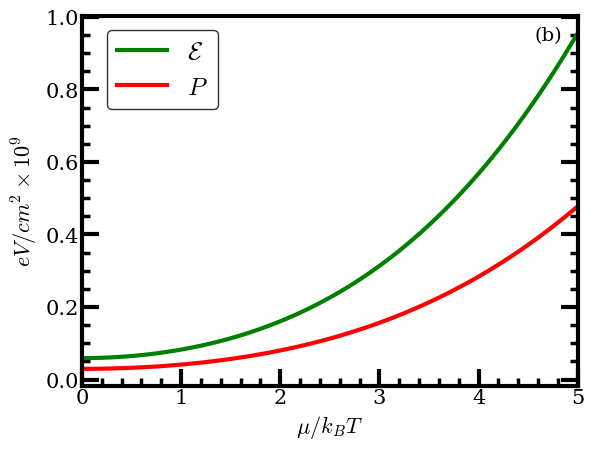}
	\caption{(a) The number density and (b) energy density and pressure of graphene with respect to $\mu/k_B T$ at $T=60$ K .} 	\label{fig:1}
\end{figure*}

\begin{figure*}  
	\centering 
	\includegraphics[scale=0.49]{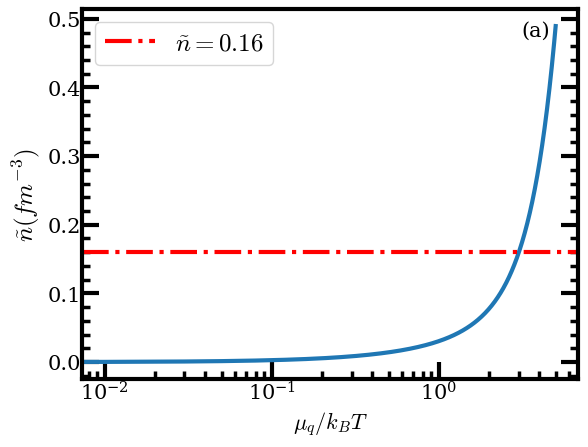}
	~ ~\includegraphics[scale=0.49]{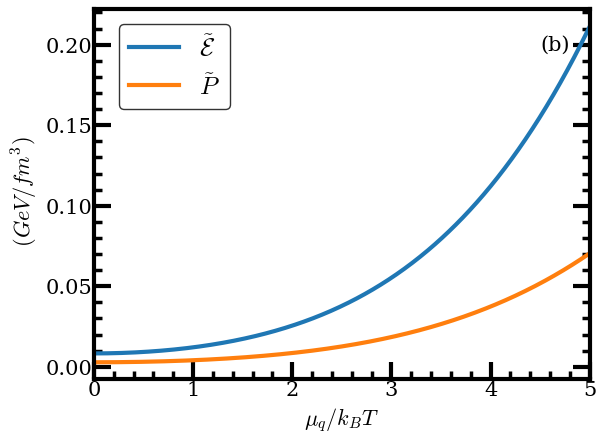}
	\caption{(a) The number density and (b) energy density and pressure of QGP with respect to $\mu_{q}/k_B T$ for QGP at $T=10^{12} \text{ K} ( \equiv100$ MeV) .} \label{fig:2}	
\end{figure*}

\begin{figure*}  
	\centering 
	\includegraphics[scale=0.5]{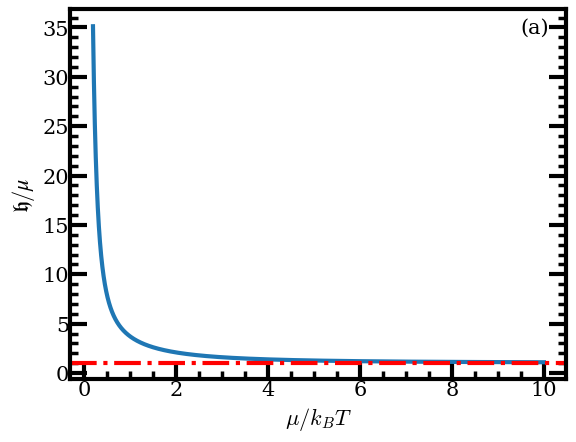}
	~ ~\includegraphics[scale=0.5]{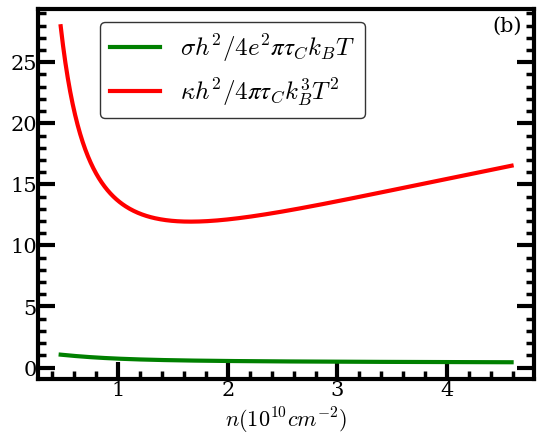}
	\caption{(a) Enthalpy density per particle $vs.$  $\mu/k_B T$ and (b) normalised $\kappa$ and $\sigma$ in graphene vs $n$ .} 
	\label{fig:3}
\end{figure*}

\begin{figure*}  
	\centering 
	\includegraphics[scale=0.49]{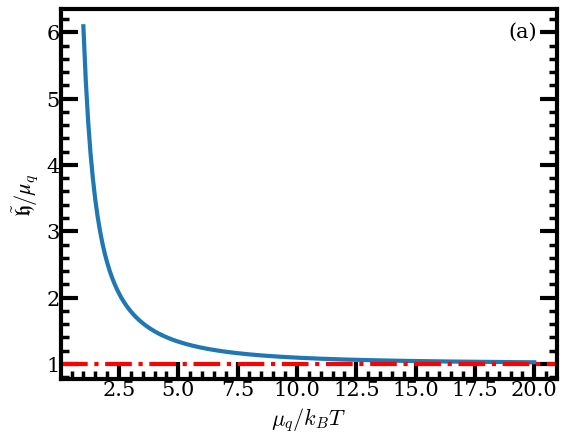}
	~ ~\includegraphics[scale=0.49]{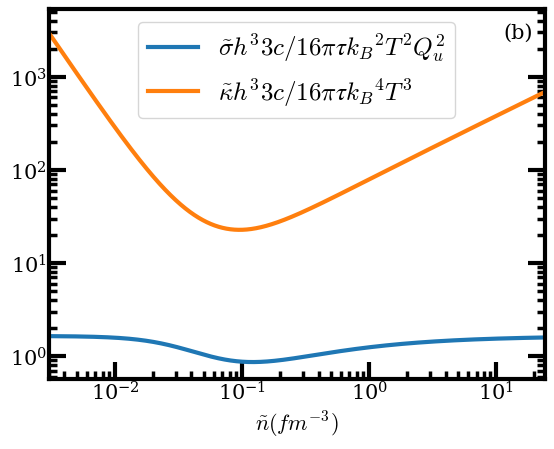}
	\caption{(a) Enthalpy density per particle $vs.$  $\mu_{q}/k_B T$ and (b) normalized $\kappa$ and $\sigma$  $vs.$ $\tilde{n}$ for QGP at $T=10^{12}\text{ K} ( \equiv100$ MeV) .} 
	\label{fig:4}
\end{figure*}

\begin{figure*}  
	\centering 
	\includegraphics[scale=0.48]{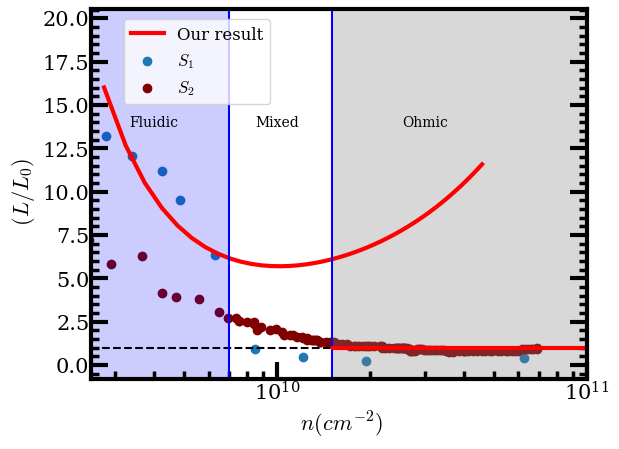}
	~ ~\includegraphics[scale=0.35]{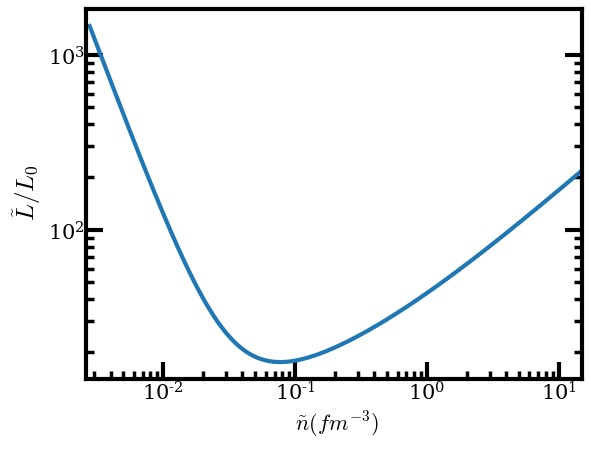}
	\caption{(a) The dependence of $\frac{L}{L_0}$ with net carrier density  $n$ in graphene   at T=60 K (b) The dependence of $\frac{\tilde{L}}{L_0}$ with net carrier density  $\tilde{n}$ in QGP   at $T=100$ MeV.} 
	\label{fig:5}
\end{figure*}

Now let us discuss the quantitative magnitude of the thermodynamic and transport variables of the graphene and QGP with the help of the expressions obtained in Sec.~(\ref{GT}) and~(\ref{QGPT}) in terms of Fermi integral functions. The systems we compare differ in spatial dimension (2D graphene $vs.$ 3D QGP), which affects our quantitative results. First, we carefully assess the causes that make the numerical magnitude of the thermodynamic or transport variables different in different spatial dimensions. The magnitude differs because of two causes of different physical origin:
\begin{itemize}
	\item density of states of the system, which is crucially dependent on the dimensionality and degeneracies of a given state (spin and valley for carriers in graphene and spin, color, and flavor for quarks in QGP)
	\item energy dispersion relation (linear for both QGP and graphene albeit with different proportionality constants $c$ and $v_{F}$)
\end{itemize} 
We have taken spin and valley degeneracy for carriers in graphene, whereas for QGP, we have assumed only the spin degeneracy. The above-mentioned causes jointly determine the density of states of the system, a quantity of paramount importance in statistical physics. The order of the Fermi integral functions that shows up in the thermodynamics and transport variables relies on the energy dependence of the density of states. For instance, consider a D-dimensional system with the energy dispersion relation given by $E=\alpha |\vec{p}|^{N}$, where $\alpha$ and $N$ are constants. This relation reduces to the dispersion relation for the electrons in metals for $\alpha=1/2m^{*}$ ($m^{*}$ is the effective mass) and $N=2$. Similarly, to obtain the energy dispersion relation of QGP and graphene, one needs to choose ($\alpha=c$, $N=1$) and ($\alpha=v_{F}$, $N=1$ ), respectively. The density of states for the dispersion $E=\alpha |\vec{p}|^{N}$ behaves as, $\frac{d^{D}p}{\hbar^{D}(2\pi)^{D}}=B\times E^{D/N-1} dE$, where $B=\left(2^{D-1}\hbar^{D}\pi^{D/2}~\Gamma(D/2) ~N \alpha^{D/N}\right)^{-1}$. Appropriate substitution of spatial dimension leads to the dependence of density of states with energy. For QGP and charge carriers in graphene the density of states varies as $E^{2} ~dE$ and $E~dE$, respectively. Accordingly, the expressions of the thermodynamic and transport variables for QGP contain Fermi integral functions of one integral order more than those of graphene (\textit{cf.} Sec.~(\ref{QGPT})). Apart from the Fermi integral functions, the exact expression of the corresponding variables of QGP and graphene are also different depending on the constant $B$, which differs in different dimensions. We now consider the normalization of the ratio of thermal and electrical conductivity in different spatial dimensions in reference to the Lorenz number. The Lorenz number is originally defined as the ratio of thermal conductivity to electrical conductivity normalized by the temperature of the metals and verified to be a characteristic number for most of the metals at room temperature. Given the possible variations in the spatial dimensionality and dispersion relations, the universality of Lorenz number $L_{0}=\frac{\pi^{2}}{3}\frac{k_{B}^{2}}{e^{2}}$ calls for special scrutiny. Nevertheless, assuming solid-like transport, it can be shown that for the parabolic dispersion relation (N=2) and in any spatial dimension D, the ratio $\kappa/(\sigma T)$ is still $L_{0}$~\cite{ma14112704}. We use the same factor $L_{0}$ to normalize both $L=\kappa/(\sigma T)$ and $\tilde{L}=\tilde{\kappa}/(\tilde{\sigma} T)$ to define the LR. Moreover, the corresponding quantities of QGP and graphene are also measured in different units, which we mention while describing their variation in Figs.~(\ref{fig:1}) to (\ref{fig:5}).

We show the variation of corresponding thermodynamic and transport parameters of graphene and QGP with respect to $\mu/k_{B}T$ and compare the numerical magnitudes. For a graphene sample, the experimental procedure involves increasing the gate voltage with positive polarity while keeping the device temperature constant. This process can be effectively mapped by plotting the variation of different sample characteristics—such as carrier density, energy density, and conductivity--against the parameter $\mu/k_{B}T$ along the $X$-axis, while keeping the temperature $T$ constant. Similarly, the thermodynamic characteristics of the QGP starting from  high energy collisions of LHC and RHIC to CBM and NICA can be mapped by changing $\mu_{q}/k_{B}T$ along the horizontal axis,  for constant $T$ and varying $\mu$ in the accessible $\mu-T$ domain (One can also visualize the same in terms of baryon chemical potential $\mu_{B}$ and baryon density $n_{B}$ by using, $\mu_{q}=\mu_{B}/3$, and $\tilde{n}=3 n_{B}$).
 At first, the variation of the density of electrons $n_{e}$, holes $n_{h}$ and net carriers $n=n_{e}-n_{h}$ is shown in  Fig.~(\ref{fig:1})(a) at  $T=60 $ K. For this purpose, Eq.~(\ref{f11}) can be used to obtain the numerical magnitudes of $n_{e}$, $n_{h}$ and $n$. The number density for the monolayer of graphene refers to number of carriers per unit area and measured in the units of cm $^{-2}$. The result shows that at a constant temperature the number density of electron $n_{e}$ (blue line) rises with a rise in $\mu/k_{B}T$ or equivalently $\mu$, whereas the number density of holes $n_{h}$ (red line) falls exponentially with increase in $\mu/k_{B}T$ in accordance with Eq.~(\ref{sf1}). The net carrier density $n$ (green line) also shows an increasing trend as $\mu$ increases. At $\mu/k_{B}T\geq 3$, the net density almost align with the total electron density $n_{e}$ and $n_{h}\xrightarrow{}0$. This suggests that in gate biased graphene devices at $T=60$ K ($k_{B}T\sim 5$ meV), for $\mu> 15$ meV, the major charge carriers are electrons, and their density is of the order of $10^{10}$ cm$^{-2}$. Therefore, at $T=60$ K, we have an electron-hole DF for $\mu\ll 15$ meV and an electron FL for $\mu> 15$ meV. In Fig.~(\ref{fig:1})(b), the energy density $\Ep$ and pressure $P$ are plotted as a function of $\mu/k_{B}T$ at $T=60$ K. We employ Eqs.~(\ref{f12}) and~(\ref{f13}) to obtain $\Ep$ and $P$ as a function of $\mu/k_{B}T$. The pressure is defined as the force per unit length in 2D and measured in the units of eV/cm$^{2}$ (SI unit is N/m or J/m$^{2}$). The energy density defined as the total thermodynamic energy per unit area of the graphene samples shares same unit as that of pressure. The total energy density (green line) increases monotonically with a change in $\mu$. At $T=60$ K, the estimated energy density of carriers lies in the range $\sim 0.1 \text{ to } 1$ eV/cm$^{2}$ for $\mu$ in the range $5 \text{ to } 25$ meV. The curve for pressure $P$ (red line) follows the same qualitative trend as energy density $\Ep$, with its magnitude being exactly half of $\Ep$ in agreement with Eq.~(\ref{apA9}). Now, let us see the changes to the thermodynamic variables of QGP by varying $\mu_{q}/k_{B}T$ at $T\approx10^{12}$ K or $\approx 100$ MeV. The net quark density $\tilde{n}$ (blue line) of QGP as a function of  $\mu_{q}/k_{B}T$ is shown in Fig.~(\ref{fig:2})(a). We apply the formula Eq.~(\ref{q2}) laid down in the Sec.~\ref{QGPT} to obtain the numerical magnitude of $\tilde{n}$ for different $\mu_{q}/k_{B}T$. The net quark density (measured in fm$^{-3}$) shows a monotonic rise with an increase in $\mu_{q}$. As a reference point, nuclear saturation density $n= 0.16$ fm$^{-3}$ is marked (red dot-dashed line). The results show that with reference to the nuclear saturation, the net quark density has a crossing point at $\mu_{q}= 300$ MeV. Fig.~(\ref{fig:2})(b) portrays the change in energy density $\tilde{\Ep}$ (blue line) and pressure $\tilde{P}$ (orange line) in relation to $\mu_{q}/k_{B}T$ at $T=10^{12}$ K. The pressure is defined as the force per unit area in 3D and measured in the units GeV/fm$^{3}$ (SI unit: N/m$^{2}$ or Pascal). The energy density known as the thermodynamic energy per unit volume shares the same unit that of pressure. Both $\tilde{\Ep}$ and $\tilde{P}$ increases monotonically with change in $\mu_{q}$ consistent with the Eq.~(\ref{q3}). The magnitude of energy density for QGP lies between $0.01 \text{ to } 0.21$ GeV/fm$^{3}$ for $\mu_{q}$ between $0 \text{ to } 500$ MeV.     


In Fig.~(\ref{fig:3})(a), we show the variation of enthalpy per net carriers normalized by the chemical potential $\frac{\mathfrak{h}}{\mu}$ (blue line) with respect to $\frac{\mu}{k_{B}T}$ at $T=60$ K. Enthalpy per net carriers is an important thermodynamic parameter in characterizing the transport phenomena in the hydrodynamic regime. $\mathfrak{h}$ appears explicitly in the expression of electrical conductivity, thermal conductivity and LR. Analyzing the $\Ep$ and $P$ curves in Fig.~(\ref{fig:1})(b) and net electron density $n$ in Fig.~(\ref{fig:1})(a), one can notice that towards $\frac{\mu}{k_{B}T}\rightarrow 0$ domain,  $\Ep$ and $P$ saturate towards finite values but $n$ tends to zero. This is the reason for the divergence in $\mathfrak{h}$ or  $\frac{\mathfrak{h}}{\mu}$ at $\frac{\mu}{k_{B}T}\rightarrow 0$. Since $\mu$ can be identified with Gibb's free energy per net carriers, the vertical axis in the plot represents the ratio of enthalpy to  Gibb's free energy. The plot displays a decreasing trend of $\frac{\mathfrak{h}}{\mu}$ with $\frac{\mu}{k_{B}T}$ and almost aligns with the line (red dash-dotted line) at unity after $\mu>30$ meV (or $\frac{\mu}{k_{B}T}>6$). This suggests that the enthalpy of the graphene significantly differs from Gibb's free energy in the DF region and almost matches with Gibb's free energy in the FL domain. In terms of Euler thermodynamic relation $Ts= \Ep + P - \mu n$, it is understandable that for $\frac{\mu}{k_{B}T} \gg 1$ or large $\mu$ and small $T$, we can write the  thermodynamic relation as,
\begin{eqnarray}
	&&\Ep +P - \mu n\approx 0 \nn\\
	\implies &&\frac{\Ep+P}{n} =\mu \nn
\end{eqnarray}
which is reflected in Fig.~(\ref{fig:3})~(a).

Next, the variation of normalized electrical $(\sigma^{\prime})$ and thermal $(\kappa^{\prime})$ conductivities of graphene with respect to $n$ at  $T=60$ K is shown in Fig.~(\ref{fig:3})(b). The heat flow in 2D is described by the heat current through a unit length (measured in J m$^{-1}$s$^{-1}$ or W m$^{-1}$). The charge flow is defined similar to heat flow and is measured in the units (SI) of A m$^{-1}$ in 2D. The units of electrical and thermal conductivities can be derived from those of charge and heat flows. In 2D, the units of $\kappa$ and $\sigma$ is W K$^{-1}$ and A V$^{-1}$, respectively. Here we make them dimensionless to show their variations in a common plot. The dimensionless $(\sigma^{\prime})$ and $(\kappa^{\prime})$ are defined as follows, from Eq.~(\ref{f28}) and (\ref{f29}).
\bea
&&\sigma^{\prime}=\frac{\sigma h^{2}}{4 e^{2}\pi\tau_{c}k_{B}T}=\frac{2n k_{B}T}{\Ep+P} (f_{2}(A^{-1})-f_{2}(A))+ (f_{1}(A^{-1})+f_{1}(A))~,\label{nsigma}\\
&&\kappa^{\prime}=\frac{\kappa h^{2}}{4\pi\tau_{c}k_{B}^{3}T^{2}}=\left(\frac{\Ep+P}{nk_{B}T}\right)^{2}\bigg[ \frac{2n k_{B}T}{\Ep+P}(f_{2}(A^{-1})-f_{2}(A))+ (f_{1}(A^{-1})+f_{1}(A))\bigg]~.\label{nkappa}
\eea
The magnitude of $\kappa^{\prime}$ (red line) shows interesting behavior as one changes the net carrier density $n$. As the carrier density decreases in the FL domain $(\frac{\mu}{k_{B}T}\gg 1)$ $\kappa^{\prime}$ first decreases and then increases significantly in the DF regime $(\frac{\mu}{k_{B}T}\ll 1)$ with $\kappa^{\prime}\xrightarrow{} \infty$ as $\mu \xrightarrow{}0$. This singular behavior can be understood from Eq.~(\ref{nkappa}), where the factor $\frac{\Ep+P}{nk_{B}T}$ diverges as $n\xrightarrow{}0$. For $\sigma^{\prime}$ (green line), it can be observed that in the same range of $n$, $\sigma^{\prime}$ is almost constant, and its magnitude lies way below $\kappa^{\prime}$. In Fig.~(\ref{fig:4})(a) we show the variation of $\frac{\tilde{\mathfrak{h}}}{\mu_{q}}$ (blue line) with change in $\mu_{q}/k_{B}T$ at $T=10^{12}$ K. Similar to the graphene case, for $\mu_{q}/k_{B}T\ll 1$, $\frac{\mathfrak{h}}{\mu_{q}}$ of quark matter significantly differs from one (shown in red dash-dotted line). The heat flow in 3D is specified by the rate of heat transport across a unit area (measured in J m$^{-2}$s$^{-1}$ or W m$^{-2}$). The charge flow is defined similar to heat flow and is measured in the units (SI) of A m$^{-2}$ in 3D. The units of $\kappa$ and $\sigma$ read as W m$^{-1}$ K$^{-1}$ and A V$^{-1}$ m$^{-1}$, respectively. We instead make them dimensionless to show their variation in a common plot. The normalized $\tilde{\sigma}^{\prime}$ and $\tilde{\kappa}^{\prime}$ for QGP are shown in Fig.~(\ref{fig:4})(b) with respect to $\mu_{q}/k_{B}T$ by using the following formula:
\bea
&& \tilde{\sigma}^{\prime}=\frac{3  h^{3}c \tilde{\sigma}}{16 Q_{u}^{2} \pi \tau k_{B}^{2} T^{2}}=  \frac{3\tilde{n} k_{B}T}{\tilde{\Ep}+\tilde{P}}(f_{3}(A^{-1})-f_{3}(A))+ (f_{2}(A^{-1})+f_{2}(A))\label{nsigmaQ}\\
&& \tilde{\kappa}^{\prime}=\frac{3  h^{3}c \tilde{\kappa}}{16\pi\tau k_{B}^{4} T^{3}}=\left(\frac{\tilde{\Ep}+\tilde{P}}{\tilde{n}k_{B}T}\right)^{2} \bigg[\frac{3\tilde{n} k_{B}T}{\tilde{\Ep}+\tilde{P}}(f_{3}(A^{-1})-f_{3}(A))+ (f_{2}(A^{-1})+f_{2}(A))\bigg].\label{nkappaQ}
\eea
The qualitative trends of both $\tilde{\kappa}^{\prime}$ and $\tilde{\sigma}^{\prime}$ are same as that of the graphene with $\tilde{\kappa}^{\prime}$ being significantly higher than $\tilde{\sigma}^{\prime}$ at low net quark densities. 

Finally, in Fig.~(\ref{fig:5}) (a) and (b), we depicted the LR of graphene with reference to $n$ at $T=60$ K and that of QGP with reference to $\tilde{n}$ at $T=10 ^{12}$ K respectively. The LR has no units; nevertheless, it has different variations depending on the spatial dimension of the system. It is easily seen by examining the factors $\frac{\Ep+P}{nk_{B}T}=3~\frac{f_{3}(A)+f_{3}(A^{-1})}{f_{2}(A)+f_{2}(A^{-1})}$ and $\frac{\tilde{\Ep}+\tilde{P}}{\tilde{n}k_{B}T}=4~\frac{f_{4}(A)+f_{4}(A^{-1})}{f_{3}(A)+f_{3}(A^{-1})}$.  These factors can be expressed in a general way as, $(\rm D+1)~\frac{f_{\rm D+1}(A)+f_{\rm D+1}(A^{-1})}{f_{\rm D}(A)+f_{\rm D}(A^{-1})}$, where D is the dimension of the system. In Fig.~(\ref{fig:5})(a) we also present the experimental data of Ref.~\cite{crossno2016observation} corresponds to $S_{1}$ (green dots) and $S_{2}$ (red dots), where $S_{1}$ is relatively cleaner sample than that of $S_{2}$. We divide the graph into three regions: fluidic, mixed, and Ohmic. The fluidic region corresponds to the DF regime ($\mu\ll k_{B}T$), and hydrodynamics is applicable here. In contrast, the Ohmic region corresponds to the FL domain  ($\mu\gg k_{B}T$) where carrier transport is diffusive and hydrodynamics is not applicable. Between the fluidic and Ohmic regions, a domain exists where carrier dynamics are neither completely hydrodynamic nor diffusive; this region is named mixed. It can be observed that the experimental data corresponding to $S_{1}$ and $S_{2}$ lie far away from the line $L/L_{0}=1$ (black dashed) for $n<10^{10}$ cm$^{-2}$ and co-linear when $n>2\times10^{10}$ cm$^{-2}$. We can see that LR obtained of Eq.~(\ref{f30}) shows a similar kind of divergence tendency as experimentally observed by Crossno et al.~\cite{crossno2016observation}. We notice that experimental data of $S_1$ and $S_2$ both have divergence tendencies in the fluidic domain, and our fluid-base theoretical curve is in good agreement with the data from $S_1$. Since our present theory does not have quantitative inputs regarding sample cleanness, instead of quantitative matching of experimental data, we highlight the qualitative divergence tendency in the fluid domain, as noticed in both theory and experiment. 

Interestingly, we have identified three domains in Fig.~(\ref{fig:5})(a) for various $\frac{\mu}{k_B T}$ values or varying $n$ at $T= 60$ K. They are $(1)$ fluidic domain ($n< 8 \times 10^9 $ cm$^{-2}$), $(2)$ mixed domain $(8~\times~10^9$ cm$^{-2} < n < 1.5 \times10^{10}$ cm$^{-2})$ and $(3)$ non fluidic or Ohmic domain $(n > 1.5 \times10^{10}$ cm$^{-2})$. Large net carrier density $(n > 1.5 \times10^{10}$ cm$^{-2})$ or large $\frac{\mu}{k_B T}$ ($i.e., \frac{\mu}{k_B T} \gg $1), traditional Ohmic domain prevails, where the LR$=1$, \textit{i.e.}, $L=L_0$, the universal constant. Metals as 3-dimensional non-relativistic matter~\cite{win2024wied} follow this constant Lorenz ratio rule, known as WF law because their Fermi energy remains within $\mu=2-10$ eV, which is located in Ohmic domain $(\frac{\mu}{k_B T}\gg 1)$. However, in a 2-dimensional graphene system, the Fermi energy $\mu$ or net charge carrier density can be reduced from $\frac{\mu}{k_B T}\gg 1$ to $\frac{\mu}{k_B T}\ll 1$. Recent experimental measurements~ \cite{Bandurin2016,levitov2016electron,krishna2017superballistic,BandurinF18,Patrick2019,sulpizio2019visualizing,ABerdyugin2019,Ella2019,ku2020imaging,crossno2016observation,block2021observation} suggest that $(\frac{\mu}{k_B T}\ll 1)$ domain show fluid property (popularly called DF). Therefore, we used fluid-based LR expression for the fluidic domain $(n< 8 \times 10^9 $ cm$^{-2}$), and an increasing trend is observed with the decreasing value of $n$. Experimental data for $n<~8~\times 10^9$ cm$^{-2}$ and $n> 1.5 \times 10^9$ cm$^{-2}$  follow respectively fluid-based and non-fluid/ Ohmic based theories. It indicates a mixed phenomenon or a transition from non-fluid to fluid phenomena within $8\times10^9$ cm$^{-2}$ $< n < 1.5 \times10^{10}  $ cm$^{-2}$ domain. One can use switching functions~\cite{win2024wied} to show a smooth transition from the non-fluid to the fluid domain; however, detecting the actual mechanism of transition may be interesting for future research.

Readers can go through earlier Refs.~\cite{AnLucas2016,tu2023wiedemann,ma14112704}, which are focused on explaining this WF law violation \cite{crossno2016observation}. In Ref.~\cite{tu2023wiedemann}, both the band gap and the case without a band gap in graphene are considered, along with bipolar diffusion. Additionally, Ref.~\cite{ma14112704} employs thermodynamic aspects through the poly-logarithmic function to address the enhancement of LR near the Dirac point. Among them, Ref.~\cite{AnLucas2016} has provided a more quantitative matching of WF law violation, and they have broadly marked that a fluid aspect is a possible reason for the violation. In this regard, we are also pointing out similar physics qualitatively matching the data from Crossno et al.'s~\cite{crossno2016observation}. The present work is intended to explore similar hydrodynamical structures in graphene and QGP systems in relation to WF law violation.

In the QGP system, Refs.~\cite{JAISWAL2015548,Sahoo:2019xjq,Rath:2019nne,singh2023effect,pradhan2023conductivity} have found a similar kind of WF law violation at $\mu_{q}=0$, expected to be produced in RHIC or LHC experiments. Fig.~(\ref{fig:5})(b) demonstrates $\frac{\tilde{L}}{L_0}$ of massless quark matter, which also supports the WF law violation due to the fluid aspect of quark matter. At high density $n\sim 2n_{0}-4n_{0}~(n_0= 0.16$ fm$^{-3})$, which is expected to be produced in CBM or NICA experiments and expected in neutron star environment, one may explore the $\frac{\tilde{L}}{L_0}$ calculations, which is not yet established. The future experimental data from CBM and NICA may play an important role in deciding the requirement of a hydrodynamic or non-fluid framework for degenerate quark or hadronic matter. Our simple massless quark matter calculation suggests that if high-density QGP carries a fluid aspect, it would still violate WF law. However, a systemic review of earlier research and future research may be required to conclude something on WF law in high-density QGP systems.     

\section{Summary and conclusions}\label{sum}
In summary, we have explored the thermoelectric transport in graphene and its equivalence with quark-gluon plasma both qualitatively and quantitatively. To obtain the thermoelectric transport coefficients in graphene in the DF regime, we have used a covariant BTE with the light speed $c$ replaced by Fermi velocity $v_{F}$. The electric field and thermal gradients in the graphene sample give rise to dissipative flows like electrical and thermal flows, which have been evaluated with the help of the BTE. We observe that the enthalpy per net carriers in graphene plays a decisive role in determining the numerical magnitude of the thermal and electrical conductivity. The enthalpy per net carriers significantly differs from the chemical potential at the Dirac point and affects the charge transport in the DF regime. Our calculation suggests a divergence of the LR at low net carrier density, consistent with the recent experimental measurements. The qualitative patterns obtained for the thermodynamic variables and transport coefficients for graphene and QGP are similar. Two factors that make the difference in the quantitative magnitude of the thermodynamic and transport coefficients of QGP and graphene are the dimensionality (three for QGP and two for graphene) and the speed ($c$ for QGP and $v_{F}$ for carriers in graphene). Now, when we follow the experimental data of the LR for the graphene, we notice that low and high carrier density obey fluid and non-fluid calculations, respectively. Since these two theoretical estimations do not merge in the intermediate density range, a mixed phenomenon or transition from non-fluid to fluid is expected.

Similar to the graphene case, we can get indirect experimental data of the LR for the QGP at low and high densities. RHIC and LHC experiments already indicated the fluid aspect of QGP at low density; therefore, the violation of WF law can also be expected here and confirmed recently~\cite{JAISWAL2015548,Sahoo:2019xjq,Rath:2019nne,singh2023effect,pradhan2023conductivity}. Future experiments at CBM and NICA may provide a better knowledge of quark matter created at high density from which the correct framework to determine LR and validity or violation of WF law can be inferred for the dense QGP.

\section{Acknowledgement}
This work was partially supported by the Ministry of Education, Govt. of India (A.D., S.N.); and the Board of Research in Nuclear Sciences and Department of Atomic Energy, Govt. of India, under Grant No. 57/14/01/2024-BRNS/313 (S.G.). The authors extend their thanks to the other members of the electron hydrodynamics club--Cho Win Aung, Thandar Zaw Win, and Narayan Prasad.

{\bf Author contributions:}\\
 Conceptualization, A.D., S.N, S.G. and S.V.; methodology, A.D. and S.G.;
validation, A.D., S.N., S.G. and S.V.; formal analysis, A.D., S.N. and S.S.K.; investigation, A.D., S.N, S.S.K., S.G. and S.V.; writing—original draft preparation, A.D. and S.N.; writing—review and editing, S.G. and
S.V.; supervision, S.G. and S.V. All authors have read and agreed to the published version of the
manuscript.\\
{\bf Funding: } This work was partially supported by the Board of Research in Nuclear Sciences and Department of Atomic Energy, Govt. of India, under Grant No. 57/14/01/2024-BRNS/313.\\
{\bf Data Availability Statement:} 
This study is purely theoretical and does not generate new datasets. The experimental data referenced for comparison are available in their respective original publications, as cited in the manuscript.\\
{\bf Code Availability Statement:} This study does not involve custom computational code.\\
\\
{\bf Declarations}\\
{\bf Conflict of interest:} The authors declare no conflicts of interest relevant to this study.
\section{appendix}
\subsection{Calculation of $\delta f_{e,h}$ as sum of bulk, shear and thermoelectrical parts}\label{ape1}
In this appendix we will briefly show the transition from Eq.~(\ref{f18}) and~(\ref{f19}) to Eq.~(\ref{f23}) and~(\ref{f24}) \textit{via} using the conservation Eqs.~(\ref{f20}) to~(\ref{f22}). The net number density $n$ and total energy density $\Ep$ being function of $\frac{\mu}{k_{B}T}$ and $\frac{1}{k_{B}T}$ can be differentiated with operator $D$ as follows:
\bea
&&Dn=\frac{\del n}{\del\left(\frac{\mu}{k_{B}T}\right)} D\frac{\mu}{k_{B}T} +  \frac{\del n}{\del\left(\frac{1}{k_{B}T}\right)} D\frac{1}{k_{B}T}~,\label{apB1}\\
&& D\Ep=\frac{\del \Ep}{\del\left(\frac{\mu}{k_{B}T}\right)} D\frac{\mu}{k_{B}T} +  \frac{\del \Ep}{\del\left(\frac{1}{k_{B}T}\right)} D\frac{1}{k_{B}T}~.\label{apB2}
\eea
Now using the conservation laws given by the Eq.~(\ref{f20}) and Eq.~(\ref{f22}) in the Eqs.~(\ref{apB1}) and~(\ref{apB2}) respectively we have,
\bea
&&\frac{\del n}{\del\left(\frac{\mu}{k_{B}T}\right)} D\frac{\mu}{k_{B}T} +  \frac{\del n}{\del\left(\frac{1}{k_{B}T}\right)} D\frac{1}{k_{B}T}=-n\na_{\mu}u^{\mu}\label{apB3},\\
&&\frac{\del \Ep}{\del\left(\frac{\mu}{k_{B}T}\right)} D\frac{\mu}{k_{B}T} +  \frac{\del \Ep}{\del\left(\frac{1}{k_{B}T}\right)} D\frac{1}{k_{B}T}=-(\Ep+P)\na_{\mu}u^{\mu}\label{apB4}.
\eea
The Eq.~(\ref{apB3}) and~(\ref{apB4}) are two linear equation and can be easily solved for $D\frac{1}{k_{B}T}$ and $D\frac{\mu}{k_{B}T}$. The solution of these equations are of the form: $D\frac{1}{k_{B}T}=x_{1}\na_{\mu}u^{\mu}$ and $D\frac{\mu}{k_{B}T}=x_{2}\na_{\mu}u^{\mu}$. The quantities $x_{1}$ and $x_{2}$ are functions of thermodynamical variables, and their exact expressions are irrelevant to our purpose. Now we eliminate all the temporal derivative terms from Eq.~(\ref{f18}) and Eq.~(\ref{f19}) by substituting $D\frac{1}{k_{B}T}=x_{1}\na_{\mu}u^{\mu}$,~ $D\frac{\mu}{k_{B}T}=x_{2}\na_{\mu}u^{\mu}$, and Eq.~(\ref{f21}). Rewriting Eq.~(\ref{f18}) in terms of spatial derivatives we have,
\bea
&&\delta f_{e}=\frac{\tau_{c}v_{F}^{2}}{u_{\mu}p^{\mu}_{e}}\Bigg[v_{F}^{-2}\big(x_{1} (u_{\al}p^{\al}_{e})^{2}-x_{2}(u_{\al}p^{\al}_{e})\big)\na_{\beta}u^{\beta}+ \frac{p^{\al}_{e}p^{\beta}_{e}}{k_{B}T}\na_{\al}u_{\beta}+ \frac{(u_{\beta}p^{\beta}_{e})p_{e}^{\al}}{k_{B}T}\frac{\na_{\al}P}{\Ep+P} + \frac{(u_{\beta}p^{\beta})p_{e}^{\al}}{k_{B}T}\frac{\rho}{\Ep+P}\tilde{E}_{\al}\nn\\
&&+(u_{\beta}p^{\beta}_{e}) p^{\al}_{e}\na_{\al}\frac{1}{k_{B}T}-p_{e}^{\al}\na_{\al}\frac{\mu}{k_{B}T} + \frac{e p_{e}^{\al}}{k_{B}T}\tilde{E}_{\al}\Bigg]f^{0}_{e}(1-f^{0}_{e})~.\label{apB5}
\eea
From the Gibbs-Duhem relation, we have,
\bea
&&n ~d\frac{\mu}{k_{B}T}=\frac{1}{k_{B}T}~dP+(\Ep+P)~d\frac{1}{k_{B}T}\nn\\
\implies&& n ~\na^{\al}\frac{\mu}{k_{B}T}=\frac{1}{k_{B}T}~\na^{\al}P+(\Ep+P)~\na^{\al}\frac{1}{k_{B}T}\nn\\
\implies&& \frac{(u_{\beta}p_{e}^{\beta})p_{e}^{\al}}{k_{B}T}\frac{\na_{\al}P}{\Ep+P}= (u_{\beta}p_{e}^{\beta})p_{e}^{\al}\left[ \frac{n}{\Ep+P}\na_{\al}\frac{\mu}{k_{B}T}- \na_{\al}\frac{1}{k_{B}T} \right]~.\label{apB6}
\eea
Using Eq.~(\ref{apB6}) in Eq.~(\ref{apB5}) we have,
\bea
&&\delta f_{e}=\frac{\tau_{c}v_{F}^{2}}{u_{\mu}p^{\mu}_{e}}\Bigg[v_{F}^{-2}\big(x_{1} (u_{\al}p^{\al}_{e})^{2}-x_{2}(u_{\al}p^{\al}_{e})\big)\na_{\beta}u^{\beta}+\frac{p^{\al}_{e}p^{\beta}_{e}}{k_{B}T}\na_{\al}u_{\beta}\nn\\
&&-p^{\al}_{e}\bigg[\frac{n}{\Ep+P}u_{\beta}p^{\beta}_{e}-1\bigg]\left(-\na_{\al}\frac{\mu}{k_{B}T}+\frac{e\tilde{E}_{\al}}{k_{B}T} \right)\Bigg]f^{0}_{e}(1-f^{0}_{e})~.\label{apB7}
\eea
Rewriting Eq.~(\ref{apB7}) using the result: $p^{\al}_{e}p^{\beta}_{e}~\na_{\al}u_{\beta}=p^{\al}_{e}p^{\beta}_{e}\sigma_{\al\beta}+\frac{1}{2}\D_{\al\nu}p^{\al}_{e}p^{\nu}_{e}(\na_{\beta}u^{\beta})$, where $\sigma_{\al\beta}\equiv \frac{1}{2}(\na_{\al}u_{\beta}+\na_{\beta}u_{\al})-\frac{1}{2}\D_{\al\beta}(\na_{\mu}u^{\mu})$, we have,
\bea
&&\delta f_{e}=\frac{\tau_{c}v_{F}^{2}}{u_{\mu}p^{\mu}_{e}}\Bigg[v_{F}^{-2}\big(x_{1} (u_{\al}p^{\al}_{e})^{2}-x_{2}(u_{\al}p^{\al}_{e})+\frac{1}{2k_{B}T}\D_{\al\nu}p^{\al}_{e}p^{\nu}_{e}\big)\na_{\beta}u^{\beta}+\frac{p^{\al}_{e}p^{\beta}_{e}}{k_{B}T}\sigma_{\al\beta}\nn\\
&&-p^{\al}_{e}\bigg[\frac{n}{\Ep+P}u_{\beta}p^{\beta}_{e}-1\bigg]\left(-\na_{\al}\frac{\mu}{k_{B}T}+\frac{e\tilde{E}_{\al}}{k_{B}T} \right)\Bigg]f^{0}_{e}(1-f^{0}_{e})~.\label{apB8}
\eea
Using similar steps one easily finds the out-of-equilibrium hole distribution $\delta f_{h}$ in terms of spatial gradients as:
\bea
&&\delta f_{h}=\frac{\tau_{c}v_{F}^{2}}{u_{\mu}p^{\mu}_{h}}\Bigg[v_{F}^{-2}\big(x_{1} (u_{\al}p^{\al}_{h})^{2}-x_{2}(u_{\al}p^{\al}_{h})+\frac{1}{2k_{B}T}\D_{\al\nu}p^{\al}_{h}p^{\nu}_{h}\big)\na_{\beta}u^{\beta}+\frac{p^{\al}_{h}p^{\beta}_{h}}{k_{B}T}\sigma_{\al\beta}\nn\\
&&-p^{\al}_{h}\bigg[\frac{n}{\Ep+P}u_{\beta}p^{\beta}_{h}+1\bigg]\left(-\na_{\al}\frac{\mu}{k_{B}T}+\frac{e\tilde{E}_{\al}}{k_{B}T} \right)\Bigg]f^{0}_{h}(1-f^{0}_{h})~.\label{apB9}
\eea
In Eq.~(\ref{apB8}) and~(\ref{apB9}), the terms proportional to $\na_{\beta}u^{\beta}$ and $\sigma_{\al\beta}$ give rise to bulk and shear viscosity in the graphene. One can readily verify that upon ignoring the terms that correspond to the viscosity in the graphene, the out-of-equilibrium distribution functions for electron and hole matches with Eq.~(\ref{f23}) and Eq.~(\ref{f24}). 

\subsection{Evaluation of thermodynamic variables and current density in terms of Fermi integral functions}\label{ape2}
Here, we  define the Fermi integral functions for electrons and holes used in the main text and briefly describe their properties. Finally, we express the thermodynamic variables in terms of these integrals.
Let us define the Fermi integrals for electrons as:
\bea
&&f_{j}(A)=\frac{1}{\Gamma(j)}\int_{0}^{\infty}\frac{x^{j-1}~dx}{e^{x}A^{-1}+1},~(A\equiv e^{\mu/k_{B}T})~.\label{apA1}
\eea
In the domain $j> 0$ and $-\infty < \frac{\mu}{k_{B}T}\leq 0$, one can represent $f_{j}(A)$ by the following series,
\bea
f_{j}(A)&=& A - \frac{A^{2}}{2^{j}} + \frac{A^{3}}{3^{j}} - ........=e^{\mu/k_{B}T}- \frac{e^{2\mu/k_{B}T}}{2^{j}} + \frac{e^{3\mu/k_{B}T}}{3^{j}} - ............~.\label{apA2}
\eea
For the region $j> 0$ and $\frac{\mu}{k_{B}T}\gg 1$, one can express $f_{j}(A)$ in the form of Sommerfeld's series as follows:
\bea
f_{j}(A)&=&\frac{1}{\Gamma(j+1)}\left(\frac{\mu}{k_{B}T}\right)^{j}+\frac{1}{\Gamma(j-1)}\frac{\pi^{2}}{6}\left(\frac{\mu}{k_{B}T}\right)^{j-2} +................~.\label{apA3}
\eea
The Fermi integrals for holes can be defined in a similar manner as follows:
\bea
&&f_{j}(A^{-1})=\frac{1}{\Gamma(j)}\int_{0}^{\infty}\frac{x^{j-1}~dx}{e^{x}A+1},~(A\equiv e^{\mu/k_{B}T})~.\label{apA4}
\eea
Different limiting expressions for the function $f_{j}(A^{-1})$ can be obtained from $f_{j}(A)$ by the replacement $\mu\xrightarrow{}-\mu$. We will write down these expressions explicitly,
\bea
f_{j}(A^{-1})&=& A^{-1} - \frac{A^{-2}}{2^{j}} + \frac{A^{-3}}{3^{j}} - ........=e^{-\mu/k_{B}T}- \frac{e^{-2\mu/k_{B}T}}{2^{j}} + \frac{e^{-3\mu/k_{B}T}}{3^{j}} - ............~,\label{apA5}
\eea
where $j> 0$ and $0 \leq \frac{\mu}{k_{B}T}<\infty $. Similarly, the Sommerfeld's series expression for the hole distribution is,
\bea
f_{j}(A^{-1})&=&\frac{1}{\Gamma(j+1)}\left(\frac{-\mu}{k_{B}T}\right)^{j}+\frac{1}{\Gamma(j-1)}\frac{\pi^{2}}{6}\left(\frac{-\mu}{k_{B}T}\right)^{j-2} +................~,\label{apA6}
\eea
where $j> 0$ and $\frac{-\mu}{k_{B}T}\gg 1$.

Now, we will move onto express the thermodynamic variables in terms of Fermi integrals. The thermodynamic integral for $n$ specified in Eqs.~(\ref{f11}) can be simplified as,
\bea
&&n=4\bigg[\int \frac{d^{2}\Vec{p}_{e}}{h^{2}} f^{0}_{e}-\int \frac{d^{2}\Vec{p}_{h}}{h^{2}} f^{0}_{h}\bigg]\nn\\
\implies&& n=4\bigg[\int_{0}^{\infty} \frac{d^{2}\Vec{p}_{e}}{h^{2}} \frac{1}{e^{(E_{e}-\mu)/k_{B}T}+1}-\int_{0}^{\infty} \frac{d^{2}\Vec{p}_{h}}{h^{2}} \frac{1}{e^{(E_{h}+\mu)/k_{B}T}+1}\bigg]\nn\\
\implies && n=\frac{8\pi}{h^{2}}\left(\frac{k_{B}T}{v_{F}}\right)^{2}\Bigg[\frac{1}{\Gamma(2)} \int_{0}^{\infty} \frac{x^{2-1}dx}{e^{x}A^{-1}+1}-\frac{1}{\Gamma(2)} \int_{0}^{\infty} \frac{x^{2-1}dx}{e^{x}A+1}  \Bigg], (\text{where }x\equiv\frac{p_{e}v_{F}}{k_{B}T})\nn\\
\implies && n=\frac{8\pi}{h^{2}}\left(\frac{k_{B}T}{v_{F}}\right)^{2}\left(f_{2}(A)-f_{2}(A^{-1}) \right)~.\label{apA7}
\eea
It can be easily seen from Eq.~(\ref{apA7}) that in the limit $\frac{\mu}{k_{B}T}\xrightarrow{}0$ we have $n=0$ and for $\frac{\mu}{k_{B}T}\gg 1$ we get,
\be
n=\frac{8\pi\mu^{2}}{(hv_{F})^{2}}\Bigg[\frac{1}{2}+\frac{\pi^{2}}{6}\left(\frac{k_{B}T}{\mu}\right)^2- \left(\frac{k_{B}T}{\mu}\right)^2 e^{-\mu/k_{B}T}+\frac{1}{4}\left(\frac{k_{B}T}{\mu}\right)^2 e^{-2\mu/k_{B}T}\Bigg]~.\label{sf1}
\ee
A quick look at Eq.~(\ref{sf1}) suggests that for $\mu\gg k_{B}T$, one can safely ignore the hole contribution (3rd and 4th terms in the square bracket) to the net number density. We can also simplify the energy density $\Ep$ given in Eq.~(\ref{f12}) as,
\bea
&&\Ep=4\bigg[\int \frac{d^{2}\Vec{p}_{e}}{h^{2}} E_{e}f^{0}_{e}+\int \frac{d^{2}\Vec{p}_{h}}{h^{2}} E_{h} f^{0}_{h}\bigg]\nn\\
\implies&& \Ep=4\bigg[\int_{0}^{\infty} \frac{d^{2}\Vec{p}_{e}}{h^{2}} \frac{E_{e}}{e^{(E_{e}-\mu)/k_{B}T}+1}+\int_{0}^{\infty} \frac{d^{2}\Vec{p}_{h}}{h^{2}} \frac{E_{h}}{e^{(E_{h}+\mu)/k_{B}T}+1}\bigg]\nn\\
\implies && \Ep=\frac{16\pi}{h^{2}}\frac{(k_{B}T)^{3}}{v_{F}^{2}}\Bigg[\frac{1}{\Gamma(3)} \int_{0}^{\infty} \frac{x^{2-1}dx}{e^{x}A^{-1}+1}+\frac{1}{\Gamma(3)} \int_{0}^{\infty} \frac{x^{2-1}dx}{e^{x}A+1}  \Bigg], (\text{where }x\equiv\frac{p_{e}v_{F}}{k_{B}T})\nn\\
\implies && \Ep=\frac{16\pi}{h^{2}}\frac{(k_{B}T)^{3}}{v_{F}^{2}}\left(f_{3}(A)+f_{3}(A^{-1}) \right)~.\label{apA8}
\eea
In the limit $\frac{\mu}{k_{B}T}\xrightarrow{}0$ we have $\Ep=\frac{32\pi}{h^{2}}\frac{(k_{B}T)^{3}}{v_{F}^{2}}\eta(3)$, where the Dirichlet eta function $\eta(j)=f_{j}(1)= \sum_{n=1}^{\infty}\frac{(-1)^{n-1}}{n^{j}}$ and in the limit $\frac{\mu}{k_{B}T}\gg 1$ we have,
\be
\Ep=\frac{16\pi}{(hv_{F})^{2}}\mu^{3}\Bigg[\frac{1}{6}+\frac{\pi^{2}}{6}\left(\frac{k_{B}T}{\mu}\right)^{2}+ \left(\frac{k_{B}T}{\mu}\right)^{3} e^{-\mu/k_{B}T}-\frac{1}{8} \left(\frac{k_{B}T}{\mu}\right)^{3} e^{-2\mu/k_{B}T}\Bigg]~.\label{sf2}
\ee
From Eq.~(\ref{sf2}), it can be easily inferred that in the parameter range $\mu\gg k_{B}T$, the hole contribution to the energy density is negligible. Similarly, pressure $P$ can be expressed as,
\bea
P&=&4\bigg[\int \frac{d^{2}\Vec{p}_{e}}{h^{2}} \frac{p_{e}^{2}}{2E_{e}}f^{0}_{e}+\int \frac{d^{2}\Vec{p}_{h}}{h^{2}} \frac{p_{h}^{2}}{2E_{h}} f^{0}_{h}\bigg]\nn\\
\implies P&=& 2 \bigg[\int \frac{d^{2}\Vec{p}_{e}}{h^{2}} E_{e}f^{0}_{e}+\int \frac{d^{2}\Vec{p}_{h}}{h^{2}} E_{h} f^{0}_{h}\bigg]=\frac{\Ep}{2}~, \label{apA9}
\eea
where we used $p_{e,h}=\frac{E_{e,h}}{v_{F}}$.
Now, we will move on to give a step-wise derivation of the expression of current density provided in Eq.~(\ref{f26}). By using the result of an integral of type  $\int d^{2}\vec{p}~ G(|\vec{p}|)~ p^{i}p^{j}=\int d^{2}p~ G(|\vec{p}|) ~\frac{p^{2}}{2}\delta^{ij}$, $(d^{2}p\equiv 2\pi p~dp)$ where $G(|\vec{p}|)$ is any arbitrary function, we can rewrite Eq.~(\ref{f25}) as follows: 
\bea
J^{i}&=&-4e~ v_{F}^{4}\tau_{c}\bigg[\int \frac{d^{2}{p}_{e}}{h^{2}} \frac{p_{e}^{2}}{2E_{e}^{2}}\delta^{ij}\left( \frac{nE_{e}}{\Ep+P}-1\right)f^{0}_{e}(1-f^{0}_{e})-\int\frac{d^{2}{p}_{h}}{h^{2}} \frac{p_{h}^{2}}{2E_{h}^{2}} \delta^{ij}\left( \frac{nE_{h}}{\Ep+P}+1\right)f^{0}_{h}(1-f^{0}_{h})\bigg] X^{j},\nn\\
&=&-\frac{4\pi e\tau_{c}}{h^{2}}\bigg[\int_{0}^{\infty} E_{e}\left( \frac{nE_{e}}{\Ep+P}-1\right)f^{0}_{e}(1-f^{0}_{e}) dE_{e}-\int_{0}^{\infty} E_{h} \left( \frac{nE_{h}}{\Ep+P}+1\right)f^{0}_{h}(1-f^{0}_{h})dE_{h}\bigg] X^{i}\nn\\
&=& -\frac{4\pi e\tau_{c}}{h^{2}} \bigg[\frac{n}{\Ep+P}\left(\int_{0}^{\infty} E_{e}^{2}~f^{0}_{e}(1-f^{0}_{e}) dE_{e}-\int_{0}^{\infty} E_{h}^{2}~f^{0}_{h}(1-f^{0}_{h}) dE_{h}\right)\nn\\
&&-\left(\int_{0}^{\infty} E_{e}~f^{0}_{e}(1-f^{0}_{e}) dE_{e}+\int_{0}^{\infty} E_{h}~f^{0}_{h}(1-f^{0}_{h}) dE_{h}\right)\bigg] X^{i}~.\label{apC1}
\eea
We can easily evaluate the integrals occurring in Eq.~(\ref{apC1}) by using the definition of Fermi integrals and the identity $\frac{\del f_{j}(A)}{\del(\mu/k_{B}T)}=f_{j-1}(A)$ as,
\bea
\int_{0}^{\infty}E_{e}^{2}~f^{0}_{e}(1-f^{0}_{e}) dE_{e}&=&\frac{\del}{\del(\mu/k_{B}T)}\int_{0}^{\infty}E_{e}^{2}~f^{0}_{e} dE_{e}=2(k_{B}T)^{3} \frac{\del f_{3}(A)}{\del(\mu/k_{B}T)}= 2(k_{B}T)^{3} f_{2}(A)\nn\\
\text{and, }\int_{0}^{\infty}E_{e}~f^{0}_{e}(1-f^{0}_{e}) dE_{e}&=&\frac{\del}{\del(\mu/k_{B}T)}\int_{0}^{\infty}E_{e}~f^{0}_{e} dE_{e}=(k_{B}T)^{2} \frac{\del f_{2}(A)}{\del(\mu/k_{B}T)}= (k_{B}T)^{2} f_{1}(A)~.\nn
\eea
Similarly, the integrals corresponding to hole distributions can be obtained by simply replacing $\mu$ by $-\mu$. The final expression for Eq.~(\ref{apC1}) can be written as,
\bea
J^{i}&=& -\frac{4\pi e\tau_{c}}{h^{2}}\Bigg[ \frac{n}{\Ep+P} 2(k_{B}T)^{3}(f_{2}(A)-f_{2}(A^{-1})-(k_{B}T^{2})(f_{1}(A)+f_{1}(A^{-1}))\Bigg]X^{i}\nn\\
&=& \frac{4\pi e\tau_{c}}{h^{2}}\frac{nk_{B}T}{\Ep+P} \Bigg[ 2(k_{B}T)^{2}(f_{2}(A^{-1})-f_{2}(A)+\frac{\Ep+P}{n}k_{B}T(f_{1}(A)+f_{1}(A^{-1}))\Bigg] X^{i}.\nn
\eea
    
\bibliography{ref}

\begin{thebibliography}{96}%
\makeatletter
\providecommand \@ifxundefined [1]{%
 \@ifx{#1\undefined}
}%
\providecommand \@ifnum [1]{%
 \ifnum #1\expandafter \@firstoftwo
 \else \expandafter \@secondoftwo
 \fi
}%
\providecommand \@ifx [1]{%
 \ifx #1\expandafter \@firstoftwo
 \else \expandafter \@secondoftwo
 \fi
}%
\providecommand \natexlab [1]{#1}%
\providecommand \enquote  [1]{``#1''}%
\providecommand \bibnamefont  [1]{#1}%
\providecommand \bibfnamefont [1]{#1}%
\providecommand \citenamefont [1]{#1}%
\providecommand \href@noop [0]{\@secondoftwo}%
\providecommand \href [0]{\begingroup \@sanitize@url \@href}%
\providecommand \@href[1]{\@@startlink{#1}\@@href}%
\providecommand \@@href[1]{\endgroup#1\@@endlink}%
\providecommand \@sanitize@url [0]{\catcode `\\12\catcode `\$12\catcode
  `\&12\catcode `\#12\catcode `\^12\catcode `\_12\catcode `\%12\relax}%
\providecommand \@@startlink[1]{}%
\providecommand \@@endlink[0]{}%
\providecommand \url  [0]{\begingroup\@sanitize@url \@url }%
\providecommand \@url [1]{\endgroup\@href {#1}{\urlprefix }}%
\providecommand \urlprefix  [0]{URL }%
\providecommand \Eprint [0]{\href }%
\providecommand \doibase [0]{https://doi.org/}%
\providecommand \selectlanguage [0]{\@gobble}%
\providecommand \bibinfo  [0]{\@secondoftwo}%
\providecommand \bibfield  [0]{\@secondoftwo}%
\providecommand \translation [1]{[#1]}%
\providecommand \BibitemOpen [0]{}%
\providecommand \bibitemStop [0]{}%
\providecommand \bibitemNoStop [0]{.\EOS\space}%
\providecommand \EOS [0]{\spacefactor3000\relax}%
\providecommand \BibitemShut  [1]{\csname bibitem#1\endcsname}%
\let\auto@bib@innerbib\@empty
\bibitem [{\citenamefont {Novoselov}\ \emph {et~al.}(2004)\citenamefont
  {Novoselov}, \citenamefont {Geim}, \citenamefont {Morozov}, \citenamefont
  {Jiang}, \citenamefont {Zhang}, \citenamefont {Dubonos}, \citenamefont
  {Grigorieva},\ and\ \citenamefont {Firsov}}]{novoselov2004}%
  \BibitemOpen
  \bibfield  {author} {\bibinfo {author} {\bibfnamefont {K.~S.}\ \bibnamefont
  {Novoselov}}, \bibinfo {author} {\bibfnamefont {A.~K.}\ \bibnamefont {Geim}},
  \bibinfo {author} {\bibfnamefont {S.~V.}\ \bibnamefont {Morozov}}, \bibinfo
  {author} {\bibfnamefont {D.}~\bibnamefont {Jiang}}, \bibinfo {author}
  {\bibfnamefont {Y.}~\bibnamefont {Zhang}}, \bibinfo {author} {\bibfnamefont
  {S.~V.}\ \bibnamefont {Dubonos}}, \bibinfo {author} {\bibfnamefont {I.~V.}\
  \bibnamefont {Grigorieva}},\ and\ \bibinfo {author} {\bibfnamefont {A.~A.}\
  \bibnamefont {Firsov}},\ }\bibfield  {title} {\bibinfo {title} {Electric
  field effect in atomically thin carbon films},\ }\href
  {https://doi.org/10.1126/science.1102896} {\bibfield  {journal} {\bibinfo
  {journal} {Science}\ }\textbf {\bibinfo {volume} {306}},\ \bibinfo {pages}
  {666} (\bibinfo {year} {2004})},\ \Eprint
  {https://arxiv.org/abs/https://www.science.org/doi/pdf/10.1126/science.1102896}
  {https://www.science.org/doi/pdf/10.1126/science.1102896} \BibitemShut
  {NoStop}%
\bibitem [{\citenamefont {Berger}\ \emph {et~al.}(2004)\citenamefont {Berger},
  \citenamefont {Song}, \citenamefont {Li}, \citenamefont {Li}, \citenamefont
  {Ogbazghi}, \citenamefont {Feng}, \citenamefont {Dai}, \citenamefont
  {Marchenkov}, \citenamefont {Conrad}, \citenamefont {First},\ and\
  \citenamefont {de~Heer}}]{berger2004}%
  \BibitemOpen
  \bibfield  {author} {\bibinfo {author} {\bibfnamefont {C.}~\bibnamefont
  {Berger}}, \bibinfo {author} {\bibfnamefont {Z.}~\bibnamefont {Song}},
  \bibinfo {author} {\bibfnamefont {T.}~\bibnamefont {Li}}, \bibinfo {author}
  {\bibfnamefont {X.}~\bibnamefont {Li}}, \bibinfo {author} {\bibfnamefont
  {A.~Y.}\ \bibnamefont {Ogbazghi}}, \bibinfo {author} {\bibfnamefont
  {R.}~\bibnamefont {Feng}}, \bibinfo {author} {\bibfnamefont {Z.}~\bibnamefont
  {Dai}}, \bibinfo {author} {\bibfnamefont {A.~N.}\ \bibnamefont {Marchenkov}},
  \bibinfo {author} {\bibfnamefont {E.~H.}\ \bibnamefont {Conrad}}, \bibinfo
  {author} {\bibfnamefont {P.~N.}\ \bibnamefont {First}},\ and\ \bibinfo
  {author} {\bibfnamefont {W.~A.}\ \bibnamefont {de~Heer}},\ }\bibfield
  {title} {\bibinfo {title} {Ultrathin epitaxial graphite: 2d electron gas
  properties and a route toward graphene-based nanoelectronics},\ }\href
  {https://doi.org/10.1021/jp040650f} {\bibfield  {journal} {\bibinfo
  {journal} {The Journal of Physical Chemistry B}\ }\textbf {\bibinfo {volume}
  {108}},\ \bibinfo {pages} {19912} (\bibinfo {year} {2004})}\BibitemShut
  {NoStop}%
\bibitem [{\citenamefont {Novoselov}\ \emph {et~al.}(2005)\citenamefont
  {Novoselov}, \citenamefont {Geim}, \citenamefont {Morozov}, \citenamefont
  {Jiang}, \citenamefont {Katsnelson}, \citenamefont {Grigorieva},
  \citenamefont {Dubonos},\ and\ \citenamefont {Firsov}}]{novoselov2005two}%
  \BibitemOpen
  \bibfield  {author} {\bibinfo {author} {\bibfnamefont {K.~S.}\ \bibnamefont
  {Novoselov}}, \bibinfo {author} {\bibfnamefont {A.~K.}\ \bibnamefont {Geim}},
  \bibinfo {author} {\bibfnamefont {S.~V.}\ \bibnamefont {Morozov}}, \bibinfo
  {author} {\bibfnamefont {D.}~\bibnamefont {Jiang}}, \bibinfo {author}
  {\bibfnamefont {M.~I.}\ \bibnamefont {Katsnelson}}, \bibinfo {author}
  {\bibfnamefont {I.~V.}\ \bibnamefont {Grigorieva}}, \bibinfo {author}
  {\bibfnamefont {S.~V.}\ \bibnamefont {Dubonos}},\ and\ \bibinfo {author}
  {\bibfnamefont {A.~A.}\ \bibnamefont {Firsov}},\ }\bibfield  {title}
  {\bibinfo {title} {Two-dimensional gas of massless dirac fermions in
  graphene},\ }\href@noop {} {\bibfield  {journal} {\bibinfo  {journal}
  {nature}\ }\textbf {\bibinfo {volume} {438}},\ \bibinfo {pages} {197}
  (\bibinfo {year} {2005})}\BibitemShut {NoStop}%
\bibitem [{\citenamefont {Zhang}\ \emph {et~al.}(2005)\citenamefont {Zhang},
  \citenamefont {Tan}, \citenamefont {Stormer},\ and\ \citenamefont
  {Kim}}]{zhang2005experimental}%
  \BibitemOpen
  \bibfield  {author} {\bibinfo {author} {\bibfnamefont {Y.}~\bibnamefont
  {Zhang}}, \bibinfo {author} {\bibfnamefont {Y.-W.}\ \bibnamefont {Tan}},
  \bibinfo {author} {\bibfnamefont {H.~L.}\ \bibnamefont {Stormer}},\ and\
  \bibinfo {author} {\bibfnamefont {P.}~\bibnamefont {Kim}},\ }\bibfield
  {title} {\bibinfo {title} {Experimental observation of the quantum hall
  effect and berry's phase in graphene},\ }\href@noop {} {\bibfield  {journal}
  {\bibinfo  {journal} {nature}\ }\textbf {\bibinfo {volume} {438}},\ \bibinfo
  {pages} {201} (\bibinfo {year} {2005})}\BibitemShut {NoStop}%
\bibitem [{\citenamefont {Tan}\ \emph {et~al.}(2007)\citenamefont {Tan},
  \citenamefont {Zhang}, \citenamefont {Bolotin}, \citenamefont {Zhao},
  \citenamefont {Adam}, \citenamefont {Hwang}, \citenamefont {Das~Sarma},
  \citenamefont {Stormer},\ and\ \citenamefont {Kim}}]{ssarma2007}%
  \BibitemOpen
  \bibfield  {author} {\bibinfo {author} {\bibfnamefont {Y.-W.}\ \bibnamefont
  {Tan}}, \bibinfo {author} {\bibfnamefont {Y.}~\bibnamefont {Zhang}}, \bibinfo
  {author} {\bibfnamefont {K.}~\bibnamefont {Bolotin}}, \bibinfo {author}
  {\bibfnamefont {Y.}~\bibnamefont {Zhao}}, \bibinfo {author} {\bibfnamefont
  {S.}~\bibnamefont {Adam}}, \bibinfo {author} {\bibfnamefont {E.~H.}\
  \bibnamefont {Hwang}}, \bibinfo {author} {\bibfnamefont {S.}~\bibnamefont
  {Das~Sarma}}, \bibinfo {author} {\bibfnamefont {H.~L.}\ \bibnamefont
  {Stormer}},\ and\ \bibinfo {author} {\bibfnamefont {P.}~\bibnamefont {Kim}},\
  }\bibfield  {title} {\bibinfo {title} {Measurement of scattering rate and
  minimum conductivity in graphene},\ }\href
  {https://doi.org/10.1103/PhysRevLett.99.246803} {\bibfield  {journal}
  {\bibinfo  {journal} {Phys. Rev. Lett.}\ }\textbf {\bibinfo {volume} {99}},\
  \bibinfo {pages} {246803} (\bibinfo {year} {2007})}\BibitemShut {NoStop}%
\bibitem [{\citenamefont {Chen}\ \emph {et~al.}(2007)\citenamefont {Chen},
  \citenamefont {Ishigami}, \citenamefont {Jang}, \citenamefont {Hines},
  \citenamefont {Fuhrer},\ and\ \citenamefont {Williams}}]{chen2007}%
  \BibitemOpen
  \bibfield  {author} {\bibinfo {author} {\bibfnamefont {J.-H.}\ \bibnamefont
  {Chen}}, \bibinfo {author} {\bibfnamefont {M.}~\bibnamefont {Ishigami}},
  \bibinfo {author} {\bibfnamefont {C.}~\bibnamefont {Jang}}, \bibinfo {author}
  {\bibfnamefont {D.}~\bibnamefont {Hines}}, \bibinfo {author} {\bibfnamefont
  {M.}~\bibnamefont {Fuhrer}},\ and\ \bibinfo {author} {\bibfnamefont
  {E.}~\bibnamefont {Williams}},\ }\bibfield  {title} {\bibinfo {title}
  {Printed graphene circuits},\ }\href
  {https://doi.org/https://doi.org/10.1002/adma.200701059} {\bibfield
  {journal} {\bibinfo  {journal} {Advanced Materials}\ }\textbf {\bibinfo
  {volume} {19}},\ \bibinfo {pages} {3623} (\bibinfo {year} {2007})},\ \Eprint
  {https://arxiv.org/abs/https://onlinelibrary.wiley.com/doi/pdf/10.1002/adma.200701059}
  {https://onlinelibrary.wiley.com/doi/pdf/10.1002/adma.200701059} \BibitemShut
  {NoStop}%
\bibitem [{\citenamefont {Jang}\ \emph {et~al.}(2008)\citenamefont {Jang},
  \citenamefont {Adam}, \citenamefont {Chen}, \citenamefont {Williams},
  \citenamefont {Das~Sarma},\ and\ \citenamefont {Fuhrer}}]{dassarmaJa2008}%
  \BibitemOpen
  \bibfield  {author} {\bibinfo {author} {\bibfnamefont {C.}~\bibnamefont
  {Jang}}, \bibinfo {author} {\bibfnamefont {S.}~\bibnamefont {Adam}}, \bibinfo
  {author} {\bibfnamefont {J.-H.}\ \bibnamefont {Chen}}, \bibinfo {author}
  {\bibfnamefont {E.~D.}\ \bibnamefont {Williams}}, \bibinfo {author}
  {\bibfnamefont {S.}~\bibnamefont {Das~Sarma}},\ and\ \bibinfo {author}
  {\bibfnamefont {M.~S.}\ \bibnamefont {Fuhrer}},\ }\bibfield  {title}
  {\bibinfo {title} {Tuning the effective fine structure constant in graphene:
  Opposing effects of dielectric screening on short- and long-range potential
  scattering},\ }\href {https://doi.org/10.1103/PhysRevLett.101.146805}
  {\bibfield  {journal} {\bibinfo  {journal} {Phys. Rev. Lett.}\ }\textbf
  {\bibinfo {volume} {101}},\ \bibinfo {pages} {146805} (\bibinfo {year}
  {2008})}\BibitemShut {NoStop}%
\bibitem [{\citenamefont {Chen}\ \emph {et~al.}(2008)\citenamefont {Chen},
  \citenamefont {Jang}, \citenamefont {Adam}, \citenamefont {Fuhrer},
  \citenamefont {Williams},\ and\ \citenamefont {Ishigami}}]{ChenJ2008}%
  \BibitemOpen
  \bibfield  {author} {\bibinfo {author} {\bibfnamefont {J.-H.}\ \bibnamefont
  {Chen}}, \bibinfo {author} {\bibfnamefont {C.}~\bibnamefont {Jang}}, \bibinfo
  {author} {\bibfnamefont {S.}~\bibnamefont {Adam}}, \bibinfo {author}
  {\bibfnamefont {M.~S.}\ \bibnamefont {Fuhrer}}, \bibinfo {author}
  {\bibfnamefont {E.~D.}\ \bibnamefont {Williams}},\ and\ \bibinfo {author}
  {\bibfnamefont {M.}~\bibnamefont {Ishigami}},\ }\bibfield  {title} {\bibinfo
  {title} {Charged-impurity scattering in graphene},\ }\href
  {https://doi.org/10.1038/nphys935} {\bibfield  {journal} {\bibinfo  {journal}
  {Nature Physics}\ }\textbf {\bibinfo {volume} {4}},\ \bibinfo {pages}
  {377–381} (\bibinfo {year} {2008})}\BibitemShut {NoStop}%
\bibitem [{\citenamefont {Cho}\ and\ \citenamefont
  {Fuhrer}(2008)}]{sungjae2008}%
  \BibitemOpen
  \bibfield  {author} {\bibinfo {author} {\bibfnamefont {S.}~\bibnamefont
  {Cho}}\ and\ \bibinfo {author} {\bibfnamefont {M.~S.}\ \bibnamefont
  {Fuhrer}},\ }\bibfield  {title} {\bibinfo {title} {Charge transport and
  inhomogeneity near the minimum conductivity point in graphene},\ }\href
  {https://doi.org/10.1103/PhysRevB.77.081402} {\bibfield  {journal} {\bibinfo
  {journal} {Phys. Rev. B}\ }\textbf {\bibinfo {volume} {77}},\ \bibinfo
  {pages} {081402} (\bibinfo {year} {2008})}\BibitemShut {NoStop}%
\bibitem [{\citenamefont {Martin}\ \emph {et~al.}(2008)\citenamefont {Martin},
  \citenamefont {Akerman}, \citenamefont {Ulbricht}, \citenamefont {Lohmann},
  \citenamefont {Smet}, \citenamefont {Von~Klitzing},\ and\ \citenamefont
  {Yacoby}}]{martin2008observation}%
  \BibitemOpen
  \bibfield  {author} {\bibinfo {author} {\bibfnamefont {J.}~\bibnamefont
  {Martin}}, \bibinfo {author} {\bibfnamefont {N.}~\bibnamefont {Akerman}},
  \bibinfo {author} {\bibfnamefont {G.}~\bibnamefont {Ulbricht}}, \bibinfo
  {author} {\bibfnamefont {T.}~\bibnamefont {Lohmann}}, \bibinfo {author}
  {\bibfnamefont {J.~v.}\ \bibnamefont {Smet}}, \bibinfo {author}
  {\bibfnamefont {K.}~\bibnamefont {Von~Klitzing}},\ and\ \bibinfo {author}
  {\bibfnamefont {A.}~\bibnamefont {Yacoby}},\ }\bibfield  {title} {\bibinfo
  {title} {Observation of electron--hole puddles in graphene using a scanning
  single-electron transistor},\ }\href@noop {} {\bibfield  {journal} {\bibinfo
  {journal} {Nature physics}\ }\textbf {\bibinfo {volume} {4}},\ \bibinfo
  {pages} {144} (\bibinfo {year} {2008})}\BibitemShut {NoStop}%
\bibitem [{\citenamefont {Ponomarenko}\ \emph {et~al.}(2009)\citenamefont
  {Ponomarenko}, \citenamefont {Yang}, \citenamefont {Mohiuddin}, \citenamefont
  {Katsnelson}, \citenamefont {Novoselov}, \citenamefont {Morozov},
  \citenamefont {Zhukov}, \citenamefont {Schedin}, \citenamefont {Hill},\ and\
  \citenamefont {Geim}}]{GeimPonomarenko2009}%
  \BibitemOpen
  \bibfield  {author} {\bibinfo {author} {\bibfnamefont {L.~A.}\ \bibnamefont
  {Ponomarenko}}, \bibinfo {author} {\bibfnamefont {R.}~\bibnamefont {Yang}},
  \bibinfo {author} {\bibfnamefont {T.~M.}\ \bibnamefont {Mohiuddin}}, \bibinfo
  {author} {\bibfnamefont {M.~I.}\ \bibnamefont {Katsnelson}}, \bibinfo
  {author} {\bibfnamefont {K.~S.}\ \bibnamefont {Novoselov}}, \bibinfo {author}
  {\bibfnamefont {S.~V.}\ \bibnamefont {Morozov}}, \bibinfo {author}
  {\bibfnamefont {A.~A.}\ \bibnamefont {Zhukov}}, \bibinfo {author}
  {\bibfnamefont {F.}~\bibnamefont {Schedin}}, \bibinfo {author} {\bibfnamefont
  {E.~W.}\ \bibnamefont {Hill}},\ and\ \bibinfo {author} {\bibfnamefont
  {A.~K.}\ \bibnamefont {Geim}},\ }\bibfield  {title} {\bibinfo {title} {Effect
  of a high-$\ensuremath{\kappa}$ environment on charge carrier mobility in
  graphene},\ }\href {https://doi.org/10.1103/PhysRevLett.102.206603}
  {\bibfield  {journal} {\bibinfo  {journal} {Phys. Rev. Lett.}\ }\textbf
  {\bibinfo {volume} {102}},\ \bibinfo {pages} {206603} (\bibinfo {year}
  {2009})}\BibitemShut {NoStop}%
\bibitem [{\citenamefont {Monteverde}\ \emph {et~al.}(2010)\citenamefont
  {Monteverde}, \citenamefont {Ojeda-Aristizabal}, \citenamefont {Weil},
  \citenamefont {Bennaceur}, \citenamefont {Ferrier}, \citenamefont {Gu\'eron},
  \citenamefont {Glattli}, \citenamefont {Bouchiat}, \citenamefont {Fuchs},\
  and\ \citenamefont {Maslov}}]{Monteverde2010}%
  \BibitemOpen
  \bibfield  {author} {\bibinfo {author} {\bibfnamefont {M.}~\bibnamefont
  {Monteverde}}, \bibinfo {author} {\bibfnamefont {C.}~\bibnamefont
  {Ojeda-Aristizabal}}, \bibinfo {author} {\bibfnamefont {R.}~\bibnamefont
  {Weil}}, \bibinfo {author} {\bibfnamefont {K.}~\bibnamefont {Bennaceur}},
  \bibinfo {author} {\bibfnamefont {M.}~\bibnamefont {Ferrier}}, \bibinfo
  {author} {\bibfnamefont {S.}~\bibnamefont {Gu\'eron}}, \bibinfo {author}
  {\bibfnamefont {C.}~\bibnamefont {Glattli}}, \bibinfo {author} {\bibfnamefont
  {H.}~\bibnamefont {Bouchiat}}, \bibinfo {author} {\bibfnamefont {J.~N.}\
  \bibnamefont {Fuchs}},\ and\ \bibinfo {author} {\bibfnamefont {D.~L.}\
  \bibnamefont {Maslov}},\ }\bibfield  {title} {\bibinfo {title} {Transport and
  elastic scattering times as probes of the nature of impurity scattering in
  single-layer and bilayer graphene},\ }\href
  {https://doi.org/10.1103/PhysRevLett.104.126801} {\bibfield  {journal}
  {\bibinfo  {journal} {Phys. Rev. Lett.}\ }\textbf {\bibinfo {volume} {104}},\
  \bibinfo {pages} {126801} (\bibinfo {year} {2010})}\BibitemShut {NoStop}%
\bibitem [{\citenamefont {Shon}\ and\ \citenamefont {Ando}(1998)}]{Ando1998}%
  \BibitemOpen
  \bibfield  {author} {\bibinfo {author} {\bibfnamefont {N.}~\bibnamefont
  {Shon}}\ and\ \bibinfo {author} {\bibfnamefont {T.}~\bibnamefont {Ando}},\
  }\bibfield  {title} {\bibinfo {title} {Quantum transport in two-dimensional
  graphite system},\ }\href {https://doi.org/10.1143/JPSJ.67.2421} {\bibfield
  {journal} {\bibinfo  {journal} {Journal of the Physical Society of Japan}\
  }\textbf {\bibinfo {volume} {67}},\ \bibinfo {pages} {2421} (\bibinfo {year}
  {1998})},\ \Eprint
  {https://arxiv.org/abs/https://doi.org/10.1143/JPSJ.67.2421}
  {https://doi.org/10.1143/JPSJ.67.2421} \BibitemShut {NoStop}%
\bibitem [{\citenamefont {Ando}(2006)}]{Ando2006}%
  \BibitemOpen
  \bibfield  {author} {\bibinfo {author} {\bibfnamefont {T.}~\bibnamefont
  {Ando}},\ }\bibfield  {title} {\bibinfo {title} {Screening effect and
  impurity scattering in monolayer graphene},\ }\href
  {https://doi.org/10.1143/JPSJ.75.074716} {\bibfield  {journal} {\bibinfo
  {journal} {Journal of the Physical Society of Japan}\ }\textbf {\bibinfo
  {volume} {75}},\ \bibinfo {pages} {074716} (\bibinfo {year} {2006})},\
  \Eprint {https://arxiv.org/abs/https://doi.org/10.1143/JPSJ.75.074716}
  {https://doi.org/10.1143/JPSJ.75.074716} \BibitemShut {NoStop}%
\bibitem [{\citenamefont {Nomura}\ and\ \citenamefont
  {MacDonald}(2006)}]{DonaldN2006}%
  \BibitemOpen
  \bibfield  {author} {\bibinfo {author} {\bibfnamefont {K.}~\bibnamefont
  {Nomura}}\ and\ \bibinfo {author} {\bibfnamefont {A.~H.}\ \bibnamefont
  {MacDonald}},\ }\bibfield  {title} {\bibinfo {title} {Quantum hall
  ferromagnetism in graphene},\ }\href
  {https://doi.org/10.1103/PhysRevLett.96.256602} {\bibfield  {journal}
  {\bibinfo  {journal} {Phys. Rev. Lett.}\ }\textbf {\bibinfo {volume} {96}},\
  \bibinfo {pages} {256602} (\bibinfo {year} {2006})}\BibitemShut {NoStop}%
\bibitem [{\citenamefont {Nomura}\ and\ \citenamefont
  {MacDonald}(2007)}]{Donald2007}%
  \BibitemOpen
  \bibfield  {author} {\bibinfo {author} {\bibfnamefont {K.}~\bibnamefont
  {Nomura}}\ and\ \bibinfo {author} {\bibfnamefont {A.~H.}\ \bibnamefont
  {MacDonald}},\ }\bibfield  {title} {\bibinfo {title} {Quantum transport of
  massless dirac fermions},\ }\href
  {https://doi.org/10.1103/PhysRevLett.98.076602} {\bibfield  {journal}
  {\bibinfo  {journal} {Phys. Rev. Lett.}\ }\textbf {\bibinfo {volume} {98}},\
  \bibinfo {pages} {076602} (\bibinfo {year} {2007})}\BibitemShut {NoStop}%
\bibitem [{\citenamefont {Adam}\ \emph {et~al.}(2008)\citenamefont {Adam},
  \citenamefont {Hwang},\ and\ \citenamefont {{Das Sarma}}}]{ADAM20081022}%
  \BibitemOpen
  \bibfield  {author} {\bibinfo {author} {\bibfnamefont {S.}~\bibnamefont
  {Adam}}, \bibinfo {author} {\bibfnamefont {E.}~\bibnamefont {Hwang}},\ and\
  \bibinfo {author} {\bibfnamefont {S.}~\bibnamefont {{Das Sarma}}},\
  }\bibfield  {title} {\bibinfo {title} {Scattering mechanisms and boltzmann
  transport in graphene},\ }\href
  {https://doi.org/https://doi.org/10.1016/j.physe.2007.09.064} {\bibfield
  {journal} {\bibinfo  {journal} {Physica E: Low-dimensional Systems and
  Nanostructures}\ }\textbf {\bibinfo {volume} {40}},\ \bibinfo {pages} {1022}
  (\bibinfo {year} {2008})}\BibitemShut {NoStop}%
\bibitem [{\citenamefont {Peres}\ \emph {et~al.}(2006)\citenamefont {Peres},
  \citenamefont {Guinea},\ and\ \citenamefont {Castro~Neto}}]{NPeres2006}%
  \BibitemOpen
  \bibfield  {author} {\bibinfo {author} {\bibfnamefont {N.~M.~R.}\
  \bibnamefont {Peres}}, \bibinfo {author} {\bibfnamefont {F.}~\bibnamefont
  {Guinea}},\ and\ \bibinfo {author} {\bibfnamefont {A.~H.}\ \bibnamefont
  {Castro~Neto}},\ }\bibfield  {title} {\bibinfo {title} {Electronic properties
  of disordered two-dimensional carbon},\ }\href
  {https://doi.org/10.1103/PhysRevB.73.125411} {\bibfield  {journal} {\bibinfo
  {journal} {Phys. Rev. B}\ }\textbf {\bibinfo {volume} {73}},\ \bibinfo
  {pages} {125411} (\bibinfo {year} {2006})}\BibitemShut {NoStop}%
\bibitem [{\citenamefont {Ziegler}(2006)}]{Ziegler2006}%
  \BibitemOpen
  \bibfield  {author} {\bibinfo {author} {\bibfnamefont {K.}~\bibnamefont
  {Ziegler}},\ }\bibfield  {title} {\bibinfo {title} {Robust transport
  properties in graphene},\ }\href
  {https://doi.org/10.1103/PhysRevLett.97.266802} {\bibfield  {journal}
  {\bibinfo  {journal} {Phys. Rev. Lett.}\ }\textbf {\bibinfo {volume} {97}},\
  \bibinfo {pages} {266802} (\bibinfo {year} {2006})}\BibitemShut {NoStop}%
\bibitem [{\citenamefont {Ostrovsky}\ \emph {et~al.}(2006)\citenamefont
  {Ostrovsky}, \citenamefont {Gornyi},\ and\ \citenamefont
  {Mirlin}}]{Ostrovsky2006}%
  \BibitemOpen
  \bibfield  {author} {\bibinfo {author} {\bibfnamefont {P.~M.}\ \bibnamefont
  {Ostrovsky}}, \bibinfo {author} {\bibfnamefont {I.~V.}\ \bibnamefont
  {Gornyi}},\ and\ \bibinfo {author} {\bibfnamefont {A.~D.}\ \bibnamefont
  {Mirlin}},\ }\bibfield  {title} {\bibinfo {title} {Electron transport in
  disordered graphene},\ }\href {https://doi.org/10.1103/PhysRevB.74.235443}
  {\bibfield  {journal} {\bibinfo  {journal} {Phys. Rev. B}\ }\textbf {\bibinfo
  {volume} {74}},\ \bibinfo {pages} {235443} (\bibinfo {year}
  {2006})}\BibitemShut {NoStop}%
\bibitem [{\citenamefont {Trushin}\ and\ \citenamefont
  {Schliemann}(2007)}]{Schliemann2007}%
  \BibitemOpen
  \bibfield  {author} {\bibinfo {author} {\bibfnamefont {M.}~\bibnamefont
  {Trushin}}\ and\ \bibinfo {author} {\bibfnamefont {J.}~\bibnamefont
  {Schliemann}},\ }\bibfield  {title} {\bibinfo {title} {Minimum electrical and
  thermal conductivity of graphene: A quasiclassical approach},\ }\href
  {https://doi.org/10.1103/PhysRevLett.99.216602} {\bibfield  {journal}
  {\bibinfo  {journal} {Phys. Rev. Lett.}\ }\textbf {\bibinfo {volume} {99}},\
  \bibinfo {pages} {216602} (\bibinfo {year} {2007})}\BibitemShut {NoStop}%
\bibitem [{\citenamefont {Adam}\ \emph {et~al.}(2007)\citenamefont {Adam},
  \citenamefont {Hwang}, \citenamefont {Galitski},\ and\ \citenamefont
  {Das~Sarma}}]{adam2007self}%
  \BibitemOpen
  \bibfield  {author} {\bibinfo {author} {\bibfnamefont {S.}~\bibnamefont
  {Adam}}, \bibinfo {author} {\bibfnamefont {E.}~\bibnamefont {Hwang}},
  \bibinfo {author} {\bibfnamefont {V.}~\bibnamefont {Galitski}},\ and\
  \bibinfo {author} {\bibfnamefont {S.}~\bibnamefont {Das~Sarma}},\ }\bibfield
  {title} {\bibinfo {title} {A self-consistent theory for graphene transport},\
  }\href@noop {} {\bibfield  {journal} {\bibinfo  {journal} {Proceedings of the
  National Academy of Sciences}\ }\textbf {\bibinfo {volume} {104}},\ \bibinfo
  {pages} {18392} (\bibinfo {year} {2007})}\BibitemShut {NoStop}%
\bibitem [{\citenamefont {Hwang}\ \emph {et~al.}(2007)\citenamefont {Hwang},
  \citenamefont {Adam},\ and\ \citenamefont {Sarma}}]{SarmaDas2007}%
  \BibitemOpen
  \bibfield  {author} {\bibinfo {author} {\bibfnamefont {E.~H.}\ \bibnamefont
  {Hwang}}, \bibinfo {author} {\bibfnamefont {S.}~\bibnamefont {Adam}},\ and\
  \bibinfo {author} {\bibfnamefont {S.~D.}\ \bibnamefont {Sarma}},\ }\bibfield
  {title} {\bibinfo {title} {Carrier transport in two-dimensional graphene
  layers},\ }\href {https://doi.org/10.1103/PhysRevLett.98.186806} {\bibfield
  {journal} {\bibinfo  {journal} {Phys. Rev. Lett.}\ }\textbf {\bibinfo
  {volume} {98}},\ \bibinfo {pages} {186806} (\bibinfo {year}
  {2007})}\BibitemShut {NoStop}%
\bibitem [{\citenamefont {Peres}\ \emph {et~al.}(2007)\citenamefont {Peres},
  \citenamefont {Lopes~dos Santos},\ and\ \citenamefont
  {Stauber}}]{PeresLo2007}%
  \BibitemOpen
  \bibfield  {author} {\bibinfo {author} {\bibfnamefont {N.~M.~R.}\
  \bibnamefont {Peres}}, \bibinfo {author} {\bibfnamefont {J.~M.~B.}\
  \bibnamefont {Lopes~dos Santos}},\ and\ \bibinfo {author} {\bibfnamefont
  {T.}~\bibnamefont {Stauber}},\ }\bibfield  {title} {\bibinfo {title}
  {Phenomenological study of the electronic transport coefficients of
  graphene},\ }\href {https://doi.org/10.1103/PhysRevB.76.073412} {\bibfield
  {journal} {\bibinfo  {journal} {Phys. Rev. B}\ }\textbf {\bibinfo {volume}
  {76}},\ \bibinfo {pages} {073412} (\bibinfo {year} {2007})}\BibitemShut
  {NoStop}%
\bibitem [{\citenamefont {Novikov}(2007)}]{Novikov2007}%
  \BibitemOpen
  \bibfield  {author} {\bibinfo {author} {\bibfnamefont {D.~S.}\ \bibnamefont
  {Novikov}},\ }\bibfield  {title} {\bibinfo {title} {{Numbers of donors and
  acceptors from transport measurements in graphene}},\ }\href
  {https://doi.org/10.1063/1.2779107} {\bibfield  {journal} {\bibinfo
  {journal} {Applied Physics Letters}\ }\textbf {\bibinfo {volume} {91}},\
  \bibinfo {pages} {102102} (\bibinfo {year} {2007})},\ \Eprint
  {https://arxiv.org/abs/https://pubs.aip.org/aip/apl/article-pdf/doi/10.1063/1.2779107/14374674/102102\_1\_online.pdf}
  {https://pubs.aip.org/aip/apl/article-pdf/doi/10.1063/1.2779107/14374674/102102\_1\_online.pdf}
  \BibitemShut {NoStop}%
\bibitem [{\citenamefont {Katsnelson}\ and\ \citenamefont
  {Geim}(2008)}]{katsnelson2008electron}%
  \BibitemOpen
  \bibfield  {author} {\bibinfo {author} {\bibfnamefont {M.~I.}\ \bibnamefont
  {Katsnelson}}\ and\ \bibinfo {author} {\bibfnamefont {A.~K.}\ \bibnamefont
  {Geim}},\ }\bibfield  {title} {\bibinfo {title} {Electron scattering on
  microscopic corrugations in graphene},\ }\href@noop {} {\bibfield  {journal}
  {\bibinfo  {journal} {Philosophical Transactions of the Royal Society A:
  Mathematical, Physical and Engineering Sciences}\ }\textbf {\bibinfo {volume}
  {366}},\ \bibinfo {pages} {195} (\bibinfo {year} {2008})}\BibitemShut
  {NoStop}%
\bibitem [{\citenamefont {Kumazaki}\ and\ \citenamefont
  {S.~Hirashima}(2006)}]{Kumazaki2006}%
  \BibitemOpen
  \bibfield  {author} {\bibinfo {author} {\bibfnamefont {H.}~\bibnamefont
  {Kumazaki}}\ and\ \bibinfo {author} {\bibfnamefont {D.}~\bibnamefont
  {S.~Hirashima}},\ }\bibfield  {title} {\bibinfo {title} {Effects of
  impurities in two-dimensional graphite},\ }\href
  {https://doi.org/10.1143/JPSJ.75.053707} {\bibfield  {journal} {\bibinfo
  {journal} {Journal of the Physical Society of Japan}\ }\textbf {\bibinfo
  {volume} {75}},\ \bibinfo {pages} {053707} (\bibinfo {year} {2006})},\
  \Eprint {https://arxiv.org/abs/https://doi.org/10.1143/JPSJ.75.053707}
  {https://doi.org/10.1143/JPSJ.75.053707} \BibitemShut {NoStop}%
\bibitem [{\citenamefont {Nilsson}\ \emph {et~al.}(2006)\citenamefont
  {Nilsson}, \citenamefont {Neto}, \citenamefont {Guinea},\ and\ \citenamefont
  {Peres}}]{Peres2006}%
  \BibitemOpen
  \bibfield  {author} {\bibinfo {author} {\bibfnamefont {J.}~\bibnamefont
  {Nilsson}}, \bibinfo {author} {\bibfnamefont {A.~H.~C.}\ \bibnamefont
  {Neto}}, \bibinfo {author} {\bibfnamefont {F.}~\bibnamefont {Guinea}},\ and\
  \bibinfo {author} {\bibfnamefont {N.~M.~R.}\ \bibnamefont {Peres}},\
  }\bibfield  {title} {\bibinfo {title} {Electronic properties of graphene
  multilayers},\ }\href {https://doi.org/10.1103/PhysRevLett.97.266801}
  {\bibfield  {journal} {\bibinfo  {journal} {Phys. Rev. Lett.}\ }\textbf
  {\bibinfo {volume} {97}},\ \bibinfo {pages} {266801} (\bibinfo {year}
  {2006})}\BibitemShut {NoStop}%
\bibitem [{\citenamefont {Efetov}\ and\ \citenamefont
  {Kim}(2010)}]{Efetov2010}%
  \BibitemOpen
  \bibfield  {author} {\bibinfo {author} {\bibfnamefont {D.~K.}\ \bibnamefont
  {Efetov}}\ and\ \bibinfo {author} {\bibfnamefont {P.}~\bibnamefont {Kim}},\
  }\bibfield  {title} {\bibinfo {title} {Controlling electron-phonon
  interactions in graphene at ultrahigh carrier densities},\ }\href
  {https://doi.org/10.1103/PhysRevLett.105.256805} {\bibfield  {journal}
  {\bibinfo  {journal} {Phys. Rev. Lett.}\ }\textbf {\bibinfo {volume} {105}},\
  \bibinfo {pages} {256805} (\bibinfo {year} {2010})}\BibitemShut {NoStop}%
\bibitem [{\citenamefont {Yi\ifmmode~\breve{g}\else \u{g}\fi{}en}\ \emph
  {et~al.}(2013)\citenamefont {Yi\ifmmode~\breve{g}\else \u{g}\fi{}en},
  \citenamefont {Tayari}, \citenamefont {Island}, \citenamefont {Porter},\ and\
  \citenamefont {Champagne}}]{Tayari2013}%
  \BibitemOpen
  \bibfield  {author} {\bibinfo {author} {\bibfnamefont {S.}~\bibnamefont
  {Yi\ifmmode~\breve{g}\else \u{g}\fi{}en}}, \bibinfo {author} {\bibfnamefont
  {V.}~\bibnamefont {Tayari}}, \bibinfo {author} {\bibfnamefont {J.~O.}\
  \bibnamefont {Island}}, \bibinfo {author} {\bibfnamefont {J.~M.}\
  \bibnamefont {Porter}},\ and\ \bibinfo {author} {\bibfnamefont {A.~R.}\
  \bibnamefont {Champagne}},\ }\bibfield  {title} {\bibinfo {title} {Electronic
  thermal conductivity measurements in intrinsic graphene},\ }\href
  {https://doi.org/10.1103/PhysRevB.87.241411} {\bibfield  {journal} {\bibinfo
  {journal} {Phys. Rev. B}\ }\textbf {\bibinfo {volume} {87}},\ \bibinfo
  {pages} {241411} (\bibinfo {year} {2013})}\BibitemShut {NoStop}%
\bibitem [{\citenamefont {Hwang}\ and\ \citenamefont
  {Das~Sarma}(2008)}]{EHwang2008}%
  \BibitemOpen
  \bibfield  {author} {\bibinfo {author} {\bibfnamefont {E.~H.}\ \bibnamefont
  {Hwang}}\ and\ \bibinfo {author} {\bibfnamefont {S.}~\bibnamefont
  {Das~Sarma}},\ }\bibfield  {title} {\bibinfo {title} {Acoustic phonon
  scattering limited carrier mobility in two-dimensional extrinsic graphene},\
  }\href {https://doi.org/10.1103/PhysRevB.77.115449} {\bibfield  {journal}
  {\bibinfo  {journal} {Phys. Rev. B}\ }\textbf {\bibinfo {volume} {77}},\
  \bibinfo {pages} {115449} (\bibinfo {year} {2008})}\BibitemShut {NoStop}%
\bibitem [{\citenamefont {Muñoz}(2012)}]{MunozP2012}%
  \BibitemOpen
  \bibfield  {author} {\bibinfo {author} {\bibfnamefont {E.}~\bibnamefont
  {Muñoz}},\ }\bibfield  {title} {\bibinfo {title} {Phonon-limited transport
  coefficients in extrinsic graphene},\ }\href
  {https://doi.org/10.1088/0953-8984/24/19/195302} {\bibfield  {journal}
  {\bibinfo  {journal} {Journal of Physics: Condensed Matter}\ }\textbf
  {\bibinfo {volume} {24}},\ \bibinfo {pages} {195302} (\bibinfo {year}
  {2012})}\BibitemShut {NoStop}%
\bibitem [{\citenamefont {Park}\ \emph {et~al.}(2014)\citenamefont {Park},
  \citenamefont {Bonini}, \citenamefont {Sohier}, \citenamefont {Samsonidze},
  \citenamefont {Kozinsky}, \citenamefont {Calandra}, \citenamefont {Mauri},\
  and\ \citenamefont {Marzari}}]{Park2014}%
  \BibitemOpen
  \bibfield  {author} {\bibinfo {author} {\bibfnamefont {C.-H.}\ \bibnamefont
  {Park}}, \bibinfo {author} {\bibfnamefont {N.}~\bibnamefont {Bonini}},
  \bibinfo {author} {\bibfnamefont {T.}~\bibnamefont {Sohier}}, \bibinfo
  {author} {\bibfnamefont {G.}~\bibnamefont {Samsonidze}}, \bibinfo {author}
  {\bibfnamefont {B.}~\bibnamefont {Kozinsky}}, \bibinfo {author}
  {\bibfnamefont {M.}~\bibnamefont {Calandra}}, \bibinfo {author}
  {\bibfnamefont {F.}~\bibnamefont {Mauri}},\ and\ \bibinfo {author}
  {\bibfnamefont {N.}~\bibnamefont {Marzari}},\ }\bibfield  {title} {\bibinfo
  {title} {Electron–phonon interactions and the intrinsic electrical
  resistivity of graphene},\ }\href {https://doi.org/10.1021/nl402696q}
  {\bibfield  {journal} {\bibinfo  {journal} {Nano Letters}\ }\textbf {\bibinfo
  {volume} {14}},\ \bibinfo {pages} {1113} (\bibinfo {year} {2014})},\ \bibinfo
  {note} {pMID: 24524418},\ \Eprint
  {https://arxiv.org/abs/https://doi.org/10.1021/nl402696q}
  {https://doi.org/10.1021/nl402696q} \BibitemShut {NoStop}%
\bibitem [{\citenamefont {Sohier}\ \emph {et~al.}(2014)\citenamefont {Sohier},
  \citenamefont {Calandra}, \citenamefont {Park}, \citenamefont {Bonini},
  \citenamefont {Marzari},\ and\ \citenamefont {Mauri}}]{Sohier2014}%
  \BibitemOpen
  \bibfield  {author} {\bibinfo {author} {\bibfnamefont {T.}~\bibnamefont
  {Sohier}}, \bibinfo {author} {\bibfnamefont {M.}~\bibnamefont {Calandra}},
  \bibinfo {author} {\bibfnamefont {C.-H.}\ \bibnamefont {Park}}, \bibinfo
  {author} {\bibfnamefont {N.}~\bibnamefont {Bonini}}, \bibinfo {author}
  {\bibfnamefont {N.}~\bibnamefont {Marzari}},\ and\ \bibinfo {author}
  {\bibfnamefont {F.}~\bibnamefont {Mauri}},\ }\bibfield  {title} {\bibinfo
  {title} {Phonon-limited resistivity of graphene by first-principles
  calculations: Electron-phonon interactions, strain-induced gauge field, and
  boltzmann equation},\ }\href {https://doi.org/10.1103/PhysRevB.90.125414}
  {\bibfield  {journal} {\bibinfo  {journal} {Phys. Rev. B}\ }\textbf {\bibinfo
  {volume} {90}},\ \bibinfo {pages} {125414} (\bibinfo {year}
  {2014})}\BibitemShut {NoStop}%
\bibitem [{\citenamefont {Kim}\ \emph {et~al.}(2016)\citenamefont {Kim},
  \citenamefont {Park},\ and\ \citenamefont {Marzari}}]{Kim2016}%
  \BibitemOpen
  \bibfield  {author} {\bibinfo {author} {\bibfnamefont {T.~Y.}\ \bibnamefont
  {Kim}}, \bibinfo {author} {\bibfnamefont {C.-H.}\ \bibnamefont {Park}},\ and\
  \bibinfo {author} {\bibfnamefont {N.}~\bibnamefont {Marzari}},\ }\bibfield
  {title} {\bibinfo {title} {The electronic thermal conductivity of graphene},\
  }\href {https://doi.org/10.1021/acs.nanolett.5b05288} {\bibfield  {journal}
  {\bibinfo  {journal} {Nano Letters}\ }\textbf {\bibinfo {volume} {16}},\
  \bibinfo {pages} {2439} (\bibinfo {year} {2016})},\ \bibinfo {note} {pMID:
  26907524},\ \Eprint
  {https://arxiv.org/abs/https://doi.org/10.1021/acs.nanolett.5b05288}
  {https://doi.org/10.1021/acs.nanolett.5b05288} \BibitemShut {NoStop}%
\bibitem [{\citenamefont {Win}\ \emph {et~al.}(2024)\citenamefont {Win},
  \citenamefont {Aung}, \citenamefont {Khandal},\ and\ \citenamefont
  {Ghosh}}]{win2024wied}%
  \BibitemOpen
  \bibfield  {author} {\bibinfo {author} {\bibfnamefont {T.~Z.}\ \bibnamefont
  {Win}}, \bibinfo {author} {\bibfnamefont {C.~W.}\ \bibnamefont {Aung}},
  \bibinfo {author} {\bibfnamefont {G.}~\bibnamefont {Khandal}},\ and\ \bibinfo
  {author} {\bibfnamefont {S.}~\bibnamefont {Ghosh}},\ }\href
  {https://arxiv.org/abs/2404.01061} {\bibinfo {title} {Wiedemann-franz law
  violation domain for graphene and nonrelativistic systems}} (\bibinfo {year}
  {2024}),\ \Eprint {https://arxiv.org/abs/2404.01061} {arXiv:2404.01061
  [cond-mat.mes-hall]} \BibitemShut {NoStop}%
\bibitem [{\citenamefont {Ashcroft}\ and\ \citenamefont
  {Mermin}(2011)}]{ashcroft2011solid}%
  \BibitemOpen
  \bibfield  {author} {\bibinfo {author} {\bibfnamefont {N.}~\bibnamefont
  {Ashcroft}}\ and\ \bibinfo {author} {\bibfnamefont {N.}~\bibnamefont
  {Mermin}},\ }\href {https://books.google.co.in/books?id=x_s_YAAACAAJ} {\emph
  {\bibinfo {title} {Solid State Physics}}}\ (\bibinfo  {publisher} {Cengage
  Learning},\ \bibinfo {year} {2011})\BibitemShut {NoStop}%
\bibitem [{\citenamefont {Ziman}(2001)}]{ziman2001electrons}%
  \BibitemOpen
  \bibfield  {author} {\bibinfo {author} {\bibfnamefont {J.}~\bibnamefont
  {Ziman}},\ }\href {https://books.google.co.in/books?id=UtEy63pjngsC} {\emph
  {\bibinfo {title} {Electrons and Phonons: The Theory of Transport Phenomena
  in Solids}}},\ International series of monographs on physics\ (\bibinfo
  {publisher} {OUP Oxford},\ \bibinfo {year} {2001})\BibitemShut {NoStop}%
\bibitem [{\citenamefont {Lucas}\ \emph {et~al.}(2016)\citenamefont {Lucas},
  \citenamefont {Crossno}, \citenamefont {Fong}, \citenamefont {Kim},\ and\
  \citenamefont {Sachdev}}]{AnLucas2016}%
  \BibitemOpen
  \bibfield  {author} {\bibinfo {author} {\bibfnamefont {A.}~\bibnamefont
  {Lucas}}, \bibinfo {author} {\bibfnamefont {J.}~\bibnamefont {Crossno}},
  \bibinfo {author} {\bibfnamefont {K.~C.}\ \bibnamefont {Fong}}, \bibinfo
  {author} {\bibfnamefont {P.}~\bibnamefont {Kim}},\ and\ \bibinfo {author}
  {\bibfnamefont {S.}~\bibnamefont {Sachdev}},\ }\bibfield  {title} {\bibinfo
  {title} {Transport in inhomogeneous quantum critical fluids and in the dirac
  fluid in graphene},\ }\href {https://doi.org/10.1103/PhysRevB.93.075426}
  {\bibfield  {journal} {\bibinfo  {journal} {Phys. Rev. B}\ }\textbf {\bibinfo
  {volume} {93}},\ \bibinfo {pages} {075426} (\bibinfo {year}
  {2016})}\BibitemShut {NoStop}%
\bibitem [{\citenamefont {Zarenia}\ \emph {et~al.}(2019)\citenamefont
  {Zarenia}, \citenamefont {Principi},\ and\ \citenamefont
  {Vignale}}]{Zarenia_2019}%
  \BibitemOpen
  \bibfield  {author} {\bibinfo {author} {\bibfnamefont {M.}~\bibnamefont
  {Zarenia}}, \bibinfo {author} {\bibfnamefont {A.}~\bibnamefont {Principi}},\
  and\ \bibinfo {author} {\bibfnamefont {G.}~\bibnamefont {Vignale}},\
  }\bibfield  {title} {\bibinfo {title} {Disorder-enabled hydrodynamics of
  charge and heat transport in monolayer graphene},\ }\href
  {https://doi.org/10.1088/2053-1583/ab1ad9} {\bibfield  {journal} {\bibinfo
  {journal} {2D Materials}\ }\textbf {\bibinfo {volume} {6}},\ \bibinfo {pages}
  {035024} (\bibinfo {year} {2019})}\BibitemShut {NoStop}%
\bibitem [{\citenamefont {Rycerz}(2021)}]{ma14112704}%
  \BibitemOpen
  \bibfield  {author} {\bibinfo {author} {\bibfnamefont {A.}~\bibnamefont
  {Rycerz}},\ }\bibfield  {title} {\bibinfo {title} {Wiedemann–franz law for
  massless dirac fermions with implications for graphene},\ }\bibfield
  {journal} {\bibinfo  {journal} {Materials}\ }\textbf {\bibinfo {volume}
  {14}},\ \href {https://doi.org/10.3390/ma14112704} {10.3390/ma14112704}
  (\bibinfo {year} {2021})\BibitemShut {NoStop}%
\bibitem [{\citenamefont {Tu}\ and\ \citenamefont
  {Das~Sarma}(2023)}]{tu2023wiedemann}%
  \BibitemOpen
  \bibfield  {author} {\bibinfo {author} {\bibfnamefont {Y.-T.}\ \bibnamefont
  {Tu}}\ and\ \bibinfo {author} {\bibfnamefont {S.}~\bibnamefont {Das~Sarma}},\
  }\bibfield  {title} {\bibinfo {title} {Wiedemann-franz law in graphene},\
  }\bibfield  {journal} {\bibinfo  {journal} {Physical Review B}\ }\textbf
  {\bibinfo {volume} {107}},\ \href
  {https://doi.org/10.1103/physrevb.107.085401} {10.1103/physrevb.107.085401}
  (\bibinfo {year} {2023})\BibitemShut {NoStop}%
\bibitem [{\citenamefont {Pongsangangan}\ \emph
  {et~al.}(2022{\natexlab{a}})\citenamefont {Pongsangangan}, \citenamefont
  {Ludwig}, \citenamefont {Stoof},\ and\ \citenamefont
  {Fritz}}]{PhysRevB.106.205126}%
  \BibitemOpen
  \bibfield  {author} {\bibinfo {author} {\bibfnamefont {K.}~\bibnamefont
  {Pongsangangan}}, \bibinfo {author} {\bibfnamefont {T.}~\bibnamefont
  {Ludwig}}, \bibinfo {author} {\bibfnamefont {H.~T.~C.}\ \bibnamefont
  {Stoof}},\ and\ \bibinfo {author} {\bibfnamefont {L.}~\bibnamefont {Fritz}},\
  }\bibfield  {title} {\bibinfo {title} {Hydrodynamics of charged
  two-dimensional dirac systems. i. thermoelectric transport},\ }\href
  {https://doi.org/10.1103/PhysRevB.106.205126} {\bibfield  {journal} {\bibinfo
   {journal} {Phys. Rev. B}\ }\textbf {\bibinfo {volume} {106}},\ \bibinfo
  {pages} {205126} (\bibinfo {year} {2022}{\natexlab{a}})}\BibitemShut
  {NoStop}%
\bibitem [{\citenamefont {Pongsangangan}\ \emph
  {et~al.}(2022{\natexlab{b}})\citenamefont {Pongsangangan}, \citenamefont
  {Grubinskas},\ and\ \citenamefont {Fritz}}]{PhysRevResearch.4.043107}%
  \BibitemOpen
  \bibfield  {author} {\bibinfo {author} {\bibfnamefont {K.}~\bibnamefont
  {Pongsangangan}}, \bibinfo {author} {\bibfnamefont {S.}~\bibnamefont
  {Grubinskas}},\ and\ \bibinfo {author} {\bibfnamefont {L.}~\bibnamefont
  {Fritz}},\ }\bibfield  {title} {\bibinfo {title} {Thermoelectric response in
  two-dimensional dirac systems: Role of particle-hole pairs},\ }\href
  {https://doi.org/10.1103/PhysRevResearch.4.043107} {\bibfield  {journal}
  {\bibinfo  {journal} {Phys. Rev. Res.}\ }\textbf {\bibinfo {volume} {4}},\
  \bibinfo {pages} {043107} (\bibinfo {year} {2022}{\natexlab{b}})}\BibitemShut
  {NoStop}%
\bibitem [{\citenamefont {Jaiswal}\ \emph {et~al.}(2015)\citenamefont
  {Jaiswal}, \citenamefont {Friman},\ and\ \citenamefont
  {Redlich}}]{JAISWAL2015548}%
  \BibitemOpen
  \bibfield  {author} {\bibinfo {author} {\bibfnamefont {A.}~\bibnamefont
  {Jaiswal}}, \bibinfo {author} {\bibfnamefont {B.}~\bibnamefont {Friman}},\
  and\ \bibinfo {author} {\bibfnamefont {K.}~\bibnamefont {Redlich}},\
  }\bibfield  {title} {\bibinfo {title} {Relativistic second-order dissipative
  hydrodynamics at finite chemical potential},\ }\href
  {https://doi.org/https://doi.org/10.1016/j.physletb.2015.11.018} {\bibfield
  {journal} {\bibinfo  {journal} {Physics Letters B}\ }\textbf {\bibinfo
  {volume} {751}},\ \bibinfo {pages} {548} (\bibinfo {year}
  {2015})}\BibitemShut {NoStop}%
\bibitem [{\citenamefont {Sahoo}\ \emph {et~al.}(2019)\citenamefont {Sahoo},
  \citenamefont {Sahoo},\ and\ \citenamefont {Tiwari}}]{Sahoo:2019xjq}%
  \BibitemOpen
  \bibfield  {author} {\bibinfo {author} {\bibfnamefont {P.}~\bibnamefont
  {Sahoo}}, \bibinfo {author} {\bibfnamefont {R.}~\bibnamefont {Sahoo}},\ and\
  \bibinfo {author} {\bibfnamefont {S.~K.}\ \bibnamefont {Tiwari}},\ }\bibfield
   {title} {\bibinfo {title} {{Wiedemann-Franz law for hot QCD matter in a
  color string percolation scenario}},\ }\href
  {https://doi.org/10.1103/PhysRevD.100.051503} {\bibfield  {journal} {\bibinfo
   {journal} {Phys. Rev. D}\ }\textbf {\bibinfo {volume} {100}},\ \bibinfo
  {pages} {051503} (\bibinfo {year} {2019})},\ \Eprint
  {https://arxiv.org/abs/1904.06961} {arXiv:1904.06961 [hep-ph]} \BibitemShut
  {NoStop}%
\bibitem [{\citenamefont {Rath}\ \emph {et~al.}(2019)\citenamefont {Rath},
  \citenamefont {Tripathy}, \citenamefont {Chatterjee}, \citenamefont {Sahoo},
  \citenamefont {Kumar~Tiwari},\ and\ \citenamefont {Nath}}]{Rath:2019nne}%
  \BibitemOpen
  \bibfield  {author} {\bibinfo {author} {\bibfnamefont {R.}~\bibnamefont
  {Rath}}, \bibinfo {author} {\bibfnamefont {S.}~\bibnamefont {Tripathy}},
  \bibinfo {author} {\bibfnamefont {B.}~\bibnamefont {Chatterjee}}, \bibinfo
  {author} {\bibfnamefont {R.}~\bibnamefont {Sahoo}}, \bibinfo {author}
  {\bibfnamefont {S.}~\bibnamefont {Kumar~Tiwari}},\ and\ \bibinfo {author}
  {\bibfnamefont {A.}~\bibnamefont {Nath}},\ }\bibfield  {title} {\bibinfo
  {title} {{Violation of Wiedemann-Franz Law for Hot Hadronic Matter created at
  NICA, FAIR and RHIC Energies using Non-extensive Statistics}},\ }\href
  {https://doi.org/10.1140/epja/i2019-12814-3} {\bibfield  {journal} {\bibinfo
  {journal} {Eur. Phys. J. A}\ }\textbf {\bibinfo {volume} {55}},\ \bibinfo
  {pages} {125} (\bibinfo {year} {2019})},\ \Eprint
  {https://arxiv.org/abs/1902.07922} {arXiv:1902.07922 [hep-ph]} \BibitemShut
  {NoStop}%
\bibitem [{\citenamefont {Singh}\ \emph {et~al.}(2023)\citenamefont {Singh},
  \citenamefont {Dey}, \citenamefont {Sahoo},\ and\ \citenamefont
  {Ghosh}}]{singh2023effect}%
  \BibitemOpen
  \bibfield  {author} {\bibinfo {author} {\bibfnamefont {K.}~\bibnamefont
  {Singh}}, \bibinfo {author} {\bibfnamefont {J.}~\bibnamefont {Dey}}, \bibinfo
  {author} {\bibfnamefont {R.}~\bibnamefont {Sahoo}},\ and\ \bibinfo {author}
  {\bibfnamefont {S.}~\bibnamefont {Ghosh}},\ }\bibfield  {title} {\bibinfo
  {title} {Effect of time-varying electromagnetic field on wiedemann-franz law
  in a hot hadronic matter},\ }\href@noop {} {\bibfield  {journal} {\bibinfo
  {journal} {Physical Review D}\ }\textbf {\bibinfo {volume} {108}},\ \bibinfo
  {pages} {094007} (\bibinfo {year} {2023})}\BibitemShut {NoStop}%
\bibitem [{\citenamefont {Pradhan}\ \emph {et~al.}(2023)\citenamefont
  {Pradhan}, \citenamefont {Sahu}, \citenamefont {Scaria},\ and\ \citenamefont
  {Sahoo}}]{pradhan2023conductivity}%
  \BibitemOpen
  \bibfield  {author} {\bibinfo {author} {\bibfnamefont {K.~K.}\ \bibnamefont
  {Pradhan}}, \bibinfo {author} {\bibfnamefont {D.}~\bibnamefont {Sahu}},
  \bibinfo {author} {\bibfnamefont {R.}~\bibnamefont {Scaria}},\ and\ \bibinfo
  {author} {\bibfnamefont {R.}~\bibnamefont {Sahoo}},\ }\bibfield  {title}
  {\bibinfo {title} {Conductivity, diffusivity, and violation of the
  wiedemann-franz law in a hadron resonance gas with van der waals
  interactions},\ }\href@noop {} {\bibfield  {journal} {\bibinfo  {journal}
  {Physical Review C}\ }\textbf {\bibinfo {volume} {107}},\ \bibinfo {pages}
  {014910} (\bibinfo {year} {2023})}\BibitemShut {NoStop}%
\bibitem [{\citenamefont {Bandurin}\ \emph {et~al.}(2016)\citenamefont
  {Bandurin}, \citenamefont {Torre}, \citenamefont {Kumar}, \citenamefont
  {Shalom}, \citenamefont {Tomadin}, \citenamefont {Principi}, \citenamefont
  {Auton}, \citenamefont {Khestanova}, \citenamefont {Novoselov}, \citenamefont
  {Grigorieva}, \citenamefont {Ponomarenko}, \citenamefont {Geim},\ and\
  \citenamefont {Polini}}]{Bandurin2016}%
  \BibitemOpen
  \bibfield  {author} {\bibinfo {author} {\bibfnamefont {D.~A.}\ \bibnamefont
  {Bandurin}}, \bibinfo {author} {\bibfnamefont {I.}~\bibnamefont {Torre}},
  \bibinfo {author} {\bibfnamefont {R.~K.}\ \bibnamefont {Kumar}}, \bibinfo
  {author} {\bibfnamefont {M.~B.}\ \bibnamefont {Shalom}}, \bibinfo {author}
  {\bibfnamefont {A.}~\bibnamefont {Tomadin}}, \bibinfo {author} {\bibfnamefont
  {A.}~\bibnamefont {Principi}}, \bibinfo {author} {\bibfnamefont {G.~H.}\
  \bibnamefont {Auton}}, \bibinfo {author} {\bibfnamefont {E.}~\bibnamefont
  {Khestanova}}, \bibinfo {author} {\bibfnamefont {K.~S.}\ \bibnamefont
  {Novoselov}}, \bibinfo {author} {\bibfnamefont {I.~V.}\ \bibnamefont
  {Grigorieva}}, \bibinfo {author} {\bibfnamefont {L.~A.}\ \bibnamefont
  {Ponomarenko}}, \bibinfo {author} {\bibfnamefont {A.~K.}\ \bibnamefont
  {Geim}},\ and\ \bibinfo {author} {\bibfnamefont {M.}~\bibnamefont {Polini}},\
  }\bibfield  {title} {\bibinfo {title} {Negative local resistance caused by
  viscous electron backflow in graphene},\ }\href
  {https://doi.org/10.1126/science.aad0201} {\bibfield  {journal} {\bibinfo
  {journal} {Science}\ }\textbf {\bibinfo {volume} {351}},\ \bibinfo {pages}
  {1055} (\bibinfo {year} {2016})},\ \Eprint
  {https://arxiv.org/abs/https://www.science.org/doi/pdf/10.1126/science.aad0201}
  {https://www.science.org/doi/pdf/10.1126/science.aad0201} \BibitemShut
  {NoStop}%
\bibitem [{\citenamefont {Castro~Neto}\ \emph {et~al.}(2009)\citenamefont
  {Castro~Neto}, \citenamefont {Guinea}, \citenamefont {Peres}, \citenamefont
  {Novoselov},\ and\ \citenamefont {Geim}}]{neto2009electronic}%
  \BibitemOpen
  \bibfield  {author} {\bibinfo {author} {\bibfnamefont {A.~H.}\ \bibnamefont
  {Castro~Neto}}, \bibinfo {author} {\bibfnamefont {F.}~\bibnamefont {Guinea}},
  \bibinfo {author} {\bibfnamefont {N.~M.~R.}\ \bibnamefont {Peres}}, \bibinfo
  {author} {\bibfnamefont {K.~S.}\ \bibnamefont {Novoselov}},\ and\ \bibinfo
  {author} {\bibfnamefont {A.~K.}\ \bibnamefont {Geim}},\ }\bibfield  {title}
  {\bibinfo {title} {The electronic properties of graphene},\ }\href
  {https://doi.org/10.1103/revmodphys.81.109} {\bibfield  {journal} {\bibinfo
  {journal} {Reviews of Modern Physics}\ }\textbf {\bibinfo {volume} {81}},\
  \bibinfo {pages} {109–162} (\bibinfo {year} {2009})}\BibitemShut {NoStop}%
\bibitem [{\citenamefont {Das~Sarma}\ \emph {et~al.}(2011)\citenamefont
  {Das~Sarma}, \citenamefont {Adam}, \citenamefont {Hwang},\ and\ \citenamefont
  {Rossi}}]{RevModPhys.83.407}%
  \BibitemOpen
  \bibfield  {author} {\bibinfo {author} {\bibfnamefont {S.}~\bibnamefont
  {Das~Sarma}}, \bibinfo {author} {\bibfnamefont {S.}~\bibnamefont {Adam}},
  \bibinfo {author} {\bibfnamefont {E.~H.}\ \bibnamefont {Hwang}},\ and\
  \bibinfo {author} {\bibfnamefont {E.}~\bibnamefont {Rossi}},\ }\bibfield
  {title} {\bibinfo {title} {Electronic transport in two-dimensional
  graphene},\ }\href {https://doi.org/10.1103/RevModPhys.83.407} {\bibfield
  {journal} {\bibinfo  {journal} {Rev. Mod. Phys.}\ }\textbf {\bibinfo {volume}
  {83}},\ \bibinfo {pages} {407} (\bibinfo {year} {2011})}\BibitemShut
  {NoStop}%
\bibitem [{\citenamefont {Jaiswal}\ and\ \citenamefont
  {Roy}(2016)}]{Jaiswal:2016hex}%
  \BibitemOpen
  \bibfield  {author} {\bibinfo {author} {\bibfnamefont {A.}~\bibnamefont
  {Jaiswal}}\ and\ \bibinfo {author} {\bibfnamefont {V.}~\bibnamefont {Roy}},\
  }\bibfield  {title} {\bibinfo {title} {{Relativistic hydrodynamics in
  heavy-ion collisions: general aspects and recent developments}},\ }\href
  {https://doi.org/10.1155/2016/9623034} {\bibfield  {journal} {\bibinfo
  {journal} {Adv. High Energy Phys.}\ }\textbf {\bibinfo {volume} {2016}},\
  \bibinfo {pages} {9623034} (\bibinfo {year} {2016})},\ \Eprint
  {https://arxiv.org/abs/1605.08694} {arXiv:1605.08694 [nucl-th]} \BibitemShut
  {NoStop}%
\bibitem [{\citenamefont {Landau.}(1957{\natexlab{a}})}]{Landau1}%
  \BibitemOpen
  \bibfield  {author} {\bibinfo {author} {\bibfnamefont {L.~D.}\ \bibnamefont
  {Landau.}},\ }\bibfield  {title} {\bibinfo {title} {{The Theory of a Fermi
  Liquid}},\ }\href {http://www.jetp.ras.ru/cgi-bin/dn/e_003_06_0920.pdf}
  {\bibfield  {journal} {\bibinfo  {journal} {J. Exptl. Theoret. Phys.}\
  }\textbf {\bibinfo {volume} {3}},\ \bibinfo {pages} {920} (\bibinfo {year}
  {1957}{\natexlab{a}})}\BibitemShut {NoStop}%
\bibitem [{\citenamefont {Landau.}(1957{\natexlab{b}})}]{Landau2}%
  \BibitemOpen
  \bibfield  {author} {\bibinfo {author} {\bibfnamefont {L.~D.}\ \bibnamefont
  {Landau.}},\ }\bibfield  {title} {\bibinfo {title} {{Oscillations in a Fermi
  Liquid}},\ }\href {http://www.jetp.ras.ru/cgi-bin/dn/e_005_01_0101.pdf}
  {\bibfield  {journal} {\bibinfo  {journal} {J. Exptl. Theoret. Phys.}\
  }\textbf {\bibinfo {volume} {5}},\ \bibinfo {pages} {101} (\bibinfo {year}
  {1957}{\natexlab{b}})}\BibitemShut {NoStop}%
\bibitem [{\citenamefont {Landau.}(1959)}]{Landau3}%
  \BibitemOpen
  \bibfield  {author} {\bibinfo {author} {\bibfnamefont {L.~D.}\ \bibnamefont
  {Landau.}},\ }\bibfield  {title} {\bibinfo {title} {{On the theory of Fermi
  Liquid}},\ }\href {http://jetp.ras.ru/cgi-bin/dn/e_008_01_0070.pdf}
  {\bibfield  {journal} {\bibinfo  {journal} {J. Exptl. Theoret. Phys.}\
  }\textbf {\bibinfo {volume} {35 (8)}},\ \bibinfo {pages} {70} (\bibinfo
  {year} {1959})}\BibitemShut {NoStop}%
\bibitem [{\citenamefont {Levitov}\ and\ \citenamefont
  {Falkovich}(2016)}]{levitov2016electron}%
  \BibitemOpen
  \bibfield  {author} {\bibinfo {author} {\bibfnamefont {L.}~\bibnamefont
  {Levitov}}\ and\ \bibinfo {author} {\bibfnamefont {G.}~\bibnamefont
  {Falkovich}},\ }\bibfield  {title} {\bibinfo {title} {Electron viscosity,
  current vortices and negative nonlocal resistance in graphene},\ }\href@noop
  {} {\bibfield  {journal} {\bibinfo  {journal} {Nature Physics}\ }\textbf
  {\bibinfo {volume} {12}},\ \bibinfo {pages} {672} (\bibinfo {year}
  {2016})}\BibitemShut {NoStop}%
\bibitem [{\citenamefont {Krishna~Kumar}\ \emph {et~al.}(2017)\citenamefont
  {Krishna~Kumar}, \citenamefont {Bandurin}, \citenamefont {Pellegrino},
  \citenamefont {Cao}, \citenamefont {Principi}, \citenamefont {Guo},
  \citenamefont {Auton}, \citenamefont {Ben~Shalom}, \citenamefont
  {Ponomarenko}, \citenamefont {Falkovich} \emph
  {et~al.}}]{krishna2017superballistic}%
  \BibitemOpen
  \bibfield  {author} {\bibinfo {author} {\bibfnamefont {R.}~\bibnamefont
  {Krishna~Kumar}}, \bibinfo {author} {\bibfnamefont {D.}~\bibnamefont
  {Bandurin}}, \bibinfo {author} {\bibfnamefont {F.}~\bibnamefont
  {Pellegrino}}, \bibinfo {author} {\bibfnamefont {Y.}~\bibnamefont {Cao}},
  \bibinfo {author} {\bibfnamefont {A.}~\bibnamefont {Principi}}, \bibinfo
  {author} {\bibfnamefont {H.}~\bibnamefont {Guo}}, \bibinfo {author}
  {\bibfnamefont {G.}~\bibnamefont {Auton}}, \bibinfo {author} {\bibfnamefont
  {M.}~\bibnamefont {Ben~Shalom}}, \bibinfo {author} {\bibfnamefont
  {L.}~\bibnamefont {Ponomarenko}}, \bibinfo {author} {\bibfnamefont
  {G.}~\bibnamefont {Falkovich}}, \emph {et~al.},\ }\bibfield  {title}
  {\bibinfo {title} {Superballistic flow of viscous electron fluid through
  graphene constrictions},\ }\href@noop {} {\bibfield  {journal} {\bibinfo
  {journal} {Nature Physics}\ }\textbf {\bibinfo {volume} {13}},\ \bibinfo
  {pages} {1182} (\bibinfo {year} {2017})}\BibitemShut {NoStop}%
\bibitem [{\citenamefont {Bandurin}\ \emph {et~al.}(2018)\citenamefont
  {Bandurin}, \citenamefont {Shytov}, \citenamefont {Levitov}, \citenamefont
  {Kumar}, \citenamefont {Berdyugin}, \citenamefont {Ben~Shalom}, \citenamefont
  {Grigorieva}, \citenamefont {Geim},\ and\ \citenamefont
  {Falkovich}}]{BandurinF18}%
  \BibitemOpen
  \bibfield  {author} {\bibinfo {author} {\bibfnamefont {D.~A.}\ \bibnamefont
  {Bandurin}}, \bibinfo {author} {\bibfnamefont {A.~V.}\ \bibnamefont
  {Shytov}}, \bibinfo {author} {\bibfnamefont {L.~S.}\ \bibnamefont {Levitov}},
  \bibinfo {author} {\bibfnamefont {R.~K.}\ \bibnamefont {Kumar}}, \bibinfo
  {author} {\bibfnamefont {A.~I.}\ \bibnamefont {Berdyugin}}, \bibinfo {author}
  {\bibfnamefont {M.}~\bibnamefont {Ben~Shalom}}, \bibinfo {author}
  {\bibfnamefont {I.~V.}\ \bibnamefont {Grigorieva}}, \bibinfo {author}
  {\bibfnamefont {A.~K.}\ \bibnamefont {Geim}},\ and\ \bibinfo {author}
  {\bibfnamefont {G.}~\bibnamefont {Falkovich}},\ }\bibfield  {title} {\bibinfo
  {title} {Fluidity onset in graphene},\ }\href
  {https://www.nature.com/articles/s41467-018-07004-4} {\bibfield  {journal}
  {\bibinfo  {journal} {Nat. Commun.}\ }\textbf {\bibinfo {volume} {9}},\
  \bibinfo {pages} {4533} (\bibinfo {year} {2018})}\BibitemShut {NoStop}%
\bibitem [{\citenamefont {Gallagher}\ \emph {et~al.}(2019)\citenamefont
  {Gallagher}, \citenamefont {Yang}, \citenamefont {Lyu}, \citenamefont {Tian},
  \citenamefont {Kou}, \citenamefont {Zhang}, \citenamefont {Watanabe},
  \citenamefont {Taniguchi},\ and\ \citenamefont {Wang}}]{Patrick2019}%
  \BibitemOpen
  \bibfield  {author} {\bibinfo {author} {\bibfnamefont {P.}~\bibnamefont
  {Gallagher}}, \bibinfo {author} {\bibfnamefont {C.-S.}\ \bibnamefont {Yang}},
  \bibinfo {author} {\bibfnamefont {T.}~\bibnamefont {Lyu}}, \bibinfo {author}
  {\bibfnamefont {F.}~\bibnamefont {Tian}}, \bibinfo {author} {\bibfnamefont
  {R.}~\bibnamefont {Kou}}, \bibinfo {author} {\bibfnamefont {H.}~\bibnamefont
  {Zhang}}, \bibinfo {author} {\bibfnamefont {K.}~\bibnamefont {Watanabe}},
  \bibinfo {author} {\bibfnamefont {T.}~\bibnamefont {Taniguchi}},\ and\
  \bibinfo {author} {\bibfnamefont {F.}~\bibnamefont {Wang}},\ }\bibfield
  {title} {\bibinfo {title} {Quantum-critical conductivity of the dirac fluid
  in graphene},\ }\href {https://doi.org/10.1126/science.aat8687} {\bibfield
  {journal} {\bibinfo  {journal} {Science}\ }\textbf {\bibinfo {volume}
  {364}},\ \bibinfo {pages} {158} (\bibinfo {year} {2019})},\ \Eprint
  {https://arxiv.org/abs/https://www.science.org/doi/pdf/10.1126/science.aat8687}
  {https://www.science.org/doi/pdf/10.1126/science.aat8687} \BibitemShut
  {NoStop}%
\bibitem [{\citenamefont {Sulpizio}\ \emph {et~al.}(2019)\citenamefont
  {Sulpizio}, \citenamefont {Ella}, \citenamefont {Rozen}, \citenamefont
  {Birkbeck}, \citenamefont {Perello}, \citenamefont {Dutta}, \citenamefont
  {Ben-Shalom}, \citenamefont {Taniguchi}, \citenamefont {Watanabe},
  \citenamefont {Holder} \emph {et~al.}}]{sulpizio2019visualizing}%
  \BibitemOpen
  \bibfield  {author} {\bibinfo {author} {\bibfnamefont {J.~A.}\ \bibnamefont
  {Sulpizio}}, \bibinfo {author} {\bibfnamefont {L.}~\bibnamefont {Ella}},
  \bibinfo {author} {\bibfnamefont {A.}~\bibnamefont {Rozen}}, \bibinfo
  {author} {\bibfnamefont {J.}~\bibnamefont {Birkbeck}}, \bibinfo {author}
  {\bibfnamefont {D.~J.}\ \bibnamefont {Perello}}, \bibinfo {author}
  {\bibfnamefont {D.}~\bibnamefont {Dutta}}, \bibinfo {author} {\bibfnamefont
  {M.}~\bibnamefont {Ben-Shalom}}, \bibinfo {author} {\bibfnamefont
  {T.}~\bibnamefont {Taniguchi}}, \bibinfo {author} {\bibfnamefont
  {K.}~\bibnamefont {Watanabe}}, \bibinfo {author} {\bibfnamefont
  {T.}~\bibnamefont {Holder}}, \emph {et~al.},\ }\bibfield  {title} {\bibinfo
  {title} {Visualizing poiseuille flow of hydrodynamic electrons},\ }\href@noop
  {} {\bibfield  {journal} {\bibinfo  {journal} {Nature}\ }\textbf {\bibinfo
  {volume} {576}},\ \bibinfo {pages} {75} (\bibinfo {year} {2019})}\BibitemShut
  {NoStop}%
\bibitem [{\citenamefont {Berdyugin}\ \emph {et~al.}(2019)\citenamefont
  {Berdyugin}, \citenamefont {Xu}, \citenamefont {Pellegrino}, \citenamefont
  {Kumar}, \citenamefont {Principi}, \citenamefont {Torre}, \citenamefont
  {Shalom}, \citenamefont {Taniguchi}, \citenamefont {Watanabe}, \citenamefont
  {Grigorieva}, \citenamefont {Polini}, \citenamefont {Geim},\ and\
  \citenamefont {Bandurin}}]{ABerdyugin2019}%
  \BibitemOpen
  \bibfield  {author} {\bibinfo {author} {\bibfnamefont {A.~I.}\ \bibnamefont
  {Berdyugin}}, \bibinfo {author} {\bibfnamefont {S.~G.}\ \bibnamefont {Xu}},
  \bibinfo {author} {\bibfnamefont {F.~M.~D.}\ \bibnamefont {Pellegrino}},
  \bibinfo {author} {\bibfnamefont {R.~K.}\ \bibnamefont {Kumar}}, \bibinfo
  {author} {\bibfnamefont {A.}~\bibnamefont {Principi}}, \bibinfo {author}
  {\bibfnamefont {I.}~\bibnamefont {Torre}}, \bibinfo {author} {\bibfnamefont
  {M.~B.}\ \bibnamefont {Shalom}}, \bibinfo {author} {\bibfnamefont
  {T.}~\bibnamefont {Taniguchi}}, \bibinfo {author} {\bibfnamefont
  {K.}~\bibnamefont {Watanabe}}, \bibinfo {author} {\bibfnamefont {I.~V.}\
  \bibnamefont {Grigorieva}}, \bibinfo {author} {\bibfnamefont
  {M.}~\bibnamefont {Polini}}, \bibinfo {author} {\bibfnamefont {A.~K.}\
  \bibnamefont {Geim}},\ and\ \bibinfo {author} {\bibfnamefont {D.~A.}\
  \bibnamefont {Bandurin}},\ }\bibfield  {title} {\bibinfo {title} {Measuring
  hall viscosity of graphene’s electron fluid},\ }\href
  {https://doi.org/10.1126/science.aau0685} {\bibfield  {journal} {\bibinfo
  {journal} {Science}\ }\textbf {\bibinfo {volume} {364}},\ \bibinfo {pages}
  {162} (\bibinfo {year} {2019})},\ \Eprint
  {https://arxiv.org/abs/https://www.science.org/doi/pdf/10.1126/science.aau0685}
  {https://www.science.org/doi/pdf/10.1126/science.aau0685} \BibitemShut
  {NoStop}%
\bibitem [{\citenamefont {Ella}\ \emph {et~al.}(2019)\citenamefont {Ella},
  \citenamefont {Rozen}, \citenamefont {Birkbeck}, \citenamefont {Ben-Shalom},
  \citenamefont {Perello}, \citenamefont {Zultak}, \citenamefont {Taniguchi},
  \citenamefont {Watanabe}, \citenamefont {Geim}, \citenamefont {Ilani} \emph
  {et~al.}}]{Ella2019}%
  \BibitemOpen
  \bibfield  {author} {\bibinfo {author} {\bibfnamefont {L.}~\bibnamefont
  {Ella}}, \bibinfo {author} {\bibfnamefont {A.}~\bibnamefont {Rozen}},
  \bibinfo {author} {\bibfnamefont {J.}~\bibnamefont {Birkbeck}}, \bibinfo
  {author} {\bibfnamefont {M.}~\bibnamefont {Ben-Shalom}}, \bibinfo {author}
  {\bibfnamefont {D.}~\bibnamefont {Perello}}, \bibinfo {author} {\bibfnamefont
  {J.}~\bibnamefont {Zultak}}, \bibinfo {author} {\bibfnamefont
  {T.}~\bibnamefont {Taniguchi}}, \bibinfo {author} {\bibfnamefont
  {K.}~\bibnamefont {Watanabe}}, \bibinfo {author} {\bibfnamefont {A.~K.}\
  \bibnamefont {Geim}}, \bibinfo {author} {\bibfnamefont {S.}~\bibnamefont
  {Ilani}}, \emph {et~al.},\ }\bibfield  {title} {\bibinfo {title}
  {Simultaneous voltage and current density imaging of flowing electrons in two
  dimensions},\ }\href {https://arxiv.org/pdf/1810.10744.pdf} {\bibfield
  {journal} {\bibinfo  {journal} {Nat. Nanotechnol.}\ }\textbf {\bibinfo
  {volume} {14}},\ \bibinfo {pages} {480} (\bibinfo {year} {2019})}\BibitemShut
  {NoStop}%
\bibitem [{\citenamefont {Ku}\ \emph {et~al.}(2020)\citenamefont {Ku},
  \citenamefont {Zhou}, \citenamefont {Li}, \citenamefont {Shin}, \citenamefont
  {Shi}, \citenamefont {Burch}, \citenamefont {Anderson}, \citenamefont
  {Pierce}, \citenamefont {Xie}, \citenamefont {Hamo} \emph
  {et~al.}}]{ku2020imaging}%
  \BibitemOpen
  \bibfield  {author} {\bibinfo {author} {\bibfnamefont {M.~J.}\ \bibnamefont
  {Ku}}, \bibinfo {author} {\bibfnamefont {T.~X.}\ \bibnamefont {Zhou}},
  \bibinfo {author} {\bibfnamefont {Q.}~\bibnamefont {Li}}, \bibinfo {author}
  {\bibfnamefont {Y.~J.}\ \bibnamefont {Shin}}, \bibinfo {author}
  {\bibfnamefont {J.~K.}\ \bibnamefont {Shi}}, \bibinfo {author} {\bibfnamefont
  {C.}~\bibnamefont {Burch}}, \bibinfo {author} {\bibfnamefont {L.~E.}\
  \bibnamefont {Anderson}}, \bibinfo {author} {\bibfnamefont {A.~T.}\
  \bibnamefont {Pierce}}, \bibinfo {author} {\bibfnamefont {Y.}~\bibnamefont
  {Xie}}, \bibinfo {author} {\bibfnamefont {A.}~\bibnamefont {Hamo}}, \emph
  {et~al.},\ }\bibfield  {title} {\bibinfo {title} {Imaging viscous flow of the
  dirac fluid in graphene},\ }\href@noop {} {\bibfield  {journal} {\bibinfo
  {journal} {Nature}\ }\textbf {\bibinfo {volume} {583}},\ \bibinfo {pages}
  {537} (\bibinfo {year} {2020})}\BibitemShut {NoStop}%
\bibitem [{\citenamefont {Crossno}\ \emph {et~al.}(2016)\citenamefont
  {Crossno}, \citenamefont {Shi}, \citenamefont {Wang}, \citenamefont {Liu},
  \citenamefont {Harzheim}, \citenamefont {Lucas}, \citenamefont {Sachdev},
  \citenamefont {Kim}, \citenamefont {Taniguchi}, \citenamefont {Watanabe}
  \emph {et~al.}}]{crossno2016observation}%
  \BibitemOpen
  \bibfield  {author} {\bibinfo {author} {\bibfnamefont {J.}~\bibnamefont
  {Crossno}}, \bibinfo {author} {\bibfnamefont {J.~K.}\ \bibnamefont {Shi}},
  \bibinfo {author} {\bibfnamefont {K.}~\bibnamefont {Wang}}, \bibinfo {author}
  {\bibfnamefont {X.}~\bibnamefont {Liu}}, \bibinfo {author} {\bibfnamefont
  {A.}~\bibnamefont {Harzheim}}, \bibinfo {author} {\bibfnamefont
  {A.}~\bibnamefont {Lucas}}, \bibinfo {author} {\bibfnamefont
  {S.}~\bibnamefont {Sachdev}}, \bibinfo {author} {\bibfnamefont
  {P.}~\bibnamefont {Kim}}, \bibinfo {author} {\bibfnamefont {T.}~\bibnamefont
  {Taniguchi}}, \bibinfo {author} {\bibfnamefont {K.}~\bibnamefont {Watanabe}},
  \emph {et~al.},\ }\bibfield  {title} {\bibinfo {title} {Observation of the
  dirac fluid and the breakdown of the wiedemann-franz law in graphene},\
  }\href@noop {} {\bibfield  {journal} {\bibinfo  {journal} {Science}\ }\textbf
  {\bibinfo {volume} {351}},\ \bibinfo {pages} {1058} (\bibinfo {year}
  {2016})}\BibitemShut {NoStop}%
\bibitem [{\citenamefont {Block}\ \emph {et~al.}(2021)\citenamefont {Block},
  \citenamefont {Principi}, \citenamefont {Hesp}, \citenamefont {Cummings},
  \citenamefont {Liebel}, \citenamefont {Watanabe}, \citenamefont {Taniguchi},
  \citenamefont {Roche}, \citenamefont {Koppens}, \citenamefont {van Hulst}
  \emph {et~al.}}]{block2021observation}%
  \BibitemOpen
  \bibfield  {author} {\bibinfo {author} {\bibfnamefont {A.}~\bibnamefont
  {Block}}, \bibinfo {author} {\bibfnamefont {A.}~\bibnamefont {Principi}},
  \bibinfo {author} {\bibfnamefont {N.~C.}\ \bibnamefont {Hesp}}, \bibinfo
  {author} {\bibfnamefont {A.~W.}\ \bibnamefont {Cummings}}, \bibinfo {author}
  {\bibfnamefont {M.}~\bibnamefont {Liebel}}, \bibinfo {author} {\bibfnamefont
  {K.}~\bibnamefont {Watanabe}}, \bibinfo {author} {\bibfnamefont
  {T.}~\bibnamefont {Taniguchi}}, \bibinfo {author} {\bibfnamefont
  {S.}~\bibnamefont {Roche}}, \bibinfo {author} {\bibfnamefont {F.~H.}\
  \bibnamefont {Koppens}}, \bibinfo {author} {\bibfnamefont {N.~F.}\
  \bibnamefont {van Hulst}}, \emph {et~al.},\ }\bibfield  {title} {\bibinfo
  {title} {Observation of giant and tunable thermal diffusivity of a dirac
  fluid at room temperature},\ }\href@noop {} {\bibfield  {journal} {\bibinfo
  {journal} {Nature Nanotechnology}\ }\textbf {\bibinfo {volume} {16}},\
  \bibinfo {pages} {1195} (\bibinfo {year} {2021})}\BibitemShut {NoStop}%
\bibitem [{\citenamefont {M\"uller}\ \emph {et~al.}(2009)\citenamefont
  {M\"uller}, \citenamefont {Schmalian},\ and\ \citenamefont
  {Fritz}}]{MuSchmalian2009}%
  \BibitemOpen
  \bibfield  {author} {\bibinfo {author} {\bibfnamefont {M.}~\bibnamefont
  {M\"uller}}, \bibinfo {author} {\bibfnamefont {J.}~\bibnamefont
  {Schmalian}},\ and\ \bibinfo {author} {\bibfnamefont {L.}~\bibnamefont
  {Fritz}},\ }\bibfield  {title} {\bibinfo {title} {Graphene: A nearly perfect
  fluid},\ }\href {https://doi.org/10.1103/PhysRevLett.103.025301} {\bibfield
  {journal} {\bibinfo  {journal} {Phys. Rev. Lett.}\ }\textbf {\bibinfo
  {volume} {103}},\ \bibinfo {pages} {025301} (\bibinfo {year}
  {2009})}\BibitemShut {NoStop}%
\bibitem [{\citenamefont {Narozhny}\ \emph {et~al.}(2015)\citenamefont
  {Narozhny}, \citenamefont {Gornyi}, \citenamefont {Titov}, \citenamefont
  {Sch\"utt},\ and\ \citenamefont {Mirlin}}]{NarozhnyBN2015}%
  \BibitemOpen
  \bibfield  {author} {\bibinfo {author} {\bibfnamefont {B.~N.}\ \bibnamefont
  {Narozhny}}, \bibinfo {author} {\bibfnamefont {I.~V.}\ \bibnamefont
  {Gornyi}}, \bibinfo {author} {\bibfnamefont {M.}~\bibnamefont {Titov}},
  \bibinfo {author} {\bibfnamefont {M.}~\bibnamefont {Sch\"utt}},\ and\
  \bibinfo {author} {\bibfnamefont {A.~D.}\ \bibnamefont {Mirlin}},\ }\bibfield
   {title} {\bibinfo {title} {Hydrodynamics in graphene: Linear-response
  transport},\ }\href {https://doi.org/10.1103/PhysRevB.91.035414} {\bibfield
  {journal} {\bibinfo  {journal} {Phys. Rev. B}\ }\textbf {\bibinfo {volume}
  {91}},\ \bibinfo {pages} {035414} (\bibinfo {year} {2015})}\BibitemShut
  {NoStop}%
\bibitem [{\citenamefont {Hartnoll}\ \emph {et~al.}(2007)\citenamefont
  {Hartnoll}, \citenamefont {Kovtun}, \citenamefont {Müller},\ and\
  \citenamefont {Sachdev}}]{Hartnoll_2007}%
  \BibitemOpen
  \bibfield  {author} {\bibinfo {author} {\bibfnamefont {S.~A.}\ \bibnamefont
  {Hartnoll}}, \bibinfo {author} {\bibfnamefont {P.~K.}\ \bibnamefont
  {Kovtun}}, \bibinfo {author} {\bibfnamefont {M.}~\bibnamefont {Müller}},\
  and\ \bibinfo {author} {\bibfnamefont {S.}~\bibnamefont {Sachdev}},\
  }\bibfield  {title} {\bibinfo {title} {Theory of the nernst effect near
  quantum phase transitions in condensed matter and in dyonic black holes},\
  }\bibfield  {journal} {\bibinfo  {journal} {Physical Review B}\ }\textbf
  {\bibinfo {volume} {76}},\ \href {https://doi.org/10.1103/physrevb.76.144502}
  {10.1103/physrevb.76.144502} (\bibinfo {year} {2007})\BibitemShut {NoStop}%
\bibitem [{\citenamefont {Majumdar}\ \emph {et~al.}(2025)\citenamefont
  {Majumdar}, \citenamefont {Chadha}, \citenamefont {Pal}, \citenamefont
  {Gugnani}, \citenamefont {Ghawri}, \citenamefont {Watanabe}, \citenamefont
  {Taniguchi}, \citenamefont {Mukerjee},\ and\ \citenamefont
  {Ghosh}}]{majumdar2025universalityquantumcriticalflow}%
  \BibitemOpen
  \bibfield  {author} {\bibinfo {author} {\bibfnamefont {A.}~\bibnamefont
  {Majumdar}}, \bibinfo {author} {\bibfnamefont {N.}~\bibnamefont {Chadha}},
  \bibinfo {author} {\bibfnamefont {P.}~\bibnamefont {Pal}}, \bibinfo {author}
  {\bibfnamefont {A.}~\bibnamefont {Gugnani}}, \bibinfo {author} {\bibfnamefont
  {B.}~\bibnamefont {Ghawri}}, \bibinfo {author} {\bibfnamefont
  {K.}~\bibnamefont {Watanabe}}, \bibinfo {author} {\bibfnamefont
  {T.}~\bibnamefont {Taniguchi}}, \bibinfo {author} {\bibfnamefont
  {S.}~\bibnamefont {Mukerjee}},\ and\ \bibinfo {author} {\bibfnamefont
  {A.}~\bibnamefont {Ghosh}},\ }\href {https://arxiv.org/abs/2501.03193}
  {\bibinfo {title} {Universality in quantum critical flow of charge and heat
  in ultra-clean graphene}} (\bibinfo {year} {2025}),\ \Eprint
  {https://arxiv.org/abs/2501.03193} {arXiv:2501.03193 [cond-mat.mes-hall]}
  \BibitemShut {NoStop}%
\bibitem [{\citenamefont {Narozhny}(2022)}]{Narozhny:2022ncn}%
  \BibitemOpen
  \bibfield  {author} {\bibinfo {author} {\bibfnamefont {B.~N.}\ \bibnamefont
  {Narozhny}},\ }\bibfield  {title} {\bibinfo {title} {{Hydrodynamic approach
  to two-dimensional electron systems}},\ }\href
  {https://doi.org/10.1007/s40766-022-00036-z} {\bibfield  {journal} {\bibinfo
  {journal} {Riv. Nuovo Cim.}\ }\textbf {\bibinfo {volume} {45}},\ \bibinfo
  {pages} {661} (\bibinfo {year} {2022})},\ \Eprint
  {https://arxiv.org/abs/2207.10004} {arXiv:2207.10004 [cond-mat.mes-hall]}
  \BibitemShut {NoStop}%
\bibitem [{\citenamefont {Narozhny}(2019)}]{Narozhny2019uib}%
  \BibitemOpen
  \bibfield  {author} {\bibinfo {author} {\bibfnamefont {B.~N.}\ \bibnamefont
  {Narozhny}},\ }\bibfield  {title} {\bibinfo {title} {{Electronic
  hydrodynamics in graphene}},\ }\href
  {https://doi.org/10.1016/j.aop.2019.167979} {\bibfield  {journal} {\bibinfo
  {journal} {Annals Phys.}\ }\textbf {\bibinfo {volume} {411}},\ \bibinfo
  {pages} {167979} (\bibinfo {year} {2019})},\ \Eprint
  {https://arxiv.org/abs/1905.09686} {arXiv:1905.09686 [cond-mat.mes-hall]}
  \BibitemShut {NoStop}%
\bibitem [{\citenamefont {Lucas}\ and\ \citenamefont
  {Fong}(2018)}]{Lucasfong2018}%
  \BibitemOpen
  \bibfield  {author} {\bibinfo {author} {\bibfnamefont {A.}~\bibnamefont
  {Lucas}}\ and\ \bibinfo {author} {\bibfnamefont {K.~C.}\ \bibnamefont
  {Fong}},\ }\bibfield  {title} {\bibinfo {title} {Hydrodynamics of electrons
  in graphene},\ }\href {https://doi.org/10.1088/1361-648X/aaa274} {\bibfield
  {journal} {\bibinfo  {journal} {Journal of Physics: Condensed Matter}\
  }\textbf {\bibinfo {volume} {30}},\ \bibinfo {pages} {053001} (\bibinfo
  {year} {2018})}\BibitemShut {NoStop}%
\bibitem [{\citenamefont {M\"uller}\ and\ \citenamefont
  {Sachdev}(2008)}]{MMuller2008}%
  \BibitemOpen
  \bibfield  {author} {\bibinfo {author} {\bibfnamefont {M.}~\bibnamefont
  {M\"uller}}\ and\ \bibinfo {author} {\bibfnamefont {S.}~\bibnamefont
  {Sachdev}},\ }\bibfield  {title} {\bibinfo {title} {Collective cyclotron
  motion of the relativistic plasma in graphene},\ }\href
  {https://doi.org/10.1103/PhysRevB.78.115419} {\bibfield  {journal} {\bibinfo
  {journal} {Phys. Rev. B}\ }\textbf {\bibinfo {volume} {78}},\ \bibinfo
  {pages} {115419} (\bibinfo {year} {2008})}\BibitemShut {NoStop}%
\bibitem [{\citenamefont {Fritz}\ \emph {et~al.}(2008)\citenamefont {Fritz},
  \citenamefont {Schmalian}, \citenamefont {M\"uller},\ and\ \citenamefont
  {Sachdev}}]{Fritz2008}%
  \BibitemOpen
  \bibfield  {author} {\bibinfo {author} {\bibfnamefont {L.}~\bibnamefont
  {Fritz}}, \bibinfo {author} {\bibfnamefont {J.}~\bibnamefont {Schmalian}},
  \bibinfo {author} {\bibfnamefont {M.}~\bibnamefont {M\"uller}},\ and\
  \bibinfo {author} {\bibfnamefont {S.}~\bibnamefont {Sachdev}},\ }\bibfield
  {title} {\bibinfo {title} {Quantum critical transport in clean graphene},\
  }\href {https://doi.org/10.1103/PhysRevB.78.085416} {\bibfield  {journal}
  {\bibinfo  {journal} {Phys. Rev. B}\ }\textbf {\bibinfo {volume} {78}},\
  \bibinfo {pages} {085416} (\bibinfo {year} {2008})}\BibitemShut {NoStop}%
\bibitem [{\citenamefont {M\"uller}\ \emph {et~al.}(2008)\citenamefont
  {M\"uller}, \citenamefont {Fritz},\ and\ \citenamefont
  {Sachdev}}]{Markus2008}%
  \BibitemOpen
  \bibfield  {author} {\bibinfo {author} {\bibfnamefont {M.}~\bibnamefont
  {M\"uller}}, \bibinfo {author} {\bibfnamefont {L.}~\bibnamefont {Fritz}},\
  and\ \bibinfo {author} {\bibfnamefont {S.}~\bibnamefont {Sachdev}},\
  }\bibfield  {title} {\bibinfo {title} {Quantum-critical relativistic
  magnetotransport in graphene},\ }\href
  {https://doi.org/10.1103/PhysRevB.78.115406} {\bibfield  {journal} {\bibinfo
  {journal} {Phys. Rev. B}\ }\textbf {\bibinfo {volume} {78}},\ \bibinfo
  {pages} {115406} (\bibinfo {year} {2008})}\BibitemShut {NoStop}%
\bibitem [{\citenamefont {Foster}\ and\ \citenamefont
  {Aleiner}(2009)}]{Foster2009}%
  \BibitemOpen
  \bibfield  {author} {\bibinfo {author} {\bibfnamefont {M.~S.}\ \bibnamefont
  {Foster}}\ and\ \bibinfo {author} {\bibfnamefont {I.~L.}\ \bibnamefont
  {Aleiner}},\ }\bibfield  {title} {\bibinfo {title} {Slow imbalance relaxation
  and thermoelectric transport in graphene},\ }\href
  {https://doi.org/10.1103/PhysRevB.79.085415} {\bibfield  {journal} {\bibinfo
  {journal} {Phys. Rev. B}\ }\textbf {\bibinfo {volume} {79}},\ \bibinfo
  {pages} {085415} (\bibinfo {year} {2009})}\BibitemShut {NoStop}%
\bibitem [{\citenamefont {Mendoza}\ \emph {et~al.}(2013)\citenamefont
  {Mendoza}, \citenamefont {Herrmann},\ and\ \citenamefont
  {Succi}}]{mendoza2013hydrodynamic}%
  \BibitemOpen
  \bibfield  {author} {\bibinfo {author} {\bibfnamefont {M.}~\bibnamefont
  {Mendoza}}, \bibinfo {author} {\bibfnamefont {H.~J.}\ \bibnamefont
  {Herrmann}},\ and\ \bibinfo {author} {\bibfnamefont {S.}~\bibnamefont
  {Succi}},\ }\bibfield  {title} {\bibinfo {title} {Hydrodynamic model for
  conductivity in graphene},\ }\href@noop {} {\bibfield  {journal} {\bibinfo
  {journal} {Scientific reports}\ }\textbf {\bibinfo {volume} {3}},\ \bibinfo
  {pages} {1052} (\bibinfo {year} {2013})}\BibitemShut {NoStop}%
\bibitem [{\citenamefont {Lucas}(2016)}]{ALucas2016}%
  \BibitemOpen
  \bibfield  {author} {\bibinfo {author} {\bibfnamefont {A.}~\bibnamefont
  {Lucas}},\ }\bibfield  {title} {\bibinfo {title} {Sound waves and resonances
  in electron-hole plasma},\ }\bibfield  {journal} {\bibinfo  {journal}
  {Physical Review B}\ }\textbf {\bibinfo {volume} {93}},\ \href
  {https://doi.org/10.1103/physrevb.93.245153} {10.1103/physrevb.93.245153}
  (\bibinfo {year} {2016})\BibitemShut {NoStop}%
\bibitem [{\citenamefont {Aung}\ \emph {et~al.}(2023)\citenamefont {Aung},
  \citenamefont {Win}, \citenamefont {Khandal},\ and\ \citenamefont
  {Ghosh}}]{Aung:2023vrr}%
  \BibitemOpen
  \bibfield  {author} {\bibinfo {author} {\bibfnamefont {C.~W.}\ \bibnamefont
  {Aung}}, \bibinfo {author} {\bibfnamefont {T.~Z.}\ \bibnamefont {Win}},
  \bibinfo {author} {\bibfnamefont {G.}~\bibnamefont {Khandal}},\ and\ \bibinfo
  {author} {\bibfnamefont {S.}~\bibnamefont {Ghosh}},\ }\bibfield  {title}
  {\bibinfo {title} {{Shear viscosity expression for a graphene system in
  relaxation time approximation}},\ }\href
  {https://doi.org/10.1103/PhysRevB.108.235172} {\bibfield  {journal} {\bibinfo
   {journal} {Phys. Rev. B}\ }\textbf {\bibinfo {volume} {108}},\ \bibinfo
  {pages} {235172} (\bibinfo {year} {2023})},\ \Eprint
  {https://arxiv.org/abs/2306.14747} {arXiv:2306.14747 [nucl-th]} \BibitemShut
  {NoStop}%
\bibitem [{\citenamefont {Janicke}(1977)}]{Janicke_1977}%
  \BibitemOpen
  \bibfield  {author} {\bibinfo {author} {\bibfnamefont {L.}~\bibnamefont
  {Janicke}},\ }\bibfield  {title} {\bibinfo {title} {Non-linear
  electromagnetic waves in a relativistic plasma},\ }\href
  {https://doi.org/10.1088/0032-1028/19/3/002} {\bibfield  {journal} {\bibinfo
  {journal} {Plasma Physics}\ }\textbf {\bibinfo {volume} {19}},\ \bibinfo
  {pages} {209} (\bibinfo {year} {1977})}\BibitemShut {NoStop}%
\bibitem [{\citenamefont {Palmroth}\ \emph {et~al.}(2018)\citenamefont
  {Palmroth}, \citenamefont {Ganse}, \citenamefont {Pfau-Kempf}, \citenamefont
  {Battarbee}, \citenamefont {Turc}, \citenamefont {Brito}, \citenamefont
  {Grandin}, \citenamefont {Hoilijoki}, \citenamefont {Sandroos},\ and\
  \citenamefont {von Alfthan}}]{Palmroth_2018}%
  \BibitemOpen
  \bibfield  {author} {\bibinfo {author} {\bibfnamefont {M.}~\bibnamefont
  {Palmroth}}, \bibinfo {author} {\bibfnamefont {U.}~\bibnamefont {Ganse}},
  \bibinfo {author} {\bibfnamefont {Y.}~\bibnamefont {Pfau-Kempf}}, \bibinfo
  {author} {\bibfnamefont {M.}~\bibnamefont {Battarbee}}, \bibinfo {author}
  {\bibfnamefont {L.}~\bibnamefont {Turc}}, \bibinfo {author} {\bibfnamefont
  {T.}~\bibnamefont {Brito}}, \bibinfo {author} {\bibfnamefont
  {M.}~\bibnamefont {Grandin}}, \bibinfo {author} {\bibfnamefont
  {S.}~\bibnamefont {Hoilijoki}}, \bibinfo {author} {\bibfnamefont
  {A.}~\bibnamefont {Sandroos}},\ and\ \bibinfo {author} {\bibfnamefont
  {S.}~\bibnamefont {von Alfthan}},\ }\bibfield  {title} {\bibinfo {title}
  {Vlasov methods in space physics and astrophysics},\ }\bibfield  {journal}
  {\bibinfo  {journal} {Living Reviews in Computational Astrophysics}\ }\textbf
  {\bibinfo {volume} {4}},\ \href {https://doi.org/10.1007/s41115-018-0003-2}
  {10.1007/s41115-018-0003-2} (\bibinfo {year} {2018})\BibitemShut {NoStop}%
\bibitem [{\citenamefont {Elskens}\ and\ \citenamefont
  {Kiessling}(2020)}]{Elskens_2020}%
  \BibitemOpen
  \bibfield  {author} {\bibinfo {author} {\bibfnamefont {Y.}~\bibnamefont
  {Elskens}}\ and\ \bibinfo {author} {\bibfnamefont {M.~K.-H.}\ \bibnamefont
  {Kiessling}},\ }\bibfield  {title} {\bibinfo {title} {Microscopic foundations
  of kinetic plasma theory: The relativistic vlasov–maxwell equations and
  their radiation-reaction-corrected generalization},\ }\href
  {https://doi.org/10.1007/s10955-020-02519-x} {\bibfield  {journal} {\bibinfo
  {journal} {Journal of Statistical Physics}\ }\textbf {\bibinfo {volume}
  {180}},\ \bibinfo {pages} {749–772} (\bibinfo {year} {2020})}\BibitemShut
  {NoStop}%
\bibitem [{\citenamefont {Kushwah}\ and\ \citenamefont
  {Denicol}(2024)}]{Kushwah:2024zgd}%
  \BibitemOpen
  \bibfield  {author} {\bibinfo {author} {\bibfnamefont {K.}~\bibnamefont
  {Kushwah}}\ and\ \bibinfo {author} {\bibfnamefont {G.~S.}\ \bibnamefont
  {Denicol}},\ }\bibfield  {title} {\bibinfo {title} {{Relativistic dissipative
  magnetohydrodynamics from the Boltzmann equation for a two-component gas}},\
  }\href {https://doi.org/10.1103/PhysRevD.109.096021} {\bibfield  {journal}
  {\bibinfo  {journal} {Phys. Rev. D}\ }\textbf {\bibinfo {volume} {109}},\
  \bibinfo {pages} {096021} (\bibinfo {year} {2024})},\ \Eprint
  {https://arxiv.org/abs/2402.01597} {arXiv:2402.01597 [nucl-th]} \BibitemShut
  {NoStop}%
\bibitem [{\citenamefont {Panda}\ \emph
  {et~al.}(2021{\natexlab{a}})\citenamefont {Panda}, \citenamefont {Dash},
  \citenamefont {Biswas},\ and\ \citenamefont {Roy}}]{Panda:2020zhr}%
  \BibitemOpen
  \bibfield  {author} {\bibinfo {author} {\bibfnamefont {A.~K.}\ \bibnamefont
  {Panda}}, \bibinfo {author} {\bibfnamefont {A.}~\bibnamefont {Dash}},
  \bibinfo {author} {\bibfnamefont {R.}~\bibnamefont {Biswas}},\ and\ \bibinfo
  {author} {\bibfnamefont {V.}~\bibnamefont {Roy}},\ }\bibfield  {title}
  {\bibinfo {title} {{Relativistic non-resistive viscous magnetohydrodynamics
  from the kinetic theory: a relaxation time approach}},\ }\href
  {https://doi.org/10.1007/JHEP03(2021)216} {\bibfield  {journal} {\bibinfo
  {journal} {JHEP}\ }\textbf {\bibinfo {volume} {03}},\ \bibinfo {pages}
  {216}},\ \Eprint {https://arxiv.org/abs/2011.01606} {arXiv:2011.01606
  [nucl-th]} \BibitemShut {NoStop}%
\bibitem [{\citenamefont {Panda}\ \emph
  {et~al.}(2021{\natexlab{b}})\citenamefont {Panda}, \citenamefont {Dash},
  \citenamefont {Biswas},\ and\ \citenamefont {Roy}}]{Panda:2021pvq}%
  \BibitemOpen
  \bibfield  {author} {\bibinfo {author} {\bibfnamefont {A.~K.}\ \bibnamefont
  {Panda}}, \bibinfo {author} {\bibfnamefont {A.}~\bibnamefont {Dash}},
  \bibinfo {author} {\bibfnamefont {R.}~\bibnamefont {Biswas}},\ and\ \bibinfo
  {author} {\bibfnamefont {V.}~\bibnamefont {Roy}},\ }\bibfield  {title}
  {\bibinfo {title} {{Relativistic resistive dissipative magnetohydrodynamics
  from the relaxation time approximation}},\ }\href
  {https://doi.org/10.1103/PhysRevD.104.054004} {\bibfield  {journal} {\bibinfo
   {journal} {Phys. Rev. D}\ }\textbf {\bibinfo {volume} {104}},\ \bibinfo
  {pages} {054004} (\bibinfo {year} {2021}{\natexlab{b}})},\ \Eprint
  {https://arxiv.org/abs/2104.12179} {arXiv:2104.12179 [nucl-th]} \BibitemShut
  {NoStop}%
\bibitem [{\citenamefont {Anderson}\ and\ \citenamefont
  {Witting}(1974)}]{ANDERSON1974466}%
  \BibitemOpen
  \bibfield  {author} {\bibinfo {author} {\bibfnamefont {J.}~\bibnamefont
  {Anderson}}\ and\ \bibinfo {author} {\bibfnamefont {H.}~\bibnamefont
  {Witting}},\ }\bibfield  {title} {\bibinfo {title} {A relativistic
  relaxation-time model for the boltzmann equation},\ }\href
  {https://doi.org/https://doi.org/10.1016/0031-8914(74)90355-3} {\bibfield
  {journal} {\bibinfo  {journal} {Physica}\ }\textbf {\bibinfo {volume} {74}},\
  \bibinfo {pages} {466} (\bibinfo {year} {1974})}\BibitemShut {NoStop}%
\bibitem [{\citenamefont {Rocha}\ \emph {et~al.}(2024)\citenamefont {Rocha},
  \citenamefont {Wagner}, \citenamefont {Denicol}, \citenamefont {Noronha},\
  and\ \citenamefont {Rischke}}]{Rocha:2023ilf}%
  \BibitemOpen
  \bibfield  {author} {\bibinfo {author} {\bibfnamefont {G.~S.}\ \bibnamefont
  {Rocha}}, \bibinfo {author} {\bibfnamefont {D.}~\bibnamefont {Wagner}},
  \bibinfo {author} {\bibfnamefont {G.~S.}\ \bibnamefont {Denicol}}, \bibinfo
  {author} {\bibfnamefont {J.}~\bibnamefont {Noronha}},\ and\ \bibinfo {author}
  {\bibfnamefont {D.~H.}\ \bibnamefont {Rischke}},\ }\bibfield  {title}
  {\bibinfo {title} {{Theories of Relativistic Dissipative Fluid Dynamics}},\
  }\href {https://doi.org/10.3390/e26030189} {\bibfield  {journal} {\bibinfo
  {journal} {Entropy}\ }\textbf {\bibinfo {volume} {26}},\ \bibinfo {pages}
  {189} (\bibinfo {year} {2024})},\ \Eprint {https://arxiv.org/abs/2311.15063}
  {arXiv:2311.15063 [nucl-th]} \BibitemShut {NoStop}%
\bibitem [{\citenamefont {Rocha}\ and\ \citenamefont
  {Denicol}(2021)}]{Rocha:2021lze}%
  \BibitemOpen
  \bibfield  {author} {\bibinfo {author} {\bibfnamefont {G.~S.}\ \bibnamefont
  {Rocha}}\ and\ \bibinfo {author} {\bibfnamefont {G.~S.}\ \bibnamefont
  {Denicol}},\ }\bibfield  {title} {\bibinfo {title} {{Transient fluid dynamics
  with general matching conditions: A first study from the method of
  moments}},\ }\href {https://doi.org/10.1103/PhysRevD.104.096016} {\bibfield
  {journal} {\bibinfo  {journal} {Phys. Rev. D}\ }\textbf {\bibinfo {volume}
  {104}},\ \bibinfo {pages} {096016} (\bibinfo {year} {2021})},\ \Eprint
  {https://arxiv.org/abs/2108.02187} {arXiv:2108.02187 [nucl-th]} \BibitemShut
  {NoStop}%
\bibitem [{\citenamefont {Dwibedi}\ \emph {et~al.}(2024)\citenamefont
  {Dwibedi}, \citenamefont {Padhan}, \citenamefont {Chatterjee},\ and\
  \citenamefont {Ghosh}}]{Dwibedi:2024mff}%
  \BibitemOpen
  \bibfield  {author} {\bibinfo {author} {\bibfnamefont {A.}~\bibnamefont
  {Dwibedi}}, \bibinfo {author} {\bibfnamefont {N.}~\bibnamefont {Padhan}},
  \bibinfo {author} {\bibfnamefont {A.}~\bibnamefont {Chatterjee}},\ and\
  \bibinfo {author} {\bibfnamefont {S.}~\bibnamefont {Ghosh}},\ }\bibfield
  {title} {\bibinfo {title} {{Transport Coefficients of Relativistic Matter: A
  Detailed Formalism with a Gross Knowledge of Their Magnitude}},\ }\href
  {https://doi.org/10.3390/universe10030132} {\bibfield  {journal} {\bibinfo
  {journal} {Universe}\ }\textbf {\bibinfo {volume} {10}},\ \bibinfo {pages}
  {132} (\bibinfo {year} {2024})},\ \Eprint {https://arxiv.org/abs/2404.01421}
  {arXiv:2404.01421 [nucl-th]} \BibitemShut {NoStop}%
\bibitem [{\citenamefont {De~Groot}(1980)}]{DeGroot:1980dk}%
  \BibitemOpen
  \bibfield  {author} {\bibinfo {author} {\bibfnamefont {S.~R.}\ \bibnamefont
  {De~Groot}},\ }\href@noop {} {\emph {\bibinfo {title} {{Relativistic Kinetic
  Theory. Principles and Applications}}}},\ edited by\ \bibinfo {editor}
  {\bibfnamefont {W.~A.}\ \bibnamefont {Van~Leeuwen}}\ and\ \bibinfo {editor}
  {\bibfnamefont {C.~G.}\ \bibnamefont {Van~Weert}}\ (\bibinfo {year}
  {1980})\BibitemShut {NoStop}%
\bibitem [{\citenamefont {Cercignani}\ and\ \citenamefont
  {Kremer}(2002)}]{Cercignani2002}%
  \BibitemOpen
  \bibfield  {author} {\bibinfo {author} {\bibfnamefont {C.}~\bibnamefont
  {Cercignani}}\ and\ \bibinfo {author} {\bibfnamefont {G.~M.}\ \bibnamefont
  {Kremer}},\ }\bibinfo {title} {Thermomechanics of relativistic fluids},\ in\
  \href {https://doi.org/10.1007/978-3-0348-8165-4_4} {\emph {\bibinfo
  {booktitle} {The Relativistic Boltzmann Equation: Theory and Applications}}}\
  (\bibinfo  {publisher} {Birkh{\"a}user Basel},\ \bibinfo {address} {Basel},\
  \bibinfo {year} {2002})\ pp.\ \bibinfo {pages} {99--110}\BibitemShut
  {NoStop}%
\bibitem [{\citenamefont {Jaiswal}(2013)}]{Jaiswal:2013npa}%
  \BibitemOpen
  \bibfield  {author} {\bibinfo {author} {\bibfnamefont {A.}~\bibnamefont
  {Jaiswal}},\ }\bibfield  {title} {\bibinfo {title} {{Relativistic dissipative
  hydrodynamics from kinetic theory with relaxation time approximation}},\
  }\href {https://doi.org/10.1103/PhysRevC.87.051901} {\bibfield  {journal}
  {\bibinfo  {journal} {Phys. Rev. C}\ }\textbf {\bibinfo {volume} {87}},\
  \bibinfo {pages} {051901} (\bibinfo {year} {2013})},\ \Eprint
  {https://arxiv.org/abs/1302.6311} {arXiv:1302.6311 [nucl-th]} \BibitemShut
  {NoStop}%
\bibitem [{\citenamefont {Itoh}(1970)}]{10.1143/PTP.44.291}%
  \BibitemOpen
  \bibfield  {author} {\bibinfo {author} {\bibfnamefont {N.}~\bibnamefont
  {Itoh}},\ }\bibfield  {title} {\bibinfo {title} {{Hydrostatic Equilibrium of
  Hypothetical Quark Stars}},\ }\href {https://doi.org/10.1143/PTP.44.291}
  {\bibfield  {journal} {\bibinfo  {journal} {Progress of Theoretical Physics}\
  }\textbf {\bibinfo {volume} {44}},\ \bibinfo {pages} {291} (\bibinfo {year}
  {1970})},\ \Eprint
  {https://arxiv.org/abs/https://academic.oup.com/ptp/article-pdf/44/1/291/5357142/44-1-291.pdf}
  {https://academic.oup.com/ptp/article-pdf/44/1/291/5357142/44-1-291.pdf}
  \BibitemShut {NoStop}%
\bibitem [{\citenamefont {Lee}\ and\ \citenamefont
  {Wick}(1974)}]{PhysRevD.9.2291}%
  \BibitemOpen
  \bibfield  {author} {\bibinfo {author} {\bibfnamefont {T.~D.}\ \bibnamefont
  {Lee}}\ and\ \bibinfo {author} {\bibfnamefont {G.~C.}\ \bibnamefont {Wick}},\
  }\bibfield  {title} {\bibinfo {title} {Vacuum stability and vacuum excitation
  in a spin-0 field theory},\ }\href {https://doi.org/10.1103/PhysRevD.9.2291}
  {\bibfield  {journal} {\bibinfo  {journal} {Phys. Rev. D}\ }\textbf {\bibinfo
  {volume} {9}},\ \bibinfo {pages} {2291} (\bibinfo {year} {1974})}\BibitemShut
  {NoStop}%
\bibitem [{\citenamefont {Collins}\ and\ \citenamefont
  {Perry}(1975)}]{PhysRevLett.34.1353}%
  \BibitemOpen
  \bibfield  {author} {\bibinfo {author} {\bibfnamefont {J.~C.}\ \bibnamefont
  {Collins}}\ and\ \bibinfo {author} {\bibfnamefont {M.~J.}\ \bibnamefont
  {Perry}},\ }\bibfield  {title} {\bibinfo {title} {Superdense matter: Neutrons
  or asymptotically free quarks?},\ }\href
  {https://doi.org/10.1103/PhysRevLett.34.1353} {\bibfield  {journal} {\bibinfo
   {journal} {Phys. Rev. Lett.}\ }\textbf {\bibinfo {volume} {34}},\ \bibinfo
  {pages} {1353} (\bibinfo {year} {1975})}\BibitemShut {NoStop}%
\end{thebibliography}%
\end{document}